\documentclass[12pt]{article}

\usepackage{amssymb}
\usepackage{epsf}

\makeatletter
\def\eqnumsection{\@addtoreset{equation}{section}
    \def\theequation{\arabic{section}.\arabic{equation}}}
\def\appendixes{\par\setcounter{section}{0}
    \setcounter{subsection}{0}\setcounter{equation}{0}
    \@addtoreset{equation}{section}
    \def\thesection{Appendix \Alph{section}}
    \def\theequation{\Alph{section}.\arabic{equation}}}
\makeatother

\def\slash#1{#1 \hskip-0.45em /}

\begin{document}

\hyphenation{theo-re-ti-cal}

\begin{flushright}
DESY 01-068 \\
WUE-ITP-01-029\\
YARU-HE-01/02 \\
October 2001\\
\end{flushright}

\begin{center}
\centerline{\large\bf Branching Ratios for
 $B \to K^* \gamma$ and $B \to \rho \gamma$ Decays in}
\vspace*{2mm}
\centerline{\large\bf 
Next-to-Leading Order in the Large Energy Effective Theory}

\vspace*{1.5cm}

{\large A.~Ali}
\vskip0.2cm
Deutsches Elektronen Synchrotron DESY, Hamburg\\
\vspace*{0.3cm}
\centerline{and}
\vspace*{0.3cm}
{\large A.Ya.~Parkhomenko}
\vskip0.2cm 
Institut f$\ddot{\rm u}$r Theoretische Physik, 
Universit$\ddot{\rm a}$t W$\ddot{\rm u}$rzburg, \\ 
D-97074 W$\ddot{\rm u}$rzburg, Germany 
\vskip0.2cm
and
\vskip0.2cm 
Department of Theoretical Physics,
Yaroslavl State University \\
Sovietskaya 14, 150000 Yaroslavl, Russia
\vskip0.5cm

\vskip0.5cm
{\Large Abstract\\}
\vskip3truemm

\parbox[t]{\textwidth}{
We calculate the so-called hard spectator corrections
in ${\cal O} (\alpha_s)$ in the leading-twist approximation
to the decay widths for $B \to K^{*} \gamma$ and
$B \to \rho \gamma$  decays and their charge conjugates, using the Large
Energy Effective Theory (LEET) techniques. Combined with the
hard vertex and annihilation contributions, they are used to compute the
branching ratios for these decays in the next-to-leading order
(NLO) in the strong coupling~$\alpha_s$ and in leading power in
$\Lambda_{\rm QCD}/M_B$. These corrections are found to be large,
leading to the inference that the theoretical branching ratios for the
decays $B \to K^* \gamma$ in the  LEET approach can be reconciled with
current data only for significantly lower values of the form factors than
their estimates in the QCD sum rule and Lattice QCD approaches.
However, the form factor related uncertainties mostly cancel in the
ratios ${\cal B}(B \to \rho \gamma)/{\cal B}(B \to K^* \gamma)$ and  
$\Delta = (\Delta^{+0}+ \Delta^{-0})/2$, where
$\Delta^{\pm 0} = \Gamma (B^\pm \to \rho^\pm \gamma)/
     [2 \Gamma (B^0 (\bar B^0)\to \rho^0 \gamma)] - 1$, and hence 
their measurements will provide quantitative information on the
standard model parameters, in particular the ratio of the 
CKM matrix elements $\vert V_{td}/V_{ts}\vert$ and the inner angle
$\alpha$ of the CKM-unitarity triangle. We also calculate direct CP
asymmetries for the decays $B^\pm \to \rho^\pm \gamma$
and $B^0/\bar B^0 \to \rho^0 \gamma$ and find, in
conformity with the observations made in the existing literature, that 
the hard spectator contributions significantly reduce the asymmetries 
arising from the vertex corrections. In addition,
the sensitivity of the CP asymmetries on the underlying
parameters is found to be discomfortingly large.
}

\end{center}

\thispagestyle{empty}
\newpage
\setcounter{page}{1}
\textheight 23.0 true cm
%

\eqnumsection

\section{Introduction}
\label{sec:Introd}

There exists a lot of theoretical interest in measuring the branching
ratios for the inclusive radiative decays $B \to X_s \gamma$ and
$B \to X_d \gamma$. The corresponding exclusive
radiative decays $B \to K^{*} \gamma$ and  $B \to \rho \gamma$, and
related decays involving higher $K^*$ and $\rho$-resonances, are
experimentally more tractable but theoretically less clean.
In particular, the form factors entering in these decays have to
be determined from a non-perturbative approach such as lattice-QCD or
QCD sum rules. Alternatively, these form factors can be related to the
ones in the semileptonic decays $B \to \rho \ell
\nu_\ell$ using heavy quark symmetry and determined from data on the
semileptonic decays. As the heavy quark symmetry is broken by perturbative
QCD and non-perturbative power corrections, these effects will have to be
taken into account at some level. This should enable us in principle to
predict the branching ratios in radiative decays in a theoretically
controlled way. In the Standard Model (SM), measurements of the radiative
decays in question, as well as their semileptonic counterparts
$B \to (K,K^*, \pi,\rho) \ell^+ \ell^-$, will constrain the matrix elements
of the Cabibbo-Kobayashi-Maskawa (CKM) matrix~\cite{Cabibbo:1963yz}.
In particular, the ratios of the branching ratios ${\cal B}(B \to \rho
\gamma)/{\cal B}(B \to K^* \gamma)$ would provide independent and
complementary information on the CKM matrix element ratio
$\vert V_{td}/V_{ts}\vert$. Likewise,  the isospin-violating ratios 
$\Delta^{\pm 0} = \Gamma (B^\pm \to \rho^\pm \gamma)]/
  [2 \Gamma (B^0 (\bar B^0)\to \rho^0 \gamma)] - 1$
and the CP-asymmetry in the rate difference 
${\cal A}_{\rm CP} (\rho^\pm \gamma) = 
 [\Gamma (B^- \to \rho^- \gamma) - \Gamma (B^+ \to \rho^+\gamma)]/
 [\Gamma (B^- \to \rho^- \gamma) + \Gamma (B^+ \to \rho^+ \gamma)]$
will determine the angle~$\alpha$, which is one of the three inner angles
of the CKM-unitarity triangle. They are also sensitive to the presence of
physics beyond the SM, such as supersymmetry~\cite{Ali:2000zu,Ali:2001ej}.
It is therefore imperative to firm up theoretical predictions in
exclusive decays $B \to V \gamma^{(*)}$, with $V=\rho$ or $K^*$, for
precision tests of the~SM and to interpret data for possible new physics
effects in these decays.

To compute the branching ratios reliably, one needs to calculate 
at least the explicit $O (\alpha_s)$ improvements to the lowest order
decay widths and take into account the leading power corrections in a
well-defined theoretical framework, such as the heavy quark
effective theory (HQET). More specifically, theoretically
improved radiative decay widths for $B \to V \gamma$, 
require the calculation of the renormalization group effects in the
appropriate Wilson  coefficients in the effective
Hamiltonian~\cite{Chetyrkin:1997vx}, an explicit $O(\alpha_s)$ calculation 
of the matrix elements involving the hard vertex 
corrections~\cite{Soares:1991te,Greub:1995tb,Greub:1996tg}, annihilation
contributions~\cite{Ali:1995uy,Khodjamirian:1995uc,Grinstein:2000pc},
which are more important in the decays $B \to \rho \gamma$, 
and the so-called hard-spectator contributions involving (virtual) hard 
gluon radiative corrections off the spectator quarks in the~$B$-, $K^*$-, 
and~$\rho$-mesons~\cite{Beneke:1999br,Beneke:2001wa}. These corrections
will shift the theoretical branching ratios and induce CP-asymmetries in the
decay rates, where the latter are expected to be measurable only in the
CKM-suppressed decays $B \to \rho \gamma$ in the SM. In addition, the 
annihilation and the hard gluon radiative corrections explicitly break
isospin symmetry, leading generically to non-zero values for
$\Delta^{\pm 0}$ in $B \to \rho \gamma$ decays, and to a lesser extent
also for the $B \to K^* \gamma$ decays.
While the former have been calculated in the lowest order
in Refs.~\cite{Ali:1995uy,Khodjamirian:1995uc},
and the explicit $O (\alpha_s)$ corrections to the leading-twist
(twist-two) annihilation amplitudes are found to vanish in the chiral
limit~\cite{Grinstein:2000pc}, the commensurate contributions from the
hard spectator diagrams have to be included in the complete
$O (\alpha_s)$-improved estimates. In
this paper we compute these corrections to the leading-twist meson
distribution amplitudes, borrowing techniques from the so-called Large
Energy Effective Theory (LEET)~\cite{Dugan:1991de,Charles:1999dr}. In
doing this, we correct several errors in the derivation of the decay
widths for $B \to \rho \gamma$, presented in the earlier version of this
paper, and discuss in addition the decays $B \to K^* \gamma$ at some
length in view of its current experimental interest.
In a closely related context, a part of these corrections were calculated
some time ago by Beneke and Feldmann~\cite{Beneke:2001wa}. With the
remaining contribution
of the hard spectator corrections presented here, and in the
meanwhile also reported by Beneke, Feldmann and
Seidel~\cite{Beneke:2001at}, and by Bosch and
Buchalla~\cite{Bosch:2001gv}, the decay rates for $B \to K^* \gamma$ and
$B \to \rho \gamma$ are now quantified in the LEET approach, up to and
including the NLO corrections in $\alpha_s$ and to leading power in
$\Lambda_{\rm QCD}/M$, where $M$ is the $B$-meson mass, in the
leading-twist approximation. These predictions have to be confronted with
data, which we undertake at some length in this paper.

 We use the $O (\alpha_s)$-improved 
estimates for the decay rate for $B \to K^* \gamma$, presented here, the
corresponding theoretical
results for the inclusive decay rate for $B \to X_s \gamma$, obtained in
Refs.~\cite{Chetyrkin:1997vx,Kagan:1999ym,Gambino:2001ew},
and current data on the branching ratios for $B \to K^* \gamma$
\cite{Chen:2001fj,Tajima:2001qp,Aubert:2001} and $B \to X_s \gamma$
\cite{cleo,aleph,belle} to determine the form factor
$\xi_\perp^{(K^*)}(0)$ in the LEET
approach. This yields  $\xi_\perp^{(K^*)} (0) = 0.25 \pm 0.04$, 
which is similar, though not identical, to the result
$\xi_\perp^{(K^*)} (0) = 0.24 \pm 0.06$ obtained in
Ref.~\cite{Beneke:2001at} in the same framework, using the experimental
branching ratio for $B \to K^* \gamma$ only.  Relating the LEET-theory
form factor
$\xi_\perp^{(K^*)} (0)$ to the full QCD form factor $T_1^{(K^*)}(0)$,
with the help of the $O(\alpha_s)$-relation calculated in
Ref.~\cite{Beneke:2001wa}, yields $T_1^{(K^*)} (0) = 0.27 \pm 0.04$. 
This is to be compared with a typical estimate in  the light-cone QCD 
(LC-QCD) sum rule, 
$T_1^{(K^*)} (0) = 0.38 \pm 0.05$~\cite{Ball:1998kk,Ali:2000mm} 
and from the lattice QCD simulations
$T_1^{(K^*)} (0) = 0.32^{+0.04}_{-0.02}$~\cite{DelDebbio:1998kr}.
Thus, the form factor $T_1^{(K^*)} (0)$ in the LEET approach is found to
be smaller compared to the  values obtained in the other two methods. At
this stage, the source of this mismatch is not well understood.

What concerns the decay $B \to \rho \gamma$, we combine the current
determination of $\xi_\perp^{(K^*)}(0)$ in the LEET
approach with an estimate of the SU(3)-breaking effects in the form
factors, using a light-cone QCD sum rule
result for this purpose, $\xi_\perp^{(\rho)}/\xi_\perp^{(K^*)} =
T_1^{(\rho)}(0)/T_1^{(K^*)}(0) \simeq 0.76 \pm 0.06$~\cite{Ali:1994vd},
yielding $\xi_\perp^{(\rho)} = 0.19 \pm 0.04$. This allows us to calculate
the branching ratios for the decays $B^0 \to \rho^0 \gamma$, 
$B^+ \to \rho^+ \gamma$ and their charge conjugates. However, as we show
by explicit calculations in this paper, a parametrically more stable
quantity for this purpose is the ratio
 ${\cal B}(B \to \rho \gamma)/{\cal B}(B \to K^* \gamma)$.
 Theoretically, this ratio can be expressed as 
\begin{displaymath}
\frac{{\cal B}_{\rm th} (B \to \rho \gamma)}
     {{\cal B}_{\rm th} (B \to K^* \gamma)} =
S_\rho \left | \frac{V_{td}}{V_{ts}} \right |^2
\frac{(1 - m_\rho^2/M^2)^3}{(1 - m_{K^*}^2/M^2)^3} \,
\zeta^2 \, \left[ 1 + \Delta R (\rho/K^*) \right ],
\end{displaymath} 
where $\zeta = \xi_\perp^{(\rho)} (0) / \xi_\perp^{(K^*)} (0)$
is the ratio of the HQET/LEET form factors, $S_\rho = 1 (1/2)$ is 
the isospin weight for the $\rho^\pm$- ($\rho^0$-) meson, and the
dominant dependence on the CKM matrix elements is made explicit.
We calculate~$\Delta R (\rho/K^*)$,
to leading order in~$\alpha_s$ and~$\Lambda_{\rm QCD}/M$, including the
leading order
annihilation contributions in $B \to \rho \gamma$ decays, and study
its sensitivity to the underlying input parameters. 
Knowing $\Delta R (\rho/K^*)$ and~$\zeta$, the branching ratio for 
$B \to \rho \gamma$ can be predicted in terms of the already known
branching ratios for $B \to K^* \gamma$.   Averaged over the 
charge conjugates, we find $\bar {\cal B}_{\rm th} (B^\pm \to \rho^\pm 
\gamma) = [0.91 \pm 0.33 {(\rm th)} \pm 0.11 {\rm (exp)}] \times 10^{-6}$ 
and $\bar {\cal B}_{\rm th} (B^0/\bar{B}^0 \to \rho^0 \gamma) = 
[0.50 \pm 0.18 {(\rm th)} \pm 0.04 {\rm (exp)}] \times 10^{-6}$,
where the theoretical uncertainty is dominated by the current dispersion
on the CKM parameters and the meson wave functions. The experimental 
uncertainty enters through the  present measurements of the branching ratios 
for $B \to K^* \gamma$.
 The isospin-violating ratios~$\Delta^{\pm 0}$ and the charge
conjugate averaged ratio $\Delta = (\Delta^{+0} + \Delta^{-0})/2$
are also calculated to the stated level of theoretical accuracy. The
resulting corrections are found to be small in $\Delta$, in particular in
the allowed CKM parameter range determined from the CKM unitarity fits in
the SM.

Finally, we compute the leading order 
CP-asymmetry ${\cal A}_{\rm CP} (\rho^\pm \gamma)$ involving the decays 
$B^\pm \to \rho^\pm \gamma$ and the direct CP-asymmetry component in  
${\cal A}_{\rm CP} (\rho^0 \gamma)$,
involving the decays $B^0/\bar B^0 \to \rho^0 \gamma$. These
CP-asymmetries arise
due to the interference of the various penguin amplitudes which have
clashing weak phases, with the required strong interaction phase
provided by the~${\cal O} (\alpha_s)$ corrections entering the penguin
amplitudes via the Bander-Silverman-Soni 
(BSS) mechanism~\cite{Bander:1980jx}. We find that the hard spectator
corrections significantly reduce the CP-asymmetries calculated from the
vertex contribution alone in $B \to \rho \gamma$
decays, qualitatively in line with the observation made by Bosch and
Buchalla~\cite{Bosch:2001gv}. However, this cancellation, and the
resulting CP asymmetries, depend rather
sensitively on the ratio of the quark masses~$m_c/m_b$ and the
annihilation contributions. This parametric
dependence, combined with the scale dependence of ${\cal A}_{\rm CP}
(\rho^\pm \gamma)$ and ${\cal A}_{\rm CP} (\rho^0 \gamma)$, also
discussed in Ref.~\cite{Bosch:2001gv},
makes the prediction of direct CP-asymmetries rather unreliable, as we
show explicitly in this paper.

This paper is organized as follows: In section~\ref{sec:LEEF}, we introduce
the underlying theoretical framework (LEET) and the relations involving
the $B \to V$ ($V = \rho, K^*$) form factors resulting from the LEET
symmetry, and sketch the explicit 
$O (\alpha_s)$-breaking of these relations. The hard scattering 
amplitude involving the spectator diagrams in $B \to V \gamma$ decays are 
calculated in section~\ref{sec:HSA}. Explicit forms of the
$O (\alpha_s)$-corrected matrix elements for these decays are
given in section~\ref{sec:ME}. Numerical results for the branching ratios
for $B \to K^* \gamma$ and $B \to \rho \gamma$ are presented in 
section~\ref{sec:ratios}. Isospin-violating ratios and the charge
conjugate averaged ratio~$\Delta$ for the decays $B \to \rho \gamma$,
and the CP-violating asymmetry ${\cal A}_{\rm CP} (\rho^\pm \gamma)$
and the direct CP-asymmetry component in ${\cal A}_{\rm CP} (\rho^0
\gamma)$ are given in section~\ref{sec:asymmetries}. 
We conclude with a brief summary and some concluding remarks in
section~\ref{sec:concl}.

\section{LEET Symmetry and Symmetry Breaking in\\ Perturbative QCD}
\label{sec:LEEF}

For the sake of definiteness, we shall work out explicitly 
the decays $B \to \rho \gamma$; the differences between these and the
decays $B \to K^* \gamma$ lie mainly in the CKM matrix elements and
in the wave functions of the final-state hadrons, and they will
be specified in sections~4 and~5. The effective Hamiltonian for the 
$B \to \rho \gamma$ decays (equivalently $b \to d \gamma$ decay) at
the scale $\mu = O (m_b)$, where~$m_b$ is the $b$-quark mass, is given by
\begin{eqnarray}
{\cal H}_{\rm eff} & = & \frac{G_F}{\sqrt 2} \,
\left \{
V_{ub} V_{ud}^* \,
\left [
C_1^{(u)} (\mu) \, {\cal O}_1^{(u)} (\mu) +
C_2^{(u)} (\mu) \, {\cal O}_2^{(u)} (\mu)
\right ]
\right.
\label{eq:eff-ham} \\
&& \qquad + \,
V_{cb} V_{cd}^* \,
\left [
C_1^{(c)} (\mu) \, {\cal O}_1^{(c)} (\mu) +
C_2^{(c)} (\mu) \, {\cal O}_2^{(c)} (\mu)
\right ]
\nonumber \\
&& \qquad - \,
\left.
V_{tb} V_{td}^* \,
\left [
C_7^{\rm eff} (\mu) \, {\cal O}_7 (\mu) +
C_8^{\rm eff} (\mu) \, {\cal O}_8 (\mu)
\right ]
+ \ldots
\right \} ,
\nonumber
\end{eqnarray}
where we have shown the contributions which will be important in our
calculations. Operators~${\cal O}_1^{(q)}$ and~${\cal O}_2^{(q)}$, 
$(q=u,c)$, are the standard four-fermion operators:
\begin{equation}
{\cal O}_1^{(q)} =
(\bar d_\alpha \gamma_\mu (1 - \gamma_5) q_\beta) \,
(\bar q_\beta \gamma^\mu (1 - \gamma_5) b_\alpha) ,
\qquad
{\cal O}_2^{(q)} =
(\bar d_\alpha \gamma_\mu (1 - \gamma_5) q_\alpha) \,
(\bar q_\beta \gamma^\mu (1 - \gamma_5) b_\beta) ,
\label{eq:four-Fermi}
\end{equation}
and~${\cal O}_7$ and~${\cal O}_8$ are the  electromagnetic and
chromomagnetic penguin operators, respectively:
\begin{equation}
{\cal O}_7 = \frac{e m_b}{8 \pi^2} \,
(\bar d_\alpha \sigma^{\mu \nu} (1 + \gamma_5) b_\alpha) \,
F_{\mu \nu} ,
\qquad
{\cal O}_8 = \frac{g_s m_b}{8 \pi^2} \,
(\bar d_\alpha \sigma^{\mu \nu} (1 + \gamma_5)
T^a_{\alpha \beta} b_\beta) \, G^a_{\mu \nu} .
\label{eq:mag-penguin}
\end{equation}
Here, $e$ and $g_s$ are the electric and colour charges,
$F_{\mu \nu}$ and $G^a_{\mu \nu}$ are the electromagnetic and gluonic
field strength tensors, respectively,
$T^a_{\alpha \beta}$ are the colour $SU (N_c)$ group generators, and the
quark colour indices~$\alpha$ and~$\beta$ and gluonic colour index~$a$
are written explicitly. Note that in the operators~${\cal O}_7$
and~${\cal O}_8$ the $d$-quark mass contributions are negligible and
therefore omitted. The coefficients~$C_1^{(q)} (\mu)$ and~$C_2^{(q)} (\mu)$
in Eq.~(\ref{eq:eff-ham}) are the usual Wilson coefficients
corresponding to the operators~${\cal O}^{(q)}_1$ and~${\cal O}^{(q)}_2$
while the coefficients~$C_7^{\rm eff} (\mu)$ and~$C_8^{\rm eff}
(\mu)$ include also the effects of the QCD penguin four-fermion
operators ${\cal O}_5$ and~${\cal O}_6$ which are assumed to be
present in the effective Hamiltonian~(\ref{eq:eff-ham}) and
denoted by ellipses there. For details and numerical values of
these coefficients, see~\cite{Buchalla:1996vs} and reference
therein. We use the standard Bjorken-Drell
convention~\cite{Bjorken:1965} for the metric and the Dirac
matrices; in particular
$\gamma_5 = i \gamma^0 \gamma^1 \gamma^2 \gamma^3$,
and the totally antisymmetric Levi-Civita tensor
$\varepsilon_{\mu \nu \rho \sigma}$ is defined as
$\varepsilon_{0 1 2 3} = + 1$.

The effective Hamiltonian~(\ref{eq:eff-ham}) sandwiched between
the~$B$- and $\rho$-meson states can be expressed in terms of
matrix elements of bilinear quark currents inducing heavy-light
transitions. These matrix elements are dominated by strong
interactions at small momentum transfer and cannot be calculated
perturbatively. The general decomposition of the matrix elements on all 
possible Lorentz structures (Vector, Axial-vector and Tensor) 
admits  seven scalar functions (form factors): $V$, $A_i$, and~$T_i$ 
($i = 1, 2, 3$) of the momentum squared~$q^2$ transferred from the heavy
meson to the light one. When the energy of the final light meson~$E$
is large (the large recoil limit), one can expand the interaction of the
energetic quark in the meson with the soft gluons in terms of 
$\Lambda_{\rm QCD}/E$. 
Using then the effective heavy quark theory for the interaction of the 
heavy $b$-quark with the gluons, one can derive non-trivial relations
between the soft contributions to the form factors \cite{Charles:1999dr}.
The resulting theory (LEET) reduces the number of
independent form factors from seven in the $B \to \rho$ transitions to two
in this limit. The relations among the form factors in the
symmetry limit are broken by perturbative  QCD radiative corrections
arising from the vertex renormalization and the hard spectator interactions. 
To incorporate  both types of QCD corrections, a tentative
factorization formula for the heavy-light form factors at large
recoil and at leading order in the inverse heavy meson mass was
introduced in Ref.~\cite{Beneke:2001wa}:
\begin{equation}
f_k (q^2) = C_{\perp k} \xi_\perp +  C_{\| k} \xi_\| +
\Phi_B \otimes T_k \otimes \Phi_\rho ,
\label{eq:fact-formula}
\end{equation}
where $f_k (q^2)$ is any of the seven independent form factors in
the $B \to \rho$ transitions at hand; $\xi_\perp$ and~$\xi_\|$ are
the two independent form factors remaining in the LEET-symmetry
limit; $T_k$ is a hard-scattering kernel calculated in
$O (\alpha_s)$ containing, in general, an end-point divergence
in the decay $B \to \rho \gamma$ which must be regulated somehow;
$\Phi_B$ and~$\Phi_\rho$ are the light-cone distribution
amplitudes of the~$B$- and~$\rho$-meson convoluted with~$T_k$;
$C_k = 1 + O (\alpha_s)$ are the hard vertex renormalization
coefficients. Hard spectator corrections contribute to the
convolution term in Eq.~(\ref{eq:fact-formula}). They break
factorization, implying that their contribution can not be
absorbed in the redefinition of the first two terms, and they are
suppressed by one power of the strong coupling~$\alpha_s$ relative
to the soft contributions defined by~$\xi_\perp$ and~$\xi_\|$. To
compute the hard spectator contribution to the $B \to \rho \gamma$
decay amplitude, one has to assume distribution amplitudes for the
initial and final mesons. To leading order in the inverse
$B$-meson mass, the dominant contribution is from the
leading-twist (twist-two) light-cone distribution amplitudes of the
mesons. In this approach both the $B$- and $\rho$-mesons can be
described by two constituents only, for example, $B^- = (b \bar u)$ 
and $\rho^- = (d \bar u)$, and the higher Fock states involving in
addition gluons are ignored. We show here that the tentative factorization
Ansatz given in Eq.~(\ref{eq:fact-formula}) holds and derive the explicit
corrections to the amplitudes $B \to V \gamma$, where $V=\rho,K^*$ in the
LEET approach. We note that an $O(\alpha_s)$ proof of the validity of
Eq.~(\ref{eq:fact-formula}) has, in the meanwhile, also been 
provided by Beneke, Feldmann and  Seidel~\cite{Beneke:2001at}, and by
Bosch and Buchalla~\cite{Bosch:2001gv}.

We restrict ourselves with the following kinematics involving
quarks~\cite{Beneke:2001wa}:
the momenta of the $b$-quark and the spectator antiquark in the
$B$-meson are
\begin{equation}
p_b^\mu \simeq m_b \, v^\mu,
\qquad
l^\mu = \frac{l_+}{2} \, n_+^\mu + l_\perp^\mu +
\frac{l_-}{2} \, n_-^\mu ,
\label{eq:B-mom}
\end{equation}
and for the quark and antiquark in the $\rho$-meson
we decompose their momentum vectors as follows
\begin{equation}
k_1^\mu \simeq u \, E \, n_-^\mu + k_\perp^\mu + O (k_\perp^2) ,
\qquad
k_2^\mu \simeq \bar u \, E \, n_-^\mu - k_\perp^\mu + O (k_\perp^2) ,
\label{eq:rho-mom}
\end{equation}
where $v^\mu$ is the heavy meson velocity ($v^2 = 1$), $n_-^\mu$
and $n_+^\mu$ are the light-like vectors ($n_\pm^2 = 0$ and $(n_-
n_+) = 2$) parallel to the four-momenta~$p^\mu$ and~$q^\mu$ of the
$\rho$-meson and the photon, respectively, in the approximation when the
effects quadratic in the light meson mass are neglected, so that 
$p^2 = m_\rho^2 \simeq 0$. However, we shall keep the $\rho$-meson mass
in the phase space factor for the decay $B \to \rho \gamma$. In
the above formula~$u$ and $\bar u = 1 - u$ 
are the relative energies of the quark and antiquark, respectively.
In terms of these vectors the $B$-meson four-velocity can be decomposed 
as $v^\mu = (n_-^\mu + n_+^\mu)/2$. Due to the energy-momentum 
conservation in a two-body decay the energy of the $\rho$-meson is 
$E \simeq M/2$, where~$M$ is the $B$-meson mass, as well as the energy 
of the photon $\omega \simeq M/2$ (we assume that $q^\mu = \omega n_+^\mu$).
The four-vectors~$l_\perp^\mu$ and~$k_\perp^\mu$
describe the transverse motion of the light quarks in $B$- and
$\rho$-mesons, respectively, and are of order~$\Lambda_{\rm QCD}$.
In this approach we neglect the internal motion of the $b$-quark
in the $B$-meson, which is also of order~$\Lambda_{\rm QCD}$, and
consider the $b$-quark static in the $B$-meson rest frame (see
Eq.~(\ref{eq:B-mom})). It means that the light antiquark in the
heavy meson do not influence strongly the $B$-meson kinematics,
and its energy is also of order~$\Lambda_{\rm QCD}$ ($l_\pm \sim
\Lambda_{\rm QCD}$), i.e., it is of the same order as its
transverse momentum~$l_\perp^\mu$.

Spectator corrections to the $B \to \rho \gamma$ decay amplitude
can be calculated in the form of a convolution formula, whose leading 
($\sim \alpha_s$) term can be expressed as~\cite{Beneke:2001wa}:
\begin{equation}
\Delta {\cal M}^{\rm (HSA)} = \frac{4 \pi \alpha_s C_F}{N_c} \,
\int\limits_0^1 du \int\limits_0^\infty dl_+ \,
M^{(B)}_{j k} M^{(\rho)}_{l i}
{\cal T}_{i j k l},
\label{eq:conv-formula}
\end{equation}
where $N_c$ is the number of colours, $C_F = (N_c^2 - 1)/(2 N_c)$
is the Casimir operator eigenvalue in the fundamental representation
of the colour $SU (N_c)$ group, and ${\cal T}_{i j k l}$ is the
hard-scattering amplitude which is calculated from the
Feynman diagrams presented in the next section. The colour trace has been
performed, while the Dirac indices~$i, \, j, \, k$, and~$l$ are
written explicitly. The leading-twist two-particle light-cone
projection operators $M^{(B)}_{j k}$~\cite{Ball:1998sk,Beneke:2001wa}
and $M^{(\rho)}_{l i}$~\cite{Grozin:1997pq,Beneke:2001wa} 
of~$B$- and~$\rho$-mesons in the momentum representation are:
\begin{eqnarray}
\hspace*{-10mm} &&
M^{(B)}_{j k} = - \frac{i f_B M}{4}
\left .
\left [
\frac{1 + \slash v}{2}
\left \{
\phi^{(B)}_+ (l_+) \slash n_+ + \phi^{(B)}_- (l_+)
\left (
\slash n_- - l_+ \gamma_\perp^\mu \frac{\partial}{\partial l_\perp^\mu}
\right )
\right \} \gamma_5
\right ]_{j k}
\right |_{l = (l_+/2) n_+} \hspace*{-2mm} ,
\label{eq:PrO-B} \\
\hspace*{-10mm} &&
M^{(\rho)}_{l i} = - \frac{i}{4}
\left [
f_\perp^{(\rho)} \, \slash \varepsilon^* \slash p \, 
\phi_\perp^{(\rho)} (u) +
f_\|^{(\rho)} \, \slash p \, \frac{m}{E} \, (v \varepsilon^*) \, 
\phi_\|^{(\rho)} (u)
\right ]_{l i} ,
\label{eq:PrO-rho}
\end{eqnarray}
where $f_B$ is the $B$-meson decay constant, $f_\|^{(\rho)}$ 
and~$f_\perp^{(\rho)}$ are the longitudinal and transverse $\rho$-meson 
decay constants, respectively, and $\varepsilon_\mu$ is the $\rho$-meson 
polarization vector. These projectors include also the leading-twist 
distribution amplitudes~$\phi^{(B)}_+ (l_+)$ and~$\phi^{(B)}_- (l_+)$ 
of the $B$-meson and~$\phi_\|^{(\rho)} (u)$ and~$\phi_\perp^{(\rho)} (u)$ 
of the $\rho$-meson.

\section{Hard Spectator Contributions in $B \to V \gamma$ Decays}
\label{sec:HSA}

We now present the set of the hard-scattering amplitudes contributing 
to the spectator corrections to the $B \to V \gamma$ decays, where 
$V = \rho, K^*$. These are calculated in $O (\alpha_s)$ based on the 
Feynman diagrams which we show and discuss in this section.

\begin{figure}[bt]
\centerline{
            \epsffile[150 660 410 750]{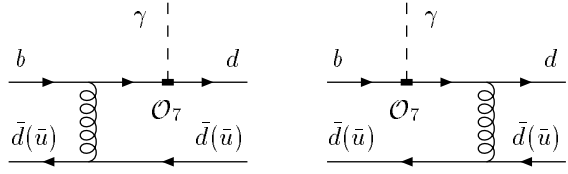}}
\caption{Feynman diagrams contributing to the spectator corrections 
         involving the ${\cal O}_7$ operator in the decay $B \to \rho 
\gamma$. The curly (dashed) line here and in subsequent figures represents
a gluon (photon).}
\label{fig:CorrO7}
\end{figure}

{\it 1. Spectator corrections due to the electromagnetic dipole
operator~${\cal O}_7$.} The corresponding diagrams are presented in
Fig.~\ref{fig:CorrO7}.
%
%
%
%
%
The explicit expression is:
\begin{eqnarray}
{\cal T}^{(1)}_{i j k l} & = &
- i \frac{G_F}{\sqrt 2} V_{td}^* V_{tb} C^{\rm eff}_7 (\mu) \,
\frac{e m_b (\mu)}{4 \pi^2} \,
\frac{[\gamma_\mu]_{k l}}{(l - k_2)^2}
\label{eq:T1} \\
& \times &
\left [
(q \sigma e^*) (1 + \gamma_5) \,
\frac{\slash p_b + \slash l - \slash k_2 + m_b}{(p_b + l - k_2)^2 - m_b^2} \,
\gamma_\mu + \gamma_\mu \,
\frac{\slash k_1 + \slash k_2 - \slash l}{(k_1 + k_2 - l)^2} \,
(q \sigma e^*) (1 + \gamma_5)
\right ]_{i j} ,
\nonumber
\end{eqnarray}
where we have used a short-hand notation
$(q \sigma e^*) = \sigma^{\mu \nu} q_\mu e^*_\nu$.

\begin{figure}[bt]
\centerline{
            \epsffile[150 660 410 750]{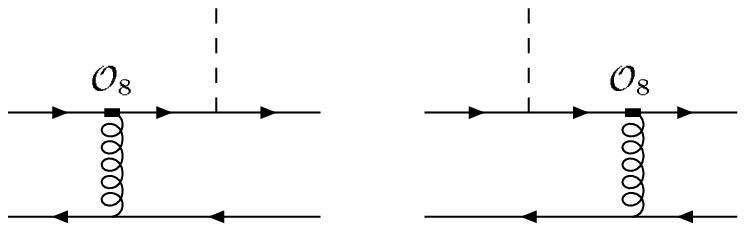}}
\unitlength=1mm
\begin{picture}(0,0)(-80,-30)
\put(0,0){\makebox(0,0)[cc]{\large $a$}}
\end{picture}
\vspace{5mm}
\centerline{
            \epsffile[150 650 410 750]{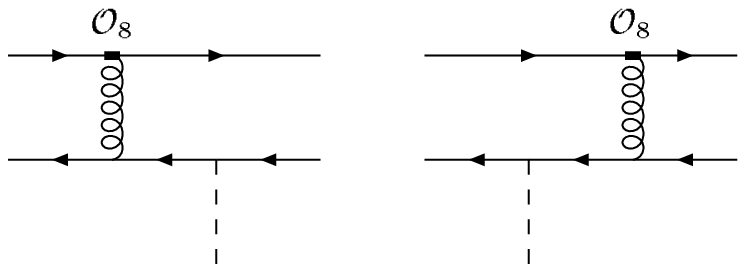}}
\unitlength=1mm
\begin{picture}(0,10)(-80,-5)
\put(0,0){\makebox(0,0)[cc]{\large $b$}}
\end{picture}
\caption{Feynman diagrams contributing to the spectator corrections       
         involving the ${\cal O}_8$ operator in the
decays $B \to V \gamma$. Row~$a$: photon is
         emitted from the flavour-changing quark line; Row~$b$: photon
         radiation off the spectator quark line.} %
\label{fig:CorrO8}
\end{figure}

{\it 2. Spectator corrections due to the chromomagnetic dipole
operator ${\cal O}_8$.} The corresponding diagrams are presented in
Fig.~\ref{fig:CorrO8}.
%
%
%
%
%
%
The top two diagrams (Fig.~\ref{fig:CorrO8}$a$) give the corrections
for the case when the photon is emitted from the flavour-changing quark
line and the result is:
\begin{eqnarray}
{\cal T}^{(2a)}_{i j k l} & = &
- i \frac{G_F}{\sqrt 2} V_{td}^* V_{tb} C_8^{\rm eff} (\mu) \,
\frac{e m_b (\mu)}{12 \pi^2} \,
[\gamma_\nu]_{k l} \,
\frac{(l - k_2)_\mu}{(l - k_2)^2}
\label{eq:T2a} \\
& \times &
\left [
\slash e^* \,
\frac{\slash p_b + \slash l - \slash k_2}{(p_b + l - k_2)^2} \,
\sigma_{\mu \nu} (1 + \gamma_5) + \sigma_{\mu \nu} (1 + \gamma_5) \,
\frac{\slash k_1 + \slash k_2 - \slash l + m_b}{(k_1 + k_2 - l)^2 - m_b^2} \,
\slash e^*
\right ]_{i j} ,
\nonumber
\end{eqnarray}
where the value of the  $b$-quark charge $Q_b = - 1/3$
is taken into account.
The second row (Fig.~\ref{fig:CorrO8}$b$) contains the diagrams
with the photon emission from the spectator quark which results
into the following hard-scattering amplitude:
\begin{eqnarray}
{\cal T}^{(2b)}_{i j k l} & = &
i \frac{G_F}{\sqrt 2} V_{td}^* V_{tb} C_8^{\rm eff} (\mu) \,
\frac{e Q_{d(u)} m_b (\mu)}{4 \pi^2} \,
[\sigma_{\mu \nu} (1 + \gamma_5)]_{i j} \,
\frac{(p_b - k_1)_\mu}{(p_b - k_1)^2}
\label{eq:T2b} \\
& \times &
\left [
\gamma_\nu \,
\frac{\slash p_b + \slash l - \slash k_1}{(p_b + l - k_1)^2} \,
\slash e^* + \slash e^* \,
\frac{\slash k_1 + \slash k_2 - \slash p_b}{(k_1 + k_2 - p_b)^2} \,
\gamma_\nu
\right ]_{k l} .
\nonumber
\end{eqnarray}
Note that this amplitude depends on the spectator quark
charge~$Q_{d (u)}$ and hence is a potential source of isospin symmetry
breaking.

\begin{figure}[bt]
\centerline{
            \epsffile[150 630 410 750]{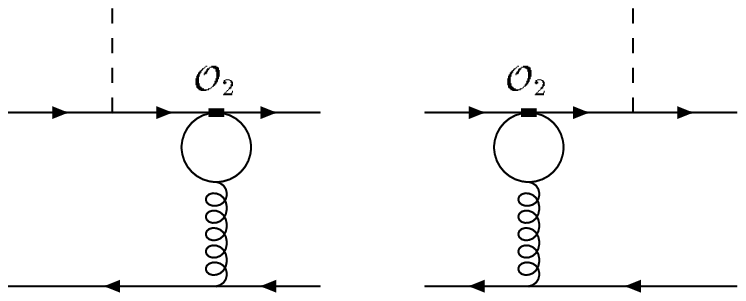}}
\unitlength=1mm
\begin{picture}(0,0)(-80,-40)
\put(0,0){\makebox(0,0)[cc]{\large $a$}}
\end{picture}
\centerline{
            \epsffile[150 630 410 750]{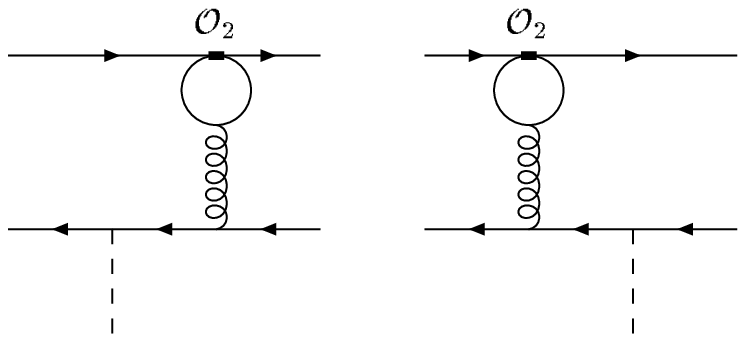}}
\unitlength=1mm
\begin{picture}(0,0)(-80,-3)
\put(0,0){\makebox(0,0)[cc]{\large $b$}}
\end{picture}
\caption{Feynman diagrams contributing to the spectator corrections       
         in $B \to V \gamma$ decays involving the ${\cal O}_2$
         operator. Row~$a$: photon emission from 
         the flavour-changing quark line; Row~$b$: 
         photon radiation off the spectator quark line.} 
\label{fig:CorrO2G}
\end{figure}

\begin{figure}[bt]
\centerline{
            \epsffile[150 630 410 750]{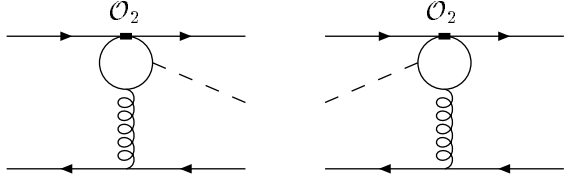}}
\caption{Feynman diagrams contributing to the spectator corrections       
         in $B \to V \gamma$ decays involving the ${\cal O}_2$ operator
         for the case when both the photon and the virtual gluon are
         emitted from the internal (loop) quark line.}
\label{fig:CorrO2Ggam}
\end{figure}

\begin{figure}[bt]
\centerline{
            \epsffile[150 630 410 750]{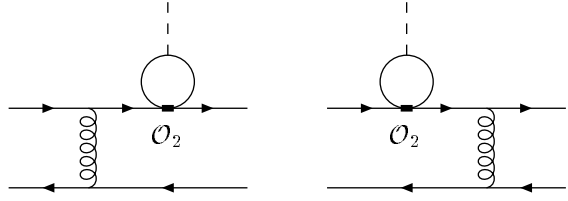}}
\caption{Feynman diagrams contributing to the spectator corrections       
         in $B \to V \gamma$ decays involving the ${\cal O}_2$ operator
         for the case when only the photon is
         emitted from the internal (loop) quark line in the 
         $b s (d) \gamma$ vertex.}
\label{fig:CorrO2gam}
\end{figure}

{\it 3. Spectator corrections involving the penguin-type diagrams
and the operator ${\cal O}_2$.}
The corresponding diagrams are presented in Figs.~\ref{fig:CorrO2G},
\ref{fig:CorrO2Ggam}, and~\ref{fig:CorrO2gam}.
%
%
%
%
The hard-scattering amplitude corresponding to the two diagrams
in Fig.~\ref{fig:CorrO2G}$a$ involving the emission of the photon from
the $b$- or $d$-quarks is as follows:
\begin{eqnarray}
{\cal T}^{(3a)}_{i j k l} & = &
\frac{G_F}{\sqrt 2} \,
\frac{e}{24 \pi^2} \,
\sum_{f = u, c} V_{fd}^* V_{fb} \, C_2^{(f)} (\mu) \,
\Delta F_1 \big ( z^{(f)}_1 \big ) \,
[\gamma_\nu]_{k l} \,
\label{eq:T3a} \\
& \times &
\left [
\left \{ \gamma_\nu -
\frac{(k_2 - l)_\nu (\slash k_2 - \slash l)}{(k_2 - l)^2}
\right \} (1 - \gamma_5)
\frac{\slash k_1 + \slash k_2 - \slash l + m_b}{(k_1 + k_2 - l)^2 - m_b^2} \,
\slash e^*
\right.
\nonumber \\
& + &
\left. \slash e^* \,
\frac{\slash p_b + \slash l - \slash k_2}{(p_b + l - k_2)^2} \,
\left \{ \gamma_\nu -
\frac{(k_2 - l)_\nu (\slash k_2 - \slash l)}{(k_2 - l)^2}
\right \} (1 - \gamma_5)
\right ]_{i j} ,
\nonumber
\end{eqnarray}
where the function $\Delta F_1 (z^{(f)})$
results from performing the integration over the momentum of the
internal quark~$f$ having the mass~$m_f$~\cite{Simma:1990nr}:
\begin{equation}
\Delta F_1 (z) =
- \frac{2}{9} - \frac{4}{3} \, \frac{Q_0 (z)}{z}
- \frac{2}{3} \, Q_0 (z) ,
\label{eq:IL-Func}
\end{equation}
and its argument is
$z^{(f)}_1 = (k_2 - l)^2 / m_f^2 \simeq - M l_+ \bar u / m_f^2$,
in the limit of the large recoil and to leading order in the inverse
$B$-meson mass. In Eq.~(\ref{eq:IL-Func}) the function~$Q_0 (z)$
is defined as follows:
\begin{equation}
Q_0 (z) = \int\limits_0^1 du \,
\ln \left [ 1 - z u (1 - u) \right ] ,
\label{eq:Q0-Func}
\end{equation}
and for the case ${\rm Im} \, z > 0$ it has the form~\cite{Simma:1990nr}:
\begin{eqnarray}
Q_0 (z) & = & - 2 -
\left [ u_+ (z) - u_- (z) \right ]
\left ( \ln \frac{u_- (z)}{u_+ (z)} + i \pi \right ) ,
\label{eq:Q0-explicit} \\
u_\pm (z) & = & \frac{1}{2}
\left ( 1 \pm \sqrt{1 - \frac{4}{z}} \right ) .
\nonumber
\end{eqnarray}
The argument~$z_1^{(f)}$ of the function ~$\Delta F_1 (z_1^{(f)})$ in
Eq.~(\ref{eq:T3a}) can be
large ($z_1^{(u)} \sim M \Lambda_{\rm QCD}/m_u^2$ for the $u$-quark),
and the asymptotic form of this function at large values of its argument 
is of interest:
\begin{equation}
\Delta F_1 (z) \bigg |_{z \to \infty} \simeq
\frac{2}{3}
\left [
\left ( 1 - \frac{6}{z^2} \right ) \,
\left [ \ln \frac{1}{z} + i \pi \right ]
+ \frac{5}{3}
+ \frac{6}{z}
+ \frac{3}{z^2}
\right ] .
\label{eq:IL-asymp}
\end{equation}
Thus, in the case of the internal $u$-quark loop, the function~$\Delta F_1
(z)$ is enhanced by the large logarithm $\ln (m_u/M)$. However, as
it has been shown in Ref.~\cite{Abud:1998zt}, summing up to all orders
in~$\alpha_s$, the penguin-like diagrams relevant for the $b \to d \gamma$
process can be safely calculated by taking the massless limit for the
$u$-quark in the penguin loop. This implies that, despite the superficial 
appearance, no large enhancement in the amplitude due to $ \ln (m_u/M)$ is
encountered.

Diagrams in Fig.~\ref{fig:CorrO2G}$b$ describing the emission
of the photon from the spectator quark line yield:
\begin{eqnarray}
{\cal T}^{(3b)}_{i j k l} & = &
- \frac{G_F}{\sqrt 2} \,
\frac{e Q_{d (u)}}{8 \pi^2} \,
\sum_{f = u, c} V_{fd}^* V_{fb} \, C_2^{(f)} (\mu) \,
\Delta F_1 \big ( z^{(f)}_0 \big )
\label{eq:T3b} \\
& \times &
\left [
\slash e^* \,
\frac{\slash k_1 + \slash k_2 - \slash p_b}{(k_1 + k_2 - p_b)^2} \,
\gamma_\nu + \gamma_\nu \,
\frac{\slash p_b + \slash l - \slash k_1}{(p_b + l - k_1)^2} \,
\slash e^*
\right ]_{k l}
\nonumber \\
& \times &
\left [
\left \{ \gamma_\nu -
\frac{(p_b - k_1)_\nu}{(p_b - k_1)^2} \, (\slash p_b - \slash k_1)
\right \} (1 - \gamma_5)
\right ]_{i j} ,
\nonumber
\end{eqnarray}
where the argument of the function~$\Delta F_1 (z^{(f)}_0)$ is
$z^{(f)}_0 = (p_b - k_1)^2 / m_f^2 \simeq M^2 \bar u / m_f^2$
in the large recoil limit.

There exists another topological class of diagrams contributing to
the spectator corrections involving the effective $b d
g^* \gamma$ vertex presented in Fig.~\ref{fig:CorrO2Ggam}.
%
%
%
%
The expression for the one-particle irreducible (OPI) $b d g^*
\gamma^*$ vertex as well as the general $b d g^* \gamma^*$ case are
known since a long time~\cite{Simma:1990nr}. For an on-shell photon
$q^2 = 0$, the OPI vertex is
simplified and can be found in Refs.~\cite{Greub:1995tb}
and~\cite{Greub:1996tg} for the four-dimensional and arbitrary
$d$-dimensional momentum spaces, respectively. For the case
considered here, the four-dimensional result derived in
Ref.~\cite{Greub:1995tb} is used.

The hard scattering amplitude corresponding to the diagrams shown in
Fig.~\ref{fig:CorrO2Ggam} is:
\begin{eqnarray}
{\cal T}^{(4)}_{i j k l} & = &
- \frac{G_F}{\sqrt 2} \,
\frac{e}{6 \pi^2} \,
\frac{
\left [ \gamma_\nu \right ]_{k l}
}{(k_2 - l)^2 \, (q [k_2 - l])} \,
\sum_{f = u, c} V_{fd}^* V_{fb} \, C_2^{(f)} (\mu) \,
\label{eq:T4} \\
& \times &
\left [ \left \{
\left [
q_\nu \, {\rm E} (k_2 - l, e^*, q) -
(q [k_2 - l]) \, {\rm E} (\nu, e^*, q)
\right. \right. \right.
\nonumber \\
& +  &
\left .
(e^* [k_2 - l]) \, {\rm E} (q, \nu, k_2 - l) -
(q [k_2 - l]) \, {\rm E} (e^*, \nu, k_2 - l)
\right ] \Delta i_5 \big ( z^{(f)}_0, z^{(f)}_1, 0 \big )
\nonumber \\
& + &
\left. \left.
\left [
(k_2 - l)^2 \, {\rm E} (\nu, e^*, q) +
(k_2 - l)_\nu \, {\rm E} (e^*, k_2 - l, q)
\right ] \Delta i_{25} \big ( z^{(f)}_0, z^{(f)}_1, 0 \big )
\right \}
(1 - \gamma_5)
\right ]_{i j} ,
\nonumber
\end{eqnarray}
where the value $Q_u = 2/3$ of the electric charge of the quark
in the loop is taken into account, and  we have used a short-hand
notation for the following expression involving products of 
$\gamma$-matrices:
\begin{equation}
{\rm E} (\mu, \nu, \rho) \equiv \frac{1}{2} \,
(\gamma_\mu \gamma_\nu \gamma_\rho - \gamma_\rho \gamma_\nu \gamma_\mu) =
- i \varepsilon_{\mu \nu \rho \sigma} \, \gamma^\sigma \, \gamma_5 .
\label{eq:E-operator}
\end{equation}
The equality shown above is valid in the four-dimensional
space only. In Eq.~(\ref{eq:T4}) the functions 
$\Delta i_5 (z_0, z_1, 0)$ and~$\Delta i_{25} (z_0, z_1, 0)$ 
are~\cite{Greub:1995tb}:
\begin{eqnarray}
\Delta i_5 (z_0, z_1, 0) & = & -1 + \frac{z_1}{z_0 - z_1} \,
\left [ Q_0 (z_0) - Q_0 (z_1) \right ] -
\frac{2}{z_0 - z_1} \, \left [ Q_- (z_0) - Q_- (z_1) \right ] ,
\qquad
\label{eq:Di5} \\
\Delta i_{25} (z_0, z_1, 0) & = & Q_0 (z_0) - Q_0 (z_1) .
\label{eq:Di25}
\end{eqnarray}
The auxiliary function~$Q_0 (z)$ is defined in Eq.~(\ref{eq:Q0-Func}),
and the other auxiliary function~$Q_- (z)$ is:
\begin{equation}
Q_- (z) = \int\limits_0^1 \frac{du}{u} \,
\ln \left [ 1 - z u (1 - u) \right ] ,
\label{eq:Qm-Func}
\end{equation}
with the explicit form for the case ${\rm Im} \, z > 0$~\cite{Simma:1990nr}:
\begin{equation}
Q_- (z) = \frac{1}{2}
\left ( \ln \frac{u_- (z)}{u_+ (z)} + i \pi \right )^2 ,
\label{eq:Qm-explicit}
\end{equation}
where the definition of $u_\pm (z)$ can be found in
Eq.~(\ref{eq:Q0-explicit}).

The arguments $z_0^{(f)} = (p_b - k_1)^2 / m_f^2 \simeq M^2 \bar u / m_f^2$
and $z_1^{(f)} = (k_2 - l)^2 / m_f^2 \simeq - M l_+ \bar u / m_f^2$,
already specified above, 
depend on the internal quark mass~$m_f$, and in the case of the $u$-quark,
they are large practically in all the region of the variables~$u$
and~$l_+$. This is not the case for the $c$-quark contribution in the
internal loop, and the charm quark mass-dependent corrections
can be important~\cite{Beneke:2001at,Bosch:2001gv}. Note that
the value of~$z_1^{(f)}$ is suppressed by the factor~$\Lambda_{\rm QCD}/M$
in comparison with~$z_0^{(f)}$ and in the framework of the large recoil
limit the corrections of order~$z_1^{(f)} / z_0^{(f)}$ can be neglected.
In this case, the functions~(\ref{eq:Di25}) and~(\ref{eq:Di5}) are
reduced, respectively, to $\Delta i_{25} (z_0, 0, 0) = Q_0 (z_0)$ and
\begin{equation}
\Delta i_5 (z_0, 0, 0) = - 1 - \frac{2}{z_0} \, Q_- (z_0) =
\left \{
\begin{array}{ll}
- 1 + (4 / z_0) \, \arctan^2 \left [ 1 / \sqrt{(4 / z_0) - 1} \right ],
& z_0 < 4, \\
- 1 - \left ( \ln \left [ u_- (z_0) / u_+ (z_0) \right ] + i \pi \right )^2
/ z_0,
& z_0 > 4.
\end{array}
\right.
\label{eq:Di50}
\end{equation}
Using the properties of the dilogarithmic function
${\rm Li}_2 (z)$, the functions~$t_\perp (u, m_q)$ in the limit
$q^2 \to 0$ (Eq.~(33) derived in Ref.~\cite{Beneke:2001at})
and~$h (u, s)$ (Eq.~(35) in Ref.~\cite{Bosch:2001gv})
are the same as the function~$\Delta i_5 (z_0, 0, 0)$, derived here, up to
overall factors given below:
\begin{displaymath}
h (\bar u, s) =
- \left. \frac{1}{2} \, t_\perp (u, m_q) \right |_{q^2 \to 0} =
\frac{2}{\bar u} \Delta i_5 (z_0, 0, 0).
\end{displaymath}
Finally, there are also diagrams where a photon is emitted from the internal
quark line due to the effective $b \to s (d) \gamma$ interaction and a
gluon is exchanged between the spectator quark and the $b$- or
$s (d)$-quarks (see Fig.~\ref{fig:CorrO2gam}).
%
%
%
%
Note that in the momentum space the amputated 
$b \to s (d) \gamma$ vertex due to the four-fermion quark
interaction (the ${\cal O}_2$ vertex) has the
form~\cite{Simma:1990nr,Greub:1996tg}:
\begin{equation}
I^{(f)}_\mu = - \frac{e}{6 \pi^2} \,
\Delta F_1 \left ( \frac{q^2}{m_f^2} \right ) \,
\left [ q_\mu \slash q - q^2 \gamma_\mu \right ] (1-\gamma_5) ,
\label{eq:photon-vert}
\end{equation}
where the function~$\Delta F_1 (z)$ is defined in
Eq.~(\ref{eq:IL-Func}). For the real photon case ($q^2 = 0$), the
amplitude contains the scalar product $(e^* I^{(f)}) \sim (e^* q)
\slash q - q^2 \slash e^*$ which is zero. This vertex gives a
non-vanishing contribution for off-shell photons, such as 
$b \to s (d) \gamma^* \to s (d) l^+ l^-$, which, however, 
is not the process we are considering in this paper.  Hence, 
for on-shell photon, the Feynman diagrams in Fig.~\ref{fig:CorrO2gam} 
do not contribute to $B \to K^* (\rho) \gamma$ (or $b \to s (d) \gamma$).

\section{$O (\alpha_s)$-Corrected Matrix Elements for 
         $B \to V \gamma$ \newline Decays}
\label{sec:ME}

The convolution of the $B$- and vector ($\rho$- or $K^*$-) meson 
projection operators displayed in~(\ref{eq:PrO-B}) and~(\ref{eq:PrO-rho}),
respectively, with the hard-scattering matrix elements derived in the
previous section can be written as:
\begin{equation}
\Delta {\cal M}_{\rm sp}^{(V)} = \frac{G_F}{\sqrt 2} \,
\frac{e \alpha_s C_F}{4 \pi N_c} \, f_B f_\perp^{(V)} M \,
\left [
(e^* \varepsilon^*) + i \, {\rm eps} (e^*, \varepsilon^*, n_-, v)
\right ] \,
\sum_{k = 1}^5 \Delta H^{(V)}_k ,
\label{eq:ME-total}
\end{equation}
where ${\rm eps} (a, b, c, d) = \varepsilon_{\mu \nu \rho \sigma}
a^\mu b^\nu c^\rho d^\sigma$ and the upper index~$V$ ($=K^*$ or $\rho$) 
characterizes the final vector meson.
The dimensionless functions~$\Delta H^{(V)}_k$ ($k = 1, 2, 3, 4, 5$)
describe the contributions of the sets of Feynman diagrams presented
in Figs.~\ref{fig:CorrO7}-\ref{fig:CorrO2gam}, respectively.
In the leading order of the inverse $B$-meson mass
($\sim \Lambda_{\rm QCD}/M$), the result reads as follows:
\begin{eqnarray}
\Delta H^{(V)}_1 (\mu) & \simeq &
V_{tp}^* V_{tb} \, C_7^{\rm eff} (\mu) \, m_b (\mu)
\left [ 
\left < l_+^{-1} \right >_+ 
\left < \bar u^{-1} \right >_\perp^{(V)} (\mu) 
+ 
\left < l_+^{-1} \right >_- 
\left < \bar u^{-2} \right >_\perp^{(V)} (\mu)
\right ] ,
\label{eq:ME1} \\
\Delta H^{(V)}_2 (\mu) & \simeq & \frac{1}{3} \,
V_{tp}^* V_{tb} \, C_8^{\rm eff} (\mu) \, m_b (\mu) \, 
\left < l_+^{-1} \right >_+ 
\left < u^{-1} \right >_\perp^{(V)} (\mu), 
\label{eq:ME2} \\
\Delta H^{(V)}_3 (\mu) & \simeq & 0 ,
\label{eq:ME3} \\
\Delta H^{(V)}_4 (\mu) & \simeq &
\frac{1}{3} \, C_2 (\mu) \, M \,
\left < l_+^{-1} \right >_+
\left [ V_{tp}^* V_{tb} \, 
\left < \bar u^{-1} \right >_\perp^{(V)} (\mu) 
+ V_{cp}^* V_{cb} \, h^{(V)} (z, \mu)
\right ] ,
\label{eq:ME4} \\
\Delta H^{(V)}_5 (\mu) & \simeq & 0 ,
\label{eq:ME5}
\end{eqnarray}
where $z = m_c^2/m_b^2$ and index~$p$ in the CKM matrix elements is 
$p = s$ for the $K^*$-meson and $p = d$ for the $\rho$-meson. 
In the above results we have used the short-hand notation for the
integrals over the mesons distribution functions:
\begin{equation}
\left < l_+^N \right >_\pm \equiv \int\limits_0^\infty dl_+ \,
l_+^N \, \phi^{(B)}_\pm (l_+) ,
\qquad
\left < f \right >_{\perp, \|}^{(V)} (\mu) \equiv
\int\limits_0^1 du \, f (u) \, \phi_{\perp, \|}^{(V)} (u, \mu) , 
\label{eq:integrals-Phi}
\end{equation}
and for convenience the following function was introduced: 
\begin{equation}
h^{(V)} (z, \mu) = \left <
\frac{\Delta i_5 (z_0^{(c)}, 0, 0) + 1}{\bar u}
\right >_{\!\!\perp}^{\!\!(V)} . 
\label{eq:h-function} 
\end{equation}
The function~$\Delta H^{(V)}_2 (\mu)$ [Eq.~(\ref{eq:ME2})] contains 
the distribution moment~$\left < u^{-1} \right >_\perp^{(V)}$, which 
in the case of the $\rho$-meson can be replaced by 
$\left < u^{-1} \right >_\perp^{(\rho)} \to 
 \left < \bar u^{-1} \right >_\perp^{(\rho)}$, as the
$\rho$-meson distribution function~$\phi_\perp^{(\rho)} (u)$
is symmetric under the 
interchange~$u \to \bar u = 1 - u$~\cite{Ball:1998sk}. 
This replacement is not valid for the case of the $K^*$-meson where a
sizable asymmetry under the interchange ~$u \to \bar u$ is present in the 
wave-function~\cite{Ball:1998sk}. The function~$\Delta H^{(V)}_4 (\mu)$ 
[Eq.~(\ref{eq:ME4})] arises from the diagrams shown in 
Fig.~\ref{fig:CorrO2Ggam} with the internal~$u$- and $c$-quarks.
The contributions of the internal quarks differ by the CKM
factor $V_{fp}^* V_{fb}$ ($f = u, c$) and the quark masses ($m_c$ and
$m_u$). If the internal quark masses are neglected, an assumption made in
the earlier version of this paper but one which we no longer invoke here,
then  $C_2^{(u)} = C_2^{(c)} = C_2$. This follows as~$\Delta H^{(V)}_4
(\mu)$ originates in the terms $\sim \Delta i_5$ from the hard-scattering
amplitude~${\cal T}^{(4)}_{i j k l}$, and the replacement
$\Delta i_5 \to - 1$ holds in the chiral
internal quark limit, $m_f = 0$. By making use of the unitarity relation
$V_{u b}^* V_{u p} + V_{c b}^* V_{c p} + V_{t b}^* V_{t p} = 0$
their sum can be expressed in terms of one independent CKM
combination $V_{t p}^* V_{t b}$ (the first term in the bracket of
Eq.~(\ref{eq:ME4})). The correction due to the non-zero $c$-quark mass,
which comes weighted by its own CKM factor $V_{c p}^* V_{c b}$,
is contained in the second term of Eq.~(\ref{eq:ME4}). A detailed
discussion of the importance of these corrections will be discussed
below. (See, also Ref.~\cite{Buchalla:1998ky}.) 

The result obtained above deserves a number of comments. First,
note that the diagrams shown in Fig.~\ref{fig:CorrO2G} involving the
operator ${\cal O}_2$ do not contribute in the large recoil limit and
to leading order in the inverse $B$-meson mass.
Second, there are no contributions from the diagrams involving
the chromomagnetic operator ${\cal O}_8$ for the case where a
photon is emitted from the spectator line (Figs.~\ref{fig:CorrO8}$b$).
It means that {\it no new contributions} to the isospin-breaking
corrections to the decay rates $B \to V \gamma$ arise from the  
hard-spectator corrections in the large recoil limit. Third, the 
contribution from the diagrams shown in Fig.~\ref{fig:CorrO7} contains
an end-point singularity of the
form~$\left < \bar u^{-2} \right >_\perp^{(V)}$
whereas the diagrams in Figs.~\ref{fig:CorrO8} and~\ref{fig:CorrO2Ggam}
give finite contributions. As argued by Beneke and Feldmann in
Ref.~\cite{Beneke:2001wa}, this end-point singularity describes
the soft-gluon physics of the matrix element and can be absorbed
into the ``soft form factor''~$\xi_\perp^{(V)}$. This removes the
singularity but introduces a factorization scheme (or renormalization
convention) for the ``soft form factor''. After adopting this procedure,
the hard-spectator corrections to the $B \to V \gamma$ decay
amplitude depends on the product of the moment of the $B$-meson
distribution~$\left < l_+^{-1} \right >_+$ with the vector meson
transverse distribution averages:
$\left < \bar u^{-1} \right >_\perp^{(V)}$,
$\left < u^{-1} \right >_\perp^{(V)}$, and 
$h^{(V)} = \left < (\Delta i_5 + 1) / \bar u \right >_\perp^{(V)}$.
These products are intrinsically non-perturbative though universal   
quantities and will have to be determined either by data from      
elsewhere or else resorting to models for the $B$-meson and the vector
meson distribution functions.

It is convenient to introduce the dimensionless 
quantity~\cite{Beneke:2001wa}
\begin{equation}
\Delta F_\perp^{(V)} (\mu) =
\frac{8 \pi^2 f_B f_\perp^{(V)} (\mu)}{N_c M \lambda_{B,+}}
\left < \bar u^{-1} \right >_\perp^{(V)} (\mu) ,
\label{eq:DFperp}
\end{equation}
where $\lambda_{B,+}^{-1} = \left < l_+^{-1} \right >_+$ is the first
negative moment of the $B$-meson distribution function~$\phi_+^{(B)} 
(l_+)$ which is typically estimated as
$\lambda_{B,+}^{-1} = (3 \pm 1)$~GeV~\cite{Grozin:1997pq,Beneke:2001wa}.
At the scale~$\mu_{\rm sp}=\sqrt{\mu_b \Lambda_H}$ of the hard-spectator
corrections, and for the central values of the parameters shown in
Table~\ref{tab:VM-parameters} with $\lambda_{B,+}^{-1} = 3$~GeV, 
this quantity is evaluated
as $\Delta F_\perp^{(K^*)} (\mu_{\rm sp}=1.52~{\rm GeV}) = 1.96$ and
$\Delta F_\perp^{(\rho)} (\mu_{\rm sp}=1.52~{\rm GeV}) = 1.64$ for the
$K^*$- and
$\rho$-meson, respectively.
In term of~$\Delta F_\perp^{(V)} (\mu)$ the hard-spectator part of
$B \to V \gamma$ decay amplitude has the form:
\begin{eqnarray}
\Delta {\cal M}_{\rm sp} & = & \frac{G_F}{\sqrt 2} \,
V_{tp}^* V_{tb} \,
\frac{\alpha_s C_F}{4 \pi} \,
\frac{e}{4 \pi^2} \, \Delta F_\perp^{(V)} (\mu)
\left [
(p P) \, (e^* \varepsilon^*) + i \,
{\rm eps} (e^*, \varepsilon^*, p, P)
\right ]
\label{eq:ME-final} \\
& \times &
\left [  C_7^{\rm eff} (\mu) 
+ \frac{1}{3} \, C_8^{\rm eff} (\mu) \, 
\frac{\left < u^{-1} \right >_\perp^{(V)}}
     {\left < \bar u^{-1} \right >_\perp^{(V)}}
+ \frac{1}{3} \, C_2 (\mu)
\left (
1 + \frac{V_{cp}^* V_{cb}}{V_{tp}^* V_{tb}} \,
\frac{h^{(V)} (z, \mu)}{\left < \bar u^{-1} \right >_\perp^{(V)} (\mu)}
\right )
\right ]~,
\nonumber
\end{eqnarray}
where $P = M v$ and $p = E n_- \simeq M n_-/2$ are the
four-momenta of the $B$- and vector meson, respectively, and,
as in Eqs.~(\ref{eq:ME1})--(\ref{eq:ME5}), index $p = s$ or~$d$
for the case of $K^*$- or $\rho$-meson.

We now proceed to give an analytic result for the function
$h^{(V)}(z,\mu)$. To that end we recall that
the leading-twist transverse distribution 
amplitude~$\phi_\perp^{(V)} (u, \mu)$ 
of a vector meson is the solution of an evolution equation and has 
the following general form~\cite{Ball:1998sk}:
\begin{equation} 
\phi_\perp^{(V)} (u, \mu) = 6 u \bar u 
\left [ 1 + 
\sum_{n = 1}^\infty a^{(V)}_{\perp n} (\mu) \, C^{3/2}_n (u - \bar u) 
\right ] , 
\label{eq:phi-perp}
\end{equation}
where $C^{3/2}_n (u - \bar u)$ are the Gegenbauer polynomials 
[$C^{3/2}_1 (u - \bar u) = 3 (u - \bar u)$, 
$C^{3/2}_2 (u - \bar u) = 3 \left [ 5 (u - \bar u)^2 -1 \right ]$/2, 
etc.] and $a_{\perp n}^{(V)} (\mu)$ are the corresponding Gegenbauer 
moments (the $\rho$-meson distribution amplitude includes the even 
moments only). These moments should be evaluated at the scale~$\mu$; 
their scale dependence is governed by~\cite{Ball:1998sk}:  
\begin{equation} 
a_{\perp n}^{(V)} (\mu) = \left ( 
\frac{\alpha_s (\mu^2)}{\alpha_s (\mu_0^2)}
\right )^{\gamma_n / \beta_0} 
a_{\perp n}^{(V)} (\mu_0), 
\qquad  
\gamma_n = 4 C_F 
\left ( \sum_{k = 1}^n \frac{1}{k} - \frac{n}{n + 1} \right ) ,
\label{eq:Geg-mom}
\end{equation}
where $\beta_0 = (11 N_c - 2 n_f)/3$ and~$\gamma_n$ is the one-loop 
anomalous dimension with $C_F = (N_c^2 - 1)/(2 N_c) = 4/3$. 
In the limit~$\mu \to \infty$ the Gegenbauer moments 
vanish,~$a_{\perp n}^{(V)} (\mu) \to 0$, and the leading-twist 
transverse distribution amplitude has its asymptotic form:
\begin{equation} 
\phi_\perp^{(V)} (u, \mu) \to \phi_\perp^{({\rm as})} (u) = 6 u \bar u . 
\label{eq:phi-perp-as}
\end{equation}
A simple model of the transverse distribution which includes 
contributions from the first~$a_{\perp 1}^{(V)} (\mu)$ and the 
second~$a_{\perp 2}^{(V)} (\mu)$ Gegenbauer moments only is used here 
in the analysis. In this approach the
quantities~$< u^{-1} >_\perp^{(V)}$ and~$< \bar u^{-1} >_\perp^{(V)}$ are:
\begin{equation}
< u^{-1} >_\perp^{(V)} = 3
\left [ 1 - a_{\perp 1}^{(V)} (\mu) + a_{\perp 2}^{(V)} (\mu) \right ],
\qquad
< \bar u^{-1} >_\perp^{(V)} = 3
\left [ 1 + a_{\perp 1}^{(V)} (\mu) + a_{\perp 2}^{(V)} (\mu) \right ] ,
\label{eq:u-ub-avar}
\end{equation}
and depend on the scale~$\mu$ due to the
coefficients~$a_{\perp n}^{(V)} (\mu)$. 
The Gegenbauer moments 
were evaluated at the scale $\mu_0 = 1$~GeV, yielding~\cite{Ball:1998sk}: 
$a_{\perp 1}^{(K^*)} (1~{\rm GeV}) = 0.20 \pm 0.05$ and 
$a_{\perp 2}^{(K^*)} (1~{\rm GeV}) = 0.04 \pm 0.04$ for the 
$K^*$-meson and $a_{\perp 1}^{(\rho)} (1~{\rm GeV}) = 0$ and 
$a_{\perp 2}^{(\rho)} (1~{\rm GeV}) = 0.20 \pm 0.10$ for the $\rho$-meson.
In the same manner, 
the function~$h^{(V)} (z, \mu)$ introduced in Eq.~(\ref{eq:h-function}) 
can be presented as an expansion on the Gegenbauer moments: 
\begin{eqnarray} 
h^{(V)} (z, \mu) 
& = & h_0 (z) + a_{\perp 1}^{(V)} (\mu) \, h_1 (z) + 
a_{\perp 2}^{(V)} (\mu) \, h_2 (z) 
\label{eq:Di5-decomp} \\ 
& = & \left [ 1 + 
3 a_{\perp 1}^{(V)} (\mu) + 6 a_{\perp 2}^{(V)} (\mu) 
\right ] 
\left < (\Delta i_5 + 1)/\bar u \right >^{(0)}_\perp  
\nonumber \\
& - & 6 \left [ 
a_{\perp 1}^{(V)} (\mu) + 5 a_{\perp 2}^{(V)} (\mu) 
\right ] 
\left < \Delta i_5 + 1 \right >^{(0)}_\perp 
+ 30 \, a_{\perp 2}^{(V)} (\mu) 
\left < \bar u \, (\Delta i_5 + 1) \right >^{(0)}_\perp , 
\nonumber 
\end{eqnarray} 
where another short-hand notation is introduced for the integral: 
\begin{equation} 
\left < f (u) \right >^{(0)}_\perp = 
\int\limits_0^1 du \, f (u) \, \phi_\perp^{({\rm as})} (u). 
\label{eq:int-p0-def}
\end{equation}
Such a decomposition allows us to define the set of functions~$h_n (z)$ 
which are dependent on the charm-to-bottom quark mass ratio 
$z = m_c^2/m_b^2$ but are independent of the parameters of the vector
meson in consideration. The analytical expressions for the integrals 
in Eq.~(\ref{eq:Di5-decomp}) are: 
\begin{eqnarray} 
\left < \frac{\Delta i_5 (\bar u/z, 0, 0) + 1}{\bar u} 
\right >^{\!\!(0)}_{\!\!\perp} 
& = & - 2 z \left \{ 6 - \ln^3 z + 6 \, Q_0 (1/z) 
\right. 
\label{eq:Di5ub-aver} \\ 
& + & \frac{3 i \pi \ln z}{\sqrt{1 - 4 z}} \, [2 + Q_0 (1/z)] 
- 6 \, (1 - 2 z) \, Q_- (1/z) 
\nonumber \\ 
& - & 6 \, [2 \ln u_+ (1/z) - i \pi] \, {\rm Li}_2 (u_+ (1/z)) 
\nonumber \\
& - & 6 \, [2 \ln u_- (1/z) + i \pi] \, {\rm Li}_2 (u_- (1/z))  
\nonumber \\ 
& + & 
\left.
12 \, [{\rm Li}_3 (u_+ (1/z)) + {\rm Li}_3 (u_- (1/z))] 
\right \} ,
\nonumber \\ 
\left < \Delta i_5 (\bar u/z, 0, 0) + 1 \right >^{(0)}_\perp & = & 
\frac{3 z}{2} \left ( 5 - 12 z \right ) 
+ 9 z \left ( 1 - 2 z \right ) Q_0 (1/z) 
\label{eq:Di5-aver} \\ 
& - & 6 z \left ( 1 - 4 z + 6 z^2 \right ) Q_- (1/z) , 
\nonumber \\ 
\left < \bar u \, [\Delta i_5 (\bar u/z, 0, 0)+ 1] \right >^{(0)}_\perp 
& = & \frac{z}{18} \, \left ( 41 + 144 z - 720 z^2 \right )
\label{eq:ubDi5-aver} \\ 
& + & 
\frac{z}{3} \left ( 5 + 34 z - 120 z^2 \right ) Q_0 (1/z) 
\nonumber \\ 
& - & 2 z \left ( 1 - 18 z^2 + 40 z^3 \right ) Q_- (1/z) , 
\nonumber 
\end{eqnarray} 
where $Q_0 (1/z)$ and $Q_- (1/z)$ are the 
functions defined in Eqs.~(\ref{eq:Q0-Func}) and~(\ref{eq:Qm-Func}),
respectively,
and the dilogarithmic~${\rm Li}_2 (z)$ and trilogarithmic~${\rm Li}_3 (z)$
functions have their usual definitions:
\begin{displaymath}
{\rm Li}_2 (z) = - \int\limits_0^z \frac{\ln (1 - t)}{t} \, dt,
\qquad
{\rm Li}_3 (z) = \int\limits_0^z \frac{{\rm Li}_2 (t)}{t} \, dt .
\end{displaymath}
The result for the charm-quark mass dependent contribution 
to~$\Delta H^{(V)}_4$ in Eq.~(\ref{eq:ME4}), 
$\langle (\Delta i_5 (z_0^{(c)}) + 1)/\bar u \rangle _\perp^{(V)}$, 
derived above in the LEET framework is finite. We concur on this point 
with the observations made in Refs.~\cite{Beneke:2001at,Bosch:2001gv}. 
Moreover, we have presented the resulting contribution in an analytic 
form.   

\begin{figure}[bt]
\centerline{
            \epsffile{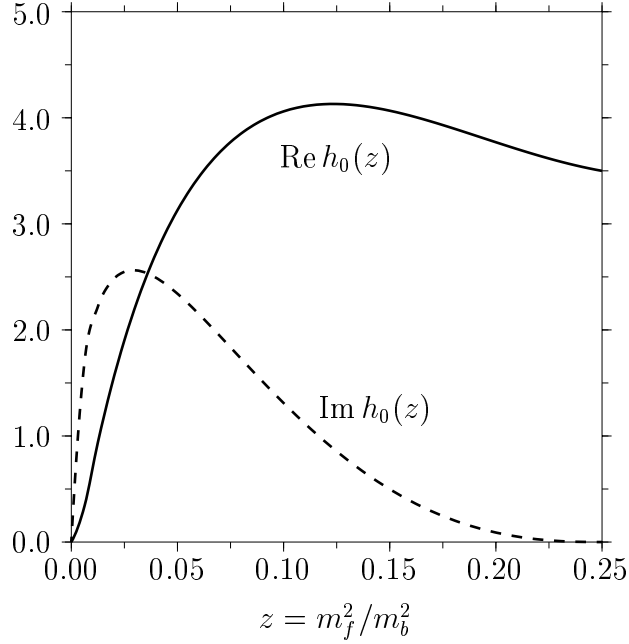}}
\caption{The function $h_0 (z)$ plotted against the
         ratio~$m_f^2/m_b^2$ where~$m_b$ is the $b$-quark mass.
         The solid curve is the real part of the function and
         the dashed curve is its imaginary part.}
\label{fig:DI5av0}
\end{figure}

\begin{figure}[bt]
\centerline{\epsfxsize=.45\textwidth
            \epsffile{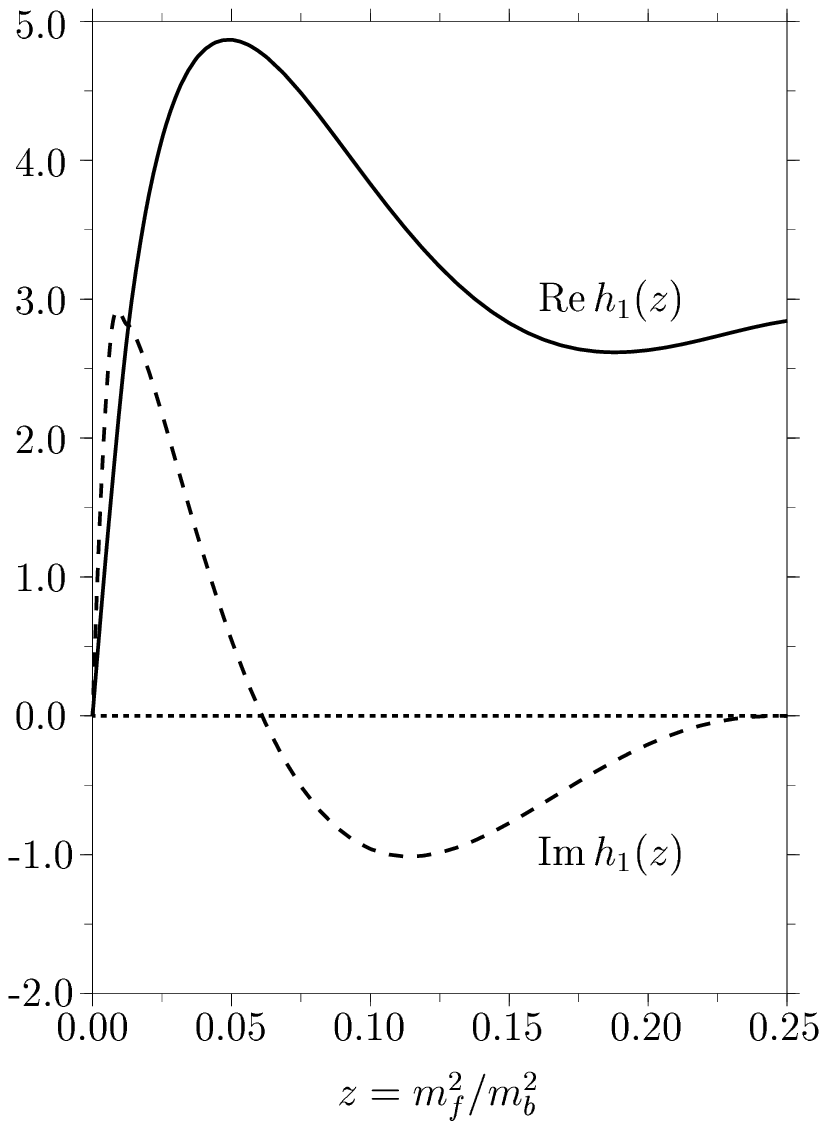} \qquad
            \epsfxsize=.45\textwidth
            \epsffile{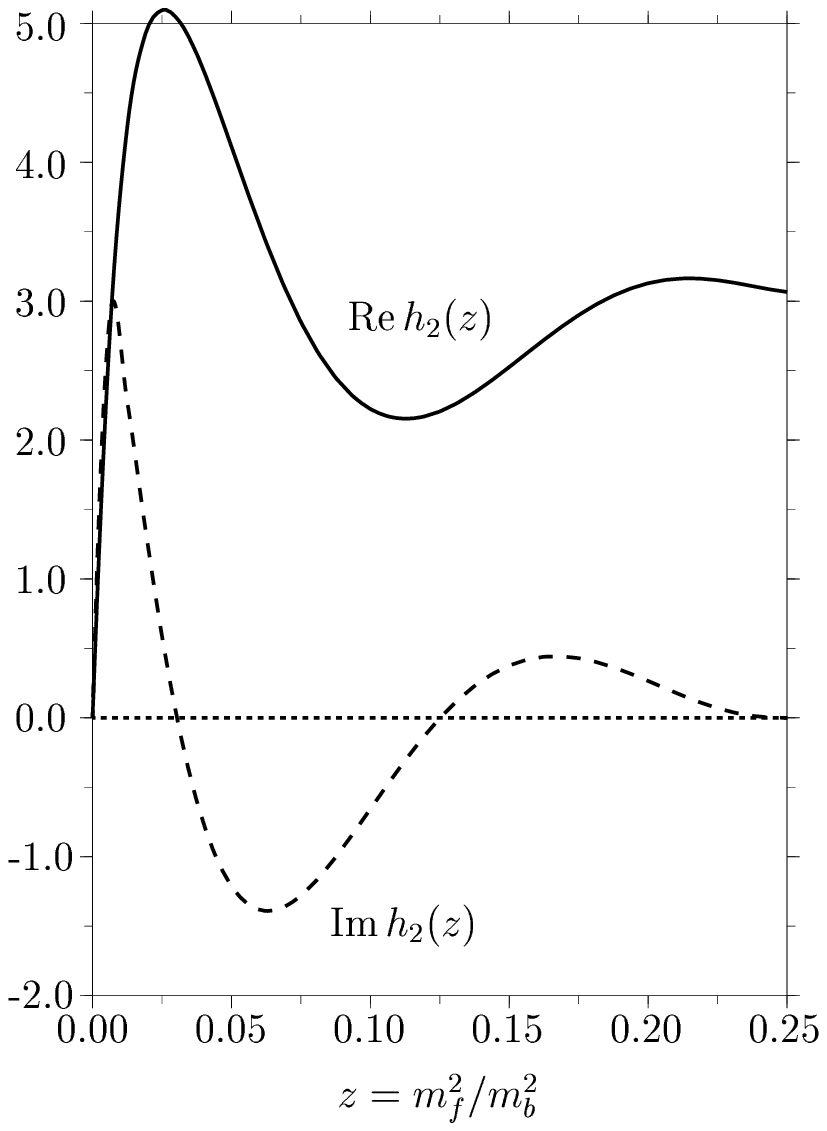}}
\caption{The functions $h_1 (z)$ (left figure) and $h_2 (z)$
         (right figure) plotted against the
         ratio~$m_f^2/m_b^2$ where~$m_b$ is the $b$-quark mass.
         The solid curves are the real parts of the functions and
         the dashed curves are their imaginary parts.}
\label{fig:DI5av12}
\end{figure}

\begin{figure}[bt]
\centerline{\epsfxsize=.45\textwidth
            \epsffile{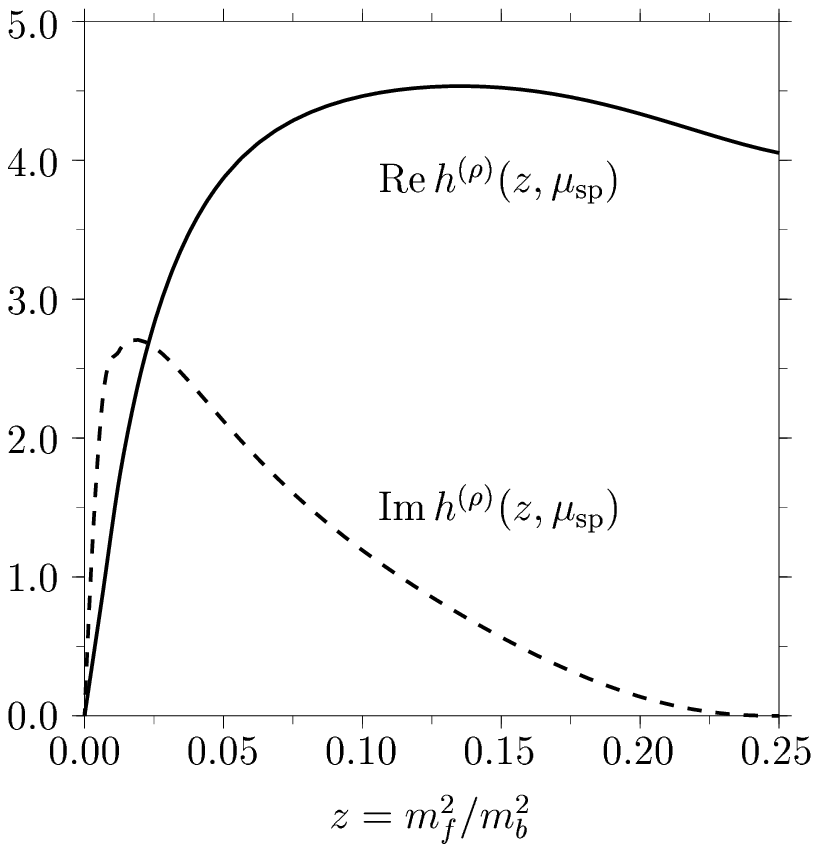} \qquad
            \epsfxsize=.45\textwidth
            \epsffile{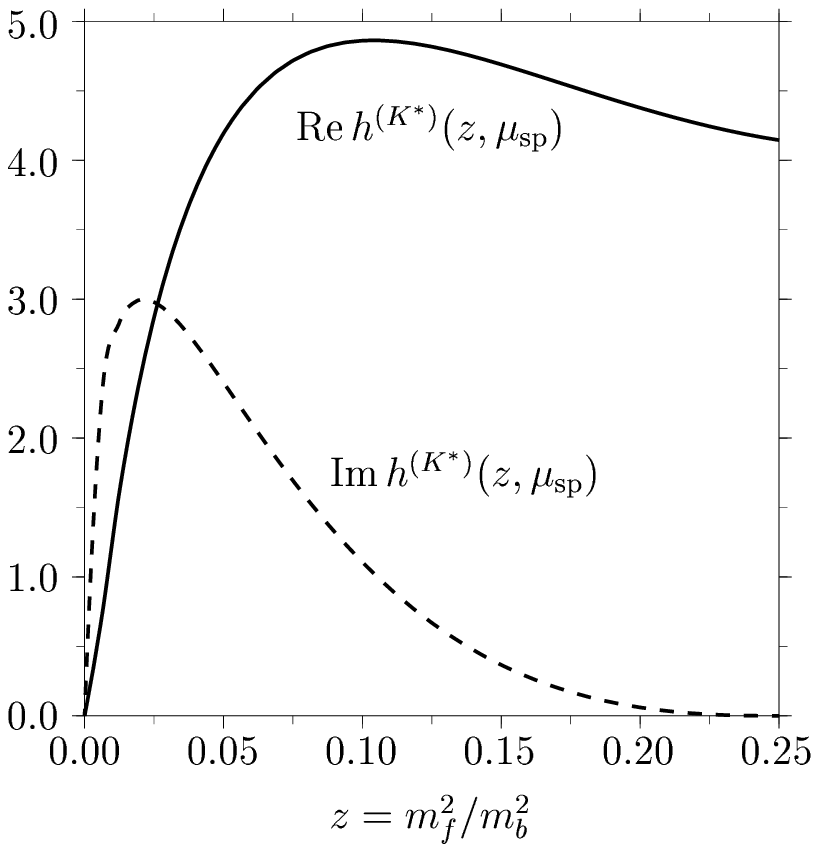}}
\caption{The functions $h^{(\rho)} (z, \mu_{\rm sp})$ for
         $B \to \rho \gamma$ (left figure) and
         $h^{(K^*)} (z, \mu_{\rm sp})$ for $B \to K^* \gamma$
         (right figure) plotted against the ratio~$m_f^2/m_b^2$
         at the mass scale of hard-spectator corrections 
         $\mu_{\rm sp} = 1.52$~GeV.
         The solid curves are the real parts of the functions and
         the dashed curves are their imaginary parts.}
\label{fig:DI5av}
\end{figure}

The real and imaginary parts of the functions~$h_n (z)$ 
are presented in Figs.~\ref{fig:DI5av0} (for $n=0$) and~\ref{fig:DI5av12}
(for $n=1$ and $n=2$). 
%
%
%
%
%
%
The dependence on $z = m_c^2/m_b^2$ of the 
function~$h^{(V)} (z, \mu)$~(\ref{eq:h-function}) 
at the mass scale~$\mu = \mu_{\rm sp} = 1.52$~GeV of hard-spectator 
corrections is presented in Fig.~\ref{fig:DI5av} for the~$\rho$- 
and $K^*$-meson. 
%
%
%
%
The values of the corresponding Gegenbauer moments used for evaluations  
are given in Table~\ref{tab:VM-parameters}.
%
%
\begin{table}[tb]
\caption{Representative values of the $K^*$- and $\rho$-meson parameters
         in the decays $B \to V \gamma$ ($V=\rho, K^*)$ at the scales 
         $\mu = \mu_{\rm sp} = 1.52$~GeV for the hard spectator
         corrections and $\mu = m_{b,{\rm pole}} = 4.65$~GeV for the vertex 
         corrections. The central values of the parameters 
         shown in Tables~\ref{tab:xiKs-B0} and~\ref{tab:Br-B-rho-gamma} 
         are used as inputs with $\sqrt z =m_c/m_b= 0.29$.} 
\label{tab:VM-parameters}
\bigskip
\begin{center}
\begin{tabular}{|l|c|c|c|c|} 
\hline \hline
& \multicolumn{2}{|c|}{$K^*$-meson} & \multicolumn{2}{|c|}{$\rho$-meson} 
\\ \hline 
& $\mu_{\rm sp}$ & $m_{b,{\rm pole}}$ & $\mu_{\rm sp}$ & 
$m_{b,{\rm pole}}$  
\\ \hline 
$\mu$, [GeV] & 1.52 & 4.65 & 1.52 & 4.65 \\ 
$a_{\perp 1}^{(V)} (\mu)$ & 0.187 & 0.164 & 0     & 0 \\ 
$a_{\perp 2}^{(V)} (\mu)$ & 0.036 & 0.029 & 0.179 & 0.143 \\ 
$h_0 (z)$ & $3.91 + i \, 1.64$ & $3.91 + i \, 1.64$ & $3.91 + i \, 1.64$ & 
$3.91 + i \, 1.64$ \\ 
$h^{(V)} (z, \mu)$ & $4.79 + i \, 1.46$ & $4.68 + i \, 1.49$ & 
$4.36 + i \, 1.45$ & $4.27 + i \, 1.48$ \\ 
$\left < \bar u^{-1} \right >_\perp^{(V)} (\mu)$ & 
3.67 & 3.58 & 3.54 & 3.43 \\  
$h^{(V)} / \left < \bar u^{-1} \right >_\perp^{(V)}$ & $1.31 + i \, 0.40$ 
& $1.31 + i \, 0.42$ & $1.23 + i \, 0.41$ & $1.25 + i \, 0.43$ \\
$f_\perp^{(V)} (\mu)$, [MeV] & 179 & 167 & 155 & 145 \\ 
$\Delta F_\perp^{(V)} (\mu)$ & 1.96 & 1.79 & 1.64 & 1.48 
\\ \hline \hline 
\end{tabular} 
\end{center} 
\end{table}
%
%
Comparison of the numerical values for the functions $h_0 (z)$ and
$h^{(V)} (z, \mu_{\rm sp})$ given in this table
shows that the influence of the Gegenbauer moments is 
more sizable in the case of the $K^*$-meson, with the real part
increasing by $\sim 20\%$ and the imaginary part decreasing by
$\sim 10\%$. In the case of the $\rho$-meson, the imaginary part 
decreases approximately by a similar amount while the real part 
increases by $\sim 10\%$.
  
The amplitude~(\ref{eq:ME-total}) is proportional to the tensor decay 
constant~$f_\perp^{(V)}$ of the vector meson which is a scale 
dependent parameter. As for the Gegenbauer moments~$a_{\perp n}^{(V)}$, 
its values were defined at the mass scale $\mu_0 = 1$~GeV for the $K^*$- 
and $\rho$-meson following Ref.~\cite{Ball:1998sk}: 
$f_\perp^{(K^*)} (1~{\rm GeV}) = (185 \pm 10)$~MeV and 
$f_\perp^{(\rho)} (1~{\rm GeV}) = (160 \pm 10)$~MeV. 
Their values at an arbitrary scale~$\mu$ can be obtained with the 
help of the evolution equation~\cite{Ball:1998sk}: 
\begin{equation} 
f_\perp^{(V)} (\mu) = \left ( 
\frac{\alpha_s (\mu^2)}{\alpha_s (\mu_0^2)}  
\right )^{4/(3 \beta_0)} f_\perp^{(V)} (\mu_0) .  
\label{eq:decay-constant}
\end{equation}
Central values of the tensor decay constants at the scales 
$\mu_{\rm sp} = 1.52$~GeV and $m_{b,{\rm pole}} = 4.65$~GeV are  
presented in Table~\ref{tab:VM-parameters}.    

The amplitude~(\ref{eq:ME-final}) allows us to get the hard-spectator
corrections to the form factors for the $B \to V$ transitions, with
$V = \rho$ or~$K^*$. The relevant form factors are defined as follows:
\begin{eqnarray}
\left < V (p, \varepsilon^*) | \bar Q \, \sigma^{\mu \nu} q_\nu b 
| \bar B (P) \right >
& = &
2 T_1^{(V)} (q^2) \, {\rm eps} (\mu, \varepsilon^*, p, P) ,
\label{eq:defFF-T} \\
\left < V (p, \varepsilon^*) | \bar Q \, \sigma^{\mu \nu} \gamma_5 
q_\nu b | \bar B (P) \right >
& = &
- i T_2^{(V)} (q^2) \, [(M^2 - m_V^2) \, \varepsilon^{* \mu} -
       (\varepsilon^* q) \, (p + P)^\mu] 
\qquad 
\label{eq:defFF-T5} \\
& &
- i T_3^{(V)} (q^2) \, (\varepsilon^* q) \,
\left [
q^\mu - \frac{q^2}{M^2 - m_V^2} \, (p + P)^\mu
\right ] ,
\nonumber
\end{eqnarray}
where $Q = d$ (for the 
$\rho$-meson) and $Q = s$ (for the $K^*$-meson) are the down quark and 
strange quark fields. Note that
only two form factors~$T_1^{(V)} (q^2)$ and~$T_2^{(V)} (q^2)$
contribute to the matrix element of the $B \to V \gamma$ decay. 
Hence, the hard spectator corrections for these form factors
from the amplitude~(\ref{eq:ME-final}) for on-shell photon 
($q^2 = 0$) are:
\begin{eqnarray}
\Delta_{\rm sp} T_1^{(\rho)} (0) & = & \Delta_{\rm sp} T_2^{(\rho)} (0)
\simeq \frac{\alpha_s C_F}{4 \pi} \, 
\frac{\Delta F_\perp^{(\rho)} (\mu)}{2}  
\label{eq:FF-SC} \\ 
& \times &
\left [ 
1 + \frac{C_8^{(0){\rm eff}} (\mu)}{3 C_7^{(0){\rm eff}} (\mu)}
+ \frac{C_2^{(0)} (\mu)}{3 C_7^{(0){\rm eff}} (\mu)}
\left (
1 + \frac{V_{cd}^* V_{cb}}{V_{td}^* V_{tb}} \, 
\frac{h^{(\rho)} (z, \mu)}
     {\left < \bar u^{-1} \right >_\perp^{(\rho)} (\mu)}
\right ) 
\right ] ~, 
\nonumber 
\end{eqnarray}
for the $\rho$-meson, and 
\begin{eqnarray}
\Delta_{\rm sp} T_1^{(K^*)} (0) & = & \Delta_{\rm sp} T_2^{(K^*)} (0) 
\simeq  
\frac{\alpha_s C_F}{4 \pi} \, \frac{\Delta F_\perp^{(K^*)} (\mu)}{2}
\label{eq:FF-KS} \\
& \times & 
\left [
1 + \frac{C_8^{(0){\rm eff}} (\mu)}{3 C_7^{(0){\rm eff}} (\mu)} \, 
\frac{\left < u^{-1} \right >_\perp^{(K^*)} (\mu)}
      {\left < \bar u^{-1} \right >_\perp^{(K^*)} (\mu)}
+ \frac{C_2^{(0)} (\mu)}{3 C_7^{(0){\rm eff}} (\mu)}
\left ( 1 - 
\frac{h^{(K^*)} (z, \mu)}
     {\left < \bar u^{-1} \right >_\perp^{(K^*)} (\mu)} 
\right )
\right ] ~,
\nonumber 
\end{eqnarray}
for the $K^*$-meson, in which the asymmetric distribution of the 
$K^*$-meson wave-function is taken into account.  
In writing the last equation we have used the CKM-unitarity relation  
$V_{cs}^* V_{cb}/V_{ts}^* V_{tb} \simeq -1$. 
We remark that the contribution obtained for the diagrams in
Fig.~\ref{fig:CorrO7} is the same as the one presented in
Ref.~\cite{Beneke:2001wa}.

\section{Branching Ratios for the Decays $B \to K^* \gamma$ and \\
         $B \to \rho \gamma$}
\label{sec:ratios}

We shall proceed by first calculating the branching ratios for the
decays $B \to K^* \gamma$ in the LEET approach. In doing this, we
will ignore the isospin-breaking differences between the decay
widths $B^\pm \to K^{*\pm} \gamma$ and $B^0(\bar{B}^0) \to K^{*0}(\bar
K^{*0}) \gamma$, as they are power suppressed.
A recent calculation shows that such isospin-breaking terms 
can lead to a difference at $(4 - 8)\%$ level in the
amplitudes~\cite{Kagan:2001zk}. Since present data is not precise enough 
to quantify isospin-violations in the decays $B \to K^* \gamma$, and
the effect is any case small, we average the data over 
the charged and neutral decay modes to get a statistically more
significant result for the form factor $\xi^{(K^*)} (0)$ 
[equivalently $T_1^{(K^*)} (0)$]. As we shall see, the
exclusive branching ratios have significant parametric uncertainties,
which translate into commensurate theoretical dispersion on the form
factors. To reduce some of these uncertainties, we shall also calculate
the ratio of the exclusive to inclusive decay widths
$R(K^*\gamma/X_s\gamma)\equiv \Gamma(B \to K^* \gamma)/\Gamma(B \to X_s
\gamma)$, and extract the form factor from the experimentally measured
value for $R(K^*\gamma/X_s\gamma)$. 

The branching ratios for the decays $B^\pm \to \rho^\pm \gamma$
and $B^0 (\bar B^0) \to \rho^0 \gamma$ can be related to those of
the experimentally measured $B$-decay modes 
$B^+ \to K^{* +} \gamma$ and $B^0 \to K^{* 0} \gamma$, using
SU(3)-symmetry breaking effects and taking into account other differences
in the decay amplitudes of which the differing CKM structure is the most
important. An important
difference in the $B \to \rho$ and $B \to K^*$ transitions is that the 
annihilation contribution is important in the former, leading to
significant isospin violations in the decay rates for $B^\pm \to \rho^\pm 
\gamma$ and $B^0 \to \rho^0 \gamma$. We shall take these isospin
violations in $B \to \rho \gamma$ decays into account.
The branching ratios for the $B \to \rho \gamma$ decay modes will be
obtained from the expressions
\begin{eqnarray}
{\cal B} (B^\pm \to \rho^\pm \gamma) & = & 
\frac{\Gamma_{\rm th} (B^\pm \to \rho^\pm \gamma)}
     {\Gamma_{\rm th} (B \to K^* \gamma)} \,
{\cal B}_{\rm exp} (B^\pm \to K^{* \pm} \gamma) ,
\label{eq:BrBpm} \\
{\cal B} (B^0 (\bar B^0) \to \rho^0 \gamma) & = &
\frac{\Gamma_{\rm th} (B^0 (\bar B^0) \to \rho^0 \gamma)}
     {\Gamma_{\rm th} (B \to K^* \gamma)} \,
{\cal B}_{\rm exp} (B^0(\bar B^0) \to K^{* 0}(\bar K^{*0}) \gamma) .
\label{eq:BrB0}
\end{eqnarray}
As we shall see, the theoretical ratios of the branching ratios on the
r.h.s. of these equations can be obtained with smaller parametric
uncertainties.

\subsection{$B \to K^* \gamma$ Decays}
\label{ssec:B-Ks-gamma}

The present measurements of the branching ratios for $B \to K^*
\gamma$ decays from the CLEO, BABAR, and BELLE collaborations are
summarized in Table~\ref{tab:Br-exp}. They yield the following world
averages:
\begin{eqnarray}
{\cal B}_{\rm exp} (B^\pm \to K^{* \pm} \gamma) =
(3.82 \pm 0.47) \times 10^{-5},
\label{eq:Br-Ks-exp} \\
{\cal B}_{\rm exp} (B^0 (\bar B^0) \to K^{* 0}(\bar K^{*0}) \gamma) =
(4.44 \pm 0.35) \times 10^{-5}.
\nonumber
\end{eqnarray}
%
%
\begin{table}[tb]
\caption{Experimental branching ratios for the decays  
         $B^0(\bar{B}^0) \to K^{*0}(\bar K^{*0}) \gamma$ and $B^\pm \to
K^{*\pm}
\gamma$.}
\label{tab:Br-exp}
\begin{center}
\begin{tabular}{|l|l|l|} 
\hline \hline 
Experiment & ${\cal B}_{\rm exp} (B^0(\bar {B}^0) \to K^{*0} (\bar K^{*0}) 
+ \gamma)$ & 
${\cal B}_{\rm exp} (B^\pm \to K^{*\pm} + \gamma)$ 
\\ \hline
CLEO \protect\cite{Chen:2001fj} & 
$(4.55^{+ 0.72}_{-0.68} \pm 0.34) \times 10^{-5}$ & 
$(3.76^{+ 0.89}_{-0.83} \pm 0.28) \times 10^{-5}$ \\ 
BELLE \protect\cite{Tajima:2001qp} & 
$(4.96 \pm 0.67 \pm 0.45) \times 10^{-5}$ &  
$(3.89 \pm 0.93 \pm 0.41) \times 10^{-5}$ \\ 
BABAR \protect\cite{Aubert:2001} & 
$(4.23 \pm 0.40 \pm 0.22) \times 10^{-5}$ & 
$(3.83 \pm 0.62 \pm 0.22) \times 10^{-5}$ \\
\hline \hline 
\end{tabular}
\end{center}
\end{table}
%
%
Since we are ignoring the isospin differences in the decay widths of
$B \to K^* \gamma$ decays, the branching ratios for $B^\pm \to K^{*\pm}
\gamma$ and $B^0(\bar B^{0}) \to K^{*0}(\bar K^{*0}) \gamma$ differ
essentially by the differing lifetimes of the $B^\pm$ and $B^0$
mesons. Thus, generically, the branching ratio can be expressed as: 
\begin{eqnarray}
{\cal B}_{\rm th} (B \to K^{*} \gamma) & = & 
\tau_B \, \Gamma_{\rm th} (B \to K^* \gamma) 
\label{eq:DW(B-Kgam)} \\
& = & 
\tau_B \,\frac{G_F^2 \alpha |V_{tb} V_{ts}^*|^2}{32 \pi^4} \, 
m_{b, {\rm pole}}^2 \, M^3 \, \left [ \xi_\perp^{(K^*)} \right ]^2  
\left ( 1 - \frac{m_{K^*}^2}{M^2} \right )^3 
\left | C^{(0){\rm eff}}_7 +  A^{(1)}(\mu) \right |^2 , 
\nonumber 
\end{eqnarray}
where~$G_F$ is the Fermi coupling constant, 
$\alpha = \alpha(0)=1/137$ is the fine-structure constant,
$m_{b, {\rm pole}}$ is the pole $b$-quark mass, 
$M$~and $m_{K^*}$ are the $B$- and $K^*$-meson masses, 
and~$\tau_B$ is the lifetime of the~$B^0$- or $B^+$-meson which
have the following world averages 
(in picoseconds)~\cite{BlifetimeWG:2001}:
\begin{equation}
\tau_{B^0} = (1.546 \pm 0.018)~{\rm ps}, \qquad 
\tau_{B^+} = (1.647 \pm 0.016)~{\rm ps}. 
\end{equation}
The product of the CKM-matrix~$|V_{tb} V_{ts}^*|$ can be 
estimated from the unitarity fit of the quantity~\cite{Groom:2000in}: 
\begin{equation}
\left | \frac{V_{tb} V_{ts}^*}{V_{cb}} \right | = 0.976 \pm 0.010~, 
\label{eq:lam-ts-ratio}
\end{equation}
and the present measurements of the CKM matrix element 
$|V_{cb}| = 0.0406 \pm 0.0019$~\cite{Groom:2000in}. This yields  
\begin{equation}
\left | V_{tb} V_{ts}^* \right | = 0.0396 \pm 0.0020~. 
\label{eq:lam-ts}
\end{equation}
The quantity $\xi_\perp^{(K^*)}$ is the value of the 
$T_1^{(K^*)} (q^2)$ form factor in $B \to K^*$ transition in  
Eq.~(\ref{eq:defFF-T}) and evaluated at $q^2 = 0$ in the HQET limit.
For this study, we consider  $\xi_\perp^{(K^*)}$  as a free parameter and
we will extract its value from the current experimental data on $B \to K^*
\gamma$ decays.
Note that the quantity $\xi_\perp^{(K^*)}$ used here is normalized 
at the scale $\mu = m_{b, {\rm pole}}$ of the pole $b$-quark mass. The 
corresponding quantity in Ref.~\cite{Beneke:2001wa} is
defined at the scale $\mu = m_{b, {\rm PS}}$ involving the
potential-subtracted~(PS) $b$-quark mass~\cite{Beneke:1998rk,Beneke:1999fe}. 

The function~$ A^{(1)}$ in Eq.~(\ref{eq:DW(B-Kgam)}) can be
decomposed into the following three components: 
\begin{equation}
 A^{(1)} (\mu)  =   A_{C_7}^{(1)} (\mu) + 
 A_{\rm ver}^{(1)} (\mu) +  A_{\rm sp}^{(1)K^*} (\mu_{\rm sp})~. 
\label{eq:A1tb} 
\end{equation}
Here, $ A^{(1)}_{C_7}$ and $ A^{(1)}_{\rm ver}$ are the
$O (\alpha_s)$ (i.e. NLO) corrrections due to the Wilson
coefficient~$C_7^{\rm eff}$
and in the $b \to s \gamma$ vertex, respectively,  and
$ A^{(1) K^*}_{\rm sp}$ is the ${\cal O} (\alpha_s)$ hard-spectator
corrections to the $B \to K^* \gamma$ amplitude computed in this paper.
Their explicit expressions are as follows:
\begin{eqnarray}
 A_{C_7}^{(1)} (\mu) & = & \frac{\alpha_s (\mu)}{4 \pi} \, 
C^{(1){\rm eff}}_7 (\mu) , 
\label{eq:A1tb-C7} \\
 A_{\rm ver}^{(1)} (\mu) & = & \frac{\alpha_s (\mu)}{4 \pi}
\left \{
\frac{32}{81} \left [ 
13 C^{(0)}_2 (\mu) + 27 C^{(0) {\rm eff}}_7 (\mu) 
- 9 \, C^{(0) {\rm eff}}_8 (\mu) 
\right ] \ln \frac{m_b}{\mu} 
\right .
\label{eq:A1tb-ver} \\ 
& - & 
\left .
\frac{20}{3} \, C^{(0) {\rm eff}}_7 (\mu)
+ \frac{4}{27} \left ( 33 - 2 \pi^2 + 6 \pi i \right ) 
C^{(0) {\rm eff}}_8 (\mu) + r_2 (z) \, C^{(0)}_2 (\mu)  
\right \}, \qquad
\nonumber \\
 A_{\rm sp}^{(1)K^*} (\mu_{\rm sp}) & = &
\frac{\alpha_s (\mu_{\rm sp})}{4 \pi} \, 
\frac{2 \Delta F^{(K^*)}_\perp (\mu_{\rm sp})}{9 \xi_\perp^{(K^*)}}
\left \{
3 C^{(0) {\rm eff}}_7 (\mu_{\rm sp})
\right.
\label{eq:A1tb-sp} \\ 
& + & 
\left.
C^{(0){\rm eff}}_8 (\mu_{\rm sp}) 
\left [ 
1 - \frac{6 a_{\perp 1}^{(K^*)} (\mu_{\rm sp})}
         {\left < \bar u^{-1} \right >_\perp^{(K^*)} (\mu_{\rm sp})} 
\right ]
+ C^{(0)}_2 (\mu_{\rm sp}) \left [ 1 - 
\frac{h^{(K^*)} (z, \mu_{\rm sp})}
     {\left < \bar u^{-1} \right >_\perp^{(K^*)} (\mu_{\rm sp})} 
\right ]
\right \} .  
\nonumber 
\end{eqnarray}
 Note, the $O (\alpha_s)$
corrections arising from the relation between the $\overline{\rm MS}$
mass~$\bar m_b (\mu)$ and the pole mass~$m_{b, {\rm pole}}$ in the
operator~${\cal O}_7$ are included in the vertex corrections. 
As indicated above, the first two contributions in $A^{(1)} (\mu)$ should
be estimated at the scale of the $b$-quark mass $\mu \sim O(m_b)$, while the
hard-spectator correction should be evaluated at the characteristic 
scale~$\mu_{\rm sp} = \sqrt{\mu \Lambda_{\rm H}}$ of the gluon virtuality, 
where~$\Lambda_{\rm H} \simeq 0.5$~GeV is a typical hadronic 
scale of order~$\Lambda_{\rm QCD}$. 
The functions~$r_2 (z)$, where $z = m_c^2 / m_b^2$,  
and the Wilson coefficients in the above equations 
can be found in Refs.~\cite{Greub:1996tg,Chetyrkin:1997vx}. We recall that 
the function $h^{(K^*)} (z)$ from the hard spectator corrections is
complex in the region $0 < z < 1/4$, likewise the 
function $r_2 (z)$ from the vertex corrections. 
The non-asymptotic corrections in the $K^*$-meson wave-function 
$6 a_{\perp 1}^{(K^*)} / \left < \bar u^{-1} \right >_\perp^{(K^*)}$
reduce the coefficient of  the anomalous chromomagnetic moment
$C_8^{(0) {\rm eff}}(\mu_{\rm sp})$ by about~20\%. 

\begin{figure}[bt]
\centerline{\epsfxsize=.45\textwidth
            \epsffile{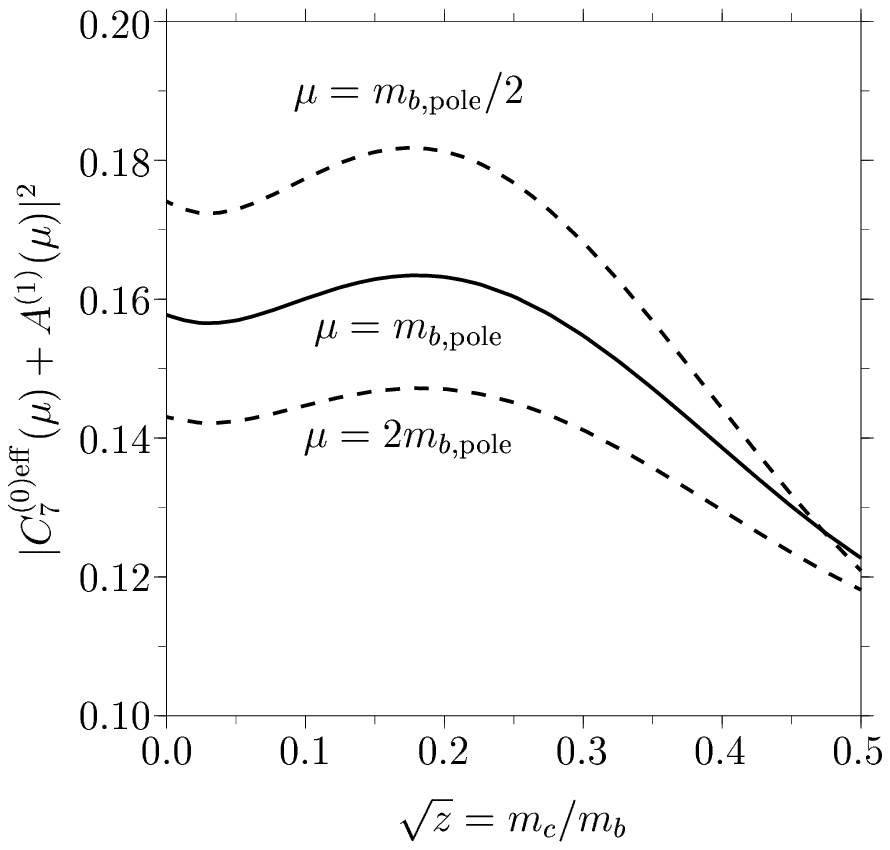} \quad
            \epsfxsize=.45\textwidth
            \epsffile{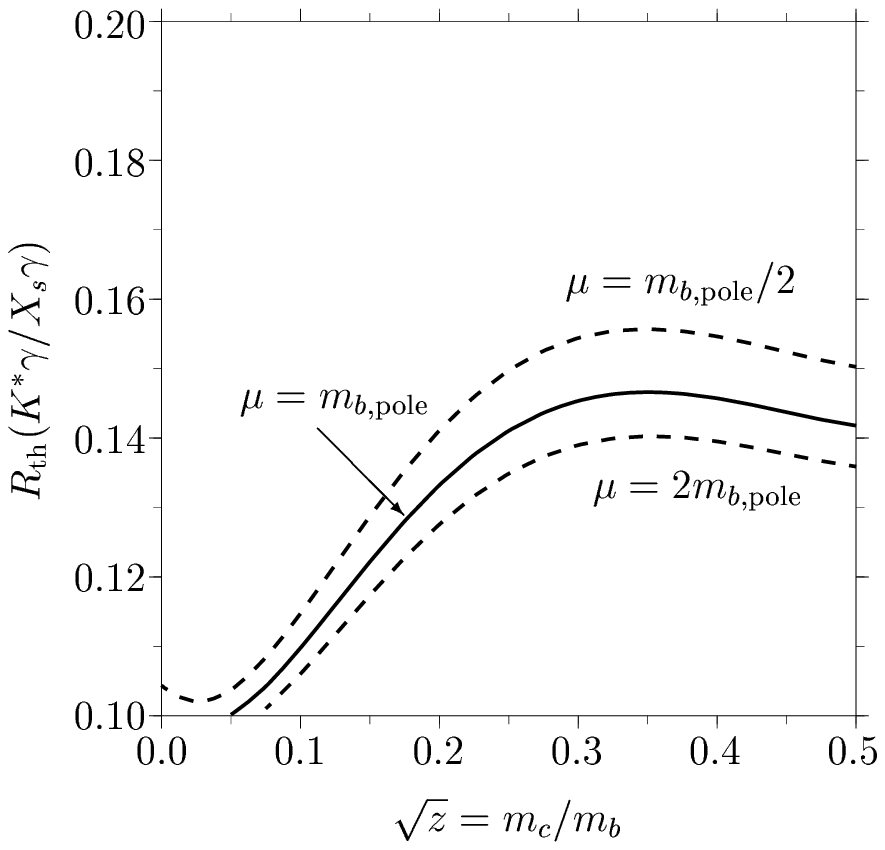}}
\caption{The effective coefficient squared in ${\cal B}(B \to K^* \gamma)$
         appearing in Eq.~(\ref{eq:DW(B-Kgam)}) (left figure)
         and the ratio of the exclusive $B \to K^* \gamma$ 
         to the inclusive $B \to X_s \gamma$ branching ratios
         defined in Eq.~(\ref{eq:Br(b->s+gam)-th}) (right 
         figure) plotted against the ratio $m_c/m_b$ for three values of 
         the scale $\mu$.}
\label{fig:mass-ratio-dependence}
\end{figure}

We now estimate numerically the importance of the $O (\alpha_s)$ 
contributions in the $B \to K^* \gamma$ decay amplitude. It is convenient 
to decompose the vertex correction $ A^{(1)}_{\rm ver} (\mu)$ and the 
hard-spectator correction $ A^{(1)K^*}_{\rm sp} (\mu)$ into
the factorizable~$ A^{(1)}_{\rm ver,f} (\mu)$ 
and~$ A^{(1)K^*}_{\rm sp,f} (\mu)$ and 
non-factorizable $ A^{(1)}_{\rm ver,nf} (\mu)$ and 
$ A^{(1)K^*}_{\rm sp,nf} (\mu)$  parts. So as not to cause any
confusion, our definition is that the first two
depend on the effective Wilson coefficient $ C^{(0) {\rm
eff}}_7$, and the last two involve the rest of the Wilson
coefficients. For the central
values of the parameters shown in Table ~\ref{tab:xiKs-B0}, and the
indicated values of the the quantities $m_c/m_b$ and $\mu$,  
the Wilson coefficient~$C^{(0) {\rm eff}}_7$ in the leading 
order and the $O (\alpha_s)$ corrections from 
Eqs.~(\ref{eq:A1tb-C7})--(\ref{eq:A1tb-sp}) assume the values 
presented in Table~\ref{tab:Ks-amplitude}. 
%
%
\begin{table}[tb]
\caption{Input parameters and individual contributions in the NLO
         amplitude for the decay $B \to K^* \gamma$ entering in
         Eqs.~(\ref{eq:DW(B-Kgam)}) and (\ref{eq:A1tb}). The entries 
         in the second and third columns refer to the choice of the 
         $b$-quark mass schemes~$\overline{\rm MS}$ and the pole mass, 
         respectively. The last two columns correspond to the  
         so-called PS-scheme $b$-quark mass used by Beneke et 
         al.~\cite{Beneke:2001at}, with $m_c/m_b=0.29$ and
         $m_c/m_b=0.22$.  Note that the total decay amplitude 
         squared, which is numerically presented in the last row, is
         truncated to the NLO accuracy.}   
\label{tab:Ks-amplitude}
\begin{center}
{\small 
\begin{tabular}{|l||c|c||c|c|}
\hline \hline 
$m_c/m_b$ & 0.29 & 0.29 & 0.29 & 0.22 \\ \hline
$\mu$ & $\bar m_b = 4.27$~GeV & $m_{b, {\rm pole}} = 4.65$~GeV & 
$m_{b, {\rm PS}} = 4.6$~GeV & $m_{b, {\rm PS}} = 4.6$~GeV  
\\ \hline 
$C^{(0) {\rm eff}}_7 (\mu)$ & 
$- 0.320$ & $- 0.315$ & $- 0.318$ & $- 0.318$ \\ \hline 
$A^{(1)}_{C_7} (\mu)$ 
& $+ 0.010$ & $+ 0.009$ & $+ 0.010$ & $+ 0.010$ \\ 
$A^{(1)}_{\rm ver,f} (\mu)$ & 
$+ 0.032$ & $+ 0.036$ & $+ 0.024$ & $+ 0.024$ \\ 
$ A^{(1)}_{\rm ver,nf} (\mu)$ & $- 0.076 - i \, 0.016$ & 
$- 0.083 - i \, 0.016$ & $- 0.085 - i \, 0.016$ & 
$- 0.103 - i \, 0.025$ \\ 
$ A^{(1)K^*}_{\rm sp,f} (\mu_{\rm sp}) $
& $- 0.038$ &  $- 0.036$ & $- 0.035$ & $- 0.035$ \\ 
$ A^{(1)K^*}_{\rm sp,nf} (\mu_{\rm sp}) $ & $- 0.016 - i \, 0.016$  
& $- 0.016 - i \, 0.015$ & $- 0.014 - i \, 0.014$ & 
$- 0.008 - i \, 0.024$ \\ \hline 
$ A^{(1)} (\mu) $ & $- 0.087 - i \, 0.032$ 
& $- 0.090 - i \, 0.031$ & $- 0.100 - i \, 0.030$ & 
$- 0.111 - i \, 0.049$ \\ \hline 
$C^{(0) {\rm eff}}_7 +  A^{(1)}(\mu)$ & $- 0.407 - i \, 0.032$ & 
$- 0.405 - i \, 0.031$ & $- 0.418 - i \, 0.030$ & 
$- 0.429 - i \, 0.049$ \\ \hline 
$|C^{(0) {\rm eff}}_7 +  A^{(1)}(\mu)|^2$ & 
0.158 & 0.156 & 0.165 & 0.172 \\ \hline 
\hline 
\end{tabular} 
}
\end{center}
\end{table}
%
%
What concerns the spectator contributions $A^{(1)K^*}_{\rm sp,f}
(\mu_{\rm sp})$ and $ A^{(1)K^*}_{\rm sp,nf} (\mu_{\rm sp})$,
the quoted numbers in this table  make use of  
the QCD sum-rules motivated value for the nonperturbative quantity 
$\xi_\perp^{(K^*)} (0) = 0.35$~\cite{Beneke:2001at}.  
The values for the amplitude presented in the second and third columns 
are obtained for the same pole quark mass ratio 
$\sqrt z = m_c/m_b = 0.29$, but with $\bar{m}_b=4.27$ GeV and
$m_{b,{\rm pole}} = 4.65$ GeV, and calculating the strong coupling
$\alpha_s(\mu)$ in the two-loop approximation. The last two columns of
this table are calculated for comparison with the numerical results
by Beneke et al.~\cite{Beneke:2001at}. They  
are presented for the two values of the quark mass ratio, 
$m_c/m_b = 0.29$ (the ratio of the pole masses) and $m_c/m_b = 0.22$ 
(the ratio of the $\overline{\rm MS}$ $c$-quark mass to the pole 
$b$-quark mass~\cite{Gambino:2001ew}), but with $m_{b,{\rm PS}}=4.6$ GeV. 
Note that for the entries in the last two columns 
the three-loop strong coupling constant $\alpha_s$ was used as well as the
effect of the 
non-leading Wilson coefficients and the correction due to the 
$m_{b,{\rm PS}}$ mass~\cite{Beneke:1998rk,Beneke:1999fe}  
were taken into account in the~$A^{(1)K^*}_{\rm sp, nf}$ 
and~$ A^{(1)}_{\rm ver, f}$ parts of the amplitude, respectively.
A comparison of the last-but-one row in this table shows that for the 
same value of the quark mass ratio $m_c/m_b$ ($=0.29$), the total
amplitude has a 
negligible  dependence on the choice of the $b$-quark mass: $\overline{\rm
MS}$, pole, or the~PS $b$-quark mass. However, decreasing the
ratio~$m_c/m_b$ from 0.29 to 0.22, the amplitude is appreciably enhanced.
The dependence of the total decay amplitude squared
$|C_7^{(0) {\rm eff}} (m_{b,{\rm pole}}) +
 A^{(1) } (\bar m_{b,{\rm pole}})|^2$ (truncated to the $O
(\alpha_s)$ accuracy) on the mass ratio~$m_c/m_b$ is presented in
Fig.~\ref{fig:mass-ratio-dependence} (left plot). We also draw attention
to the marked scale-dependence of the amplitude squared (i.e., of the
branching ratio ${\cal B}(B \to K^* \gamma$)).
%
%
%
%
It is seen that for $z > 0.2$ the amplitude squared becomes sensitive 
to this mass ratio and falls down fast enough. A similar sensitivity was 
observed in the inclusive $B \to X_s \gamma$ 
decay rate~\cite{Gambino:2001ew}.

Thus, for the central values of the input parameters, we estimate 
the amplitude squared at the scale of the pole $b$-quark mass as
\begin{equation}
|C_7^{(0) {\rm eff}} (m_{b,{\rm pole}}) + 
 A^{(1) } (m_{b,{\rm pole}}) |^2 \simeq 0.156. 
\end{equation}
In Ref.~\cite{Beneke:2001at} such a detailed analysis
of the amplitude was not shown but the result was presented in the
form of the total amplitude squared: $|C_7|_{\rm NLO}^2 =
|C^{(0) {\rm eff}}_7 (m_{b,{\rm PS}}) +  
 A^{(1) } (m_{b,{\rm PS}}) |^2 = 0.175^{+0.029}_{-0.026}$.
This value has to be compared with the entries given in the last row
(the last two columns) in Table~\ref{tab:Ks-amplitude} . For the same
input parameters, these numbers are somewhat smaller than the ones
by Beneke et al~\cite{Beneke:2001at}. The source of this discrepancy is
to be attributed to the fact that we have calculated the branching ratio
by keeping only the $O(\alpha_s)$ corrections in the decay rates, 
whereas in Ref.~\cite{Beneke:2001at}, the amplitude is corrected to 
$O(\alpha_s)$\footnote{Private communication (M. Beneke).}.

To compare our evaluation of the amplitude for $B \to K^* \gamma$
decay with the one presented in the 
paper by Bosch and Buchalla~\cite{Bosch:2001gv}, we recall that their 
calculations were done in the approach where the QCD form factor 
$T_1^{(K^*)} (0, \mu)$ was used  and not its HQET/LEET
analog~$\xi^{(K^*)}_\perp (0)$.  The two form factors are related
in $O(\alpha_s)$ via the relation~\cite{Beneke:2001wa}: 
\begin{equation} 
T_1^{(K^*)} (0, \bar m_b) = \xi_\perp^{(K^*)} (0) 
\left ( 1 + \frac{\alpha_s (\bar m_b)}{3 \pi} 
\left [ \ln \frac{m_{b, {\rm pole}}^2}{\bar m_b^2} - 1 \right ]  
+ 
\frac{\alpha_s (\bar \mu_{\rm sp})}{6 \pi}  
\frac{\Delta F_\perp^{(K^*)} (\bar \mu_{\rm sp})}
     {\xi_\perp^{(K^*)} (0)}  
\right ) 
\simeq 1.08 \, \xi_\perp^{(K^*)} (0) ,  
\label{eq:FF-connection}
\end{equation}
where $\bar m_b$ is the $\overline{\rm MS}$ $b$-quark mass and 
$\bar \mu_{\rm sp} = \sqrt{\Lambda_{\rm H} \bar m_b} \simeq 1.47$~GeV. 
We also recall that the form factor $T_1^{(K^*)} (0, \mu)$ is 
a scale-dependent quantity. The $B \to K^* \gamma$ decay amplitude 
includes this from factor in combination with the running 
$\overline{\rm MS}$ $b$-quark mass~$\bar m_b (\mu)$, and the scale 
dependence of this product is governed by:
\begin{equation} 
\bar m_b (\mu) \, T_1^{(K^*)} (0, \mu) = 
\bar m_b (\bar m_b) \, T_1^{(K^*)} (0, \bar m_b) 
\left [ 
1 + \frac{\alpha_s (\mu)}{\pi} \, C_F \, 
\ln \frac{\bar m_b^2}{\mu^2} 
\right ] . 
\label{eq:mb-T1-scale}
\end{equation}
In addition to the form factor and $b$-quark mass, the amplitude  
also contains the quantity $C_7 (\mu) = 
C_7^{(0) {\rm eff}} (\mu) +  A^{(1) } (\mu)$. The transition to 
the QCD form factor and the use of running $b$-quark mass 
changes~$C_7 (\mu)$ in 
a way that all the factorizable $O (\alpha_s)$ corrections (i.e.,terms 
proportional to $C_7^{(0) {\rm eff}} (\mu)$ in~$A^{(1)}(\mu)$) are 
absorbed into~$T_1^{(K^*)} (0, \mu)$ and~$\bar m_b (\mu)$. With this
interpretation, we give below the numerical estimates of the various
contributions to the decay amplitude in $B \to K^* \gamma$ calculated by
us and compare them with the equivalent quantities in
Ref.~\cite{Bosch:2001gv}, given in the square brackets.
For a meaningful comparison, the values of the 
input parameters are taken from~Ref.~\cite{Bosch:2001gv} with 
$m_c/m_b = 1.3~{\rm GeV}/4.2~{\rm GeV} \simeq 0.31$:  
\begin{eqnarray} 
&& C^{(0) {\rm eff}}_7 (4.2~{\rm GeV}) = - 0.321
\quad [C_7^{\rm LO} = - 0.322], 
\nonumber \\
&&  A^{(1)}_{C_7} (4.2~{\rm GeV}) = + 0.011
\quad [\Delta C_7^{\rm NLO} = + 0.011], 
\nonumber \\ 
&&  A^{(1)}_{\rm ver, nf} (4.2~{\rm GeV}) = - 0.081 - i \, 0.015 
\quad [T_{1,8}^{I} = - 0.082 - i \, 0.015],  
\label{eq:A1tnum} \\
&&  A^{(1) K^*}_{\rm sp, nf} (1.4~{\rm GeV}) = - 0.014 - i \, 0.011 
\quad [T_{1,8}^{II} = - 0.014 - i \, 0.011]. 
\nonumber \\ 
&& C^{(0) {\rm eff}}_7 (4.2~{\rm GeV}) + 
 A^{(1)} (4.2~{\rm GeV}) = - 0.405 - i \, 0.025 \quad  
[a_7 (K^*\gamma) = - 0.407 - i \, 0.026]. 
\nonumber
\end{eqnarray}
We agree with the evaluation of these quantities reported in
Ref.~\cite{Bosch:2001gv}. To be precise, we get for the amplitude 
squared at the~$\overline{\rm  MS}$ $b$-quark mass scale: 
$|C_7|^2_{\rm NLO} = |C^{(0) {\rm eff}}_7|^2 + 
2 C^{(0) {\rm eff}}_7 {\rm Re} ( A^{(1)}) \simeq 0.157$,
to be compared with the corresponding value 
$|C_7|^2_{\rm NLO} \simeq 0.158$, which can be obtained from 
Eq.~(55) of Ref.~\cite{Bosch:2001gv}.

With the numerical estimates given above, and varying the parameters in
their stated ranges, we get the following
branching ratio for $B \to K^* \gamma$ decays: 
\begin{equation} 
{\cal B}_{\rm th} (B \to K^* \gamma) \simeq 
(7.3 \pm 1.1)\times 10^{-5} \, 
\left ( \frac{\tau_B}{1.6~{\rm ps}} \right ) 
\left ( \frac{m_{b,{\rm pole}}}{4.65~{\rm GeV}} \right )^2 
\left ( \frac{\xi_\perp^{(K^*)}}{0.35} \right )^2 
= (7.3 \pm 2.7) \times 10^{-5}, 
\label{eq:Br-Ksgam}
\end{equation}
where the enlarged error in the second equation reflects the assumed error 
in the nonperturbative quantity, $\xi_\perp^{(K^*)} (0) = 0.35 \pm 0.07$.  
This estimate of the branching ratio for $B \to K^* \gamma$ is to be
compared with the corresponding one from Ref.~\cite{Beneke:2001at} where 
a value ${\cal B}_{\rm th} (B \to K^* \gamma) = 7.9^{+3.5}_{-3.0} 
\times 10^{-5}$ is obtained. (Note that after some recalculations due 
to the differences in the definition of the form factor the value 
$7.1 \times 10^{-5}$ obtained for $\tau_{B^0} = 1.56~{\rm ps}$ 
in Ref.~\cite{Bosch:2001gv} becomes the same as the one in 
Ref.~\cite{Beneke:2001at}). If we use instead a value for the
$b$-quark pole mass $m_{b, {\rm pole}} = 4.8$~GeV, the branching 
ratio we get is ${\cal B}_{\rm th} (B \to K^* \gamma) \simeq 
(7.8 \pm 2.7) \times 10^{-5}$, which is in agreement with  
the ones presented in Refs.~\cite{Beneke:2001at,Bosch:2001gv}.
All these estimates in the LEET approach are  larger than
the experimental branching ratio for $B \to K^* \gamma$, though the
attendant theoretical error, estimated as~$\pm 40\%$, does not allow
to draw a completely quantitative conclusion.  

\begin{figure}[bt]
\centerline{\epsfxsize=.45\textwidth
            \epsffile{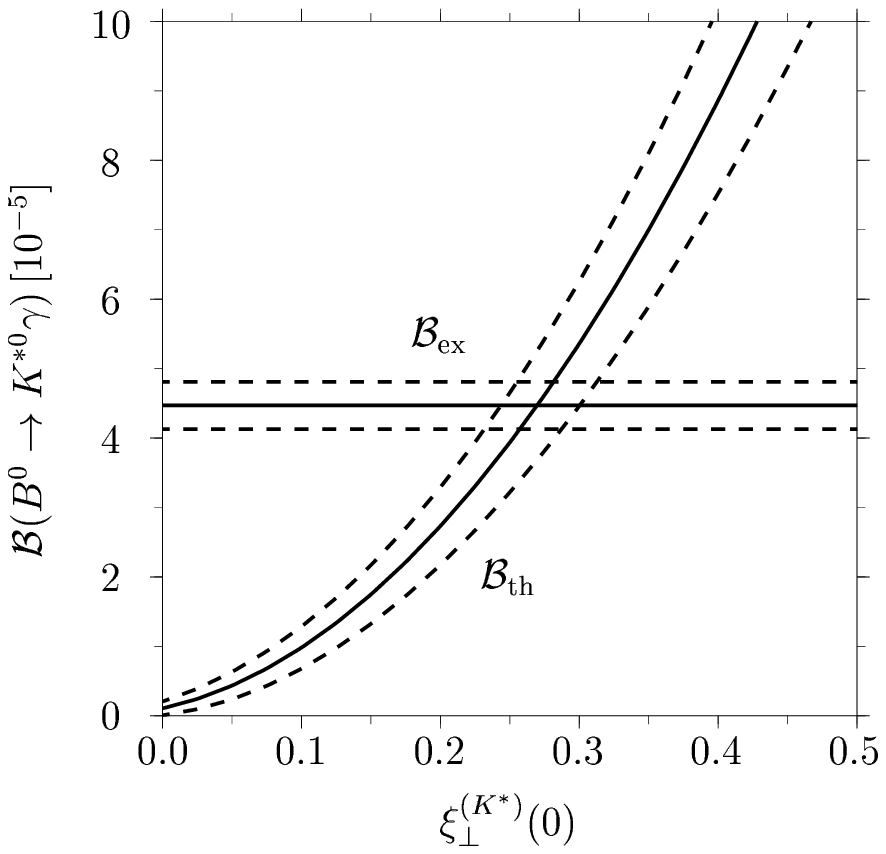} \qquad
            \epsfxsize=.45\textwidth
            \epsffile{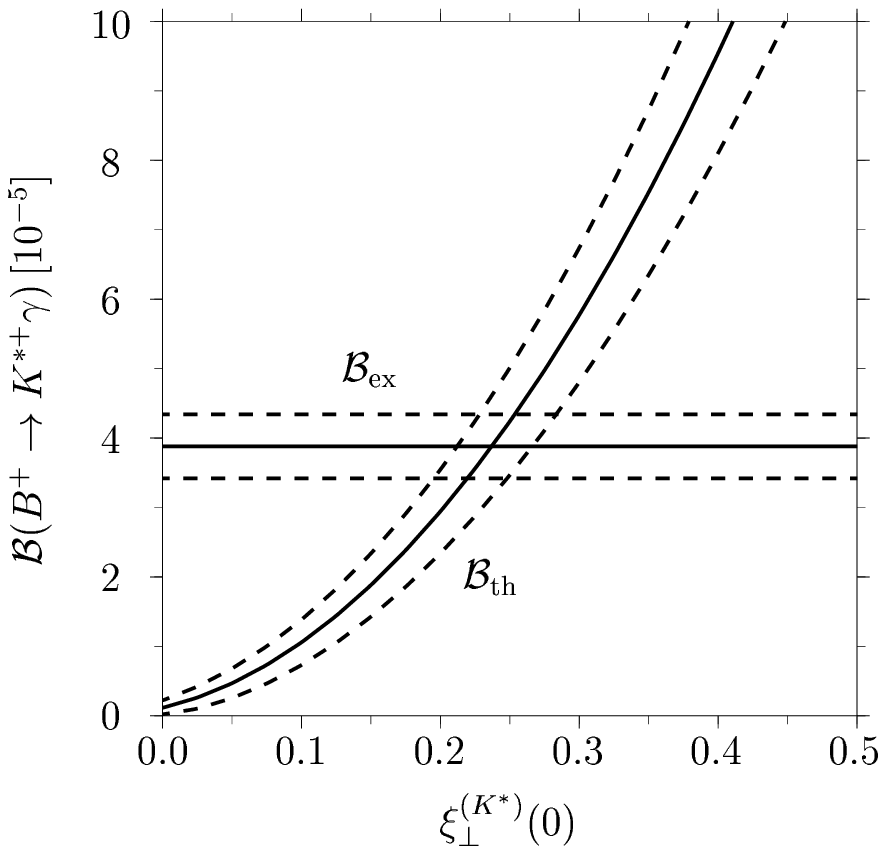}}
\caption{Branching ratios for the decays $B^0 \to K^{* 0} \gamma$
         (left figure)
         and $B^+ \to K^{* +} \gamma$ (right figure) as functions of 
         the LEET form factor $\xi_\perp^{(K^*)} (0)$.  
         Solid lines are the central experimental and theoretically 
         predicted values and the dotted lines delimit the~$\pm 1 \sigma$
         errors in experiment and theory, as discussed in the text.}  
\label{fig:Br-Ks}
\end{figure}

To quantify the price of agreement between the LEET approach and data, 
we determine the value of the LEET-form factor $\xi_\perp^{(K^*)} (0)$ 
from the current measurements. To that end, we show the
theoretically predicted branching ratios in the NLO accuracy for the
decays $B^0 \to K^{*0} \gamma$ (left figure) and $B^+ \to K^{*+} \gamma$ 
(right figure) in Fig.~\ref{fig:Br-Ks} and the corresponding 
measured branching ratios (horizontal bands),
where the solid lines are the central values 
and the dotted lines define their~$\pm 1 \sigma$ ranges.  
%
%
%
%
Theoretical uncertainties on the curves labeled as
${\cal B}_{\rm th}$ are estimated from all the parametric uncertainties,
detailed in Table~\ref{tab:xiKs-B0} for the decay
$B^0 \to K^{*0} \gamma$, where the last column contains the errors due 
to the variation of the input parameters resulting from the indicated
ranges in the second column. The pole $b$-quark mass is taken from a
recent estimate of the same in the NLO accuracy 
$m_{b, {\rm pole}} = (4.65 \pm 0.10)$~GeV~\cite{Kuhn:2001dm}, 
and the $B$-meson decay constant in the same accuracy is taken as  
$f_B = (200 \pm 20)$~MeV~\cite{Penin:2001ux,Jamin:2001fw}.
The enlarged error on the charm-to-bottom quark mass ratio $\sqrt z
= m_c/m_b = 0.27 \pm 0.06$ in Table ~\ref{tab:xiKs-B0} deserves a comment.
We recall from a recent discussion of the inclusive $B \to X_s \gamma$
branching ratio in the
literature~\cite{Gambino:2001ew} that, within the theoretical accuracy 
of the present calculations, there exists an intrinsic uncertainty in the
interpretation of the quantity~$\sqrt z$. 
It has been recently argued in ~\cite{Gambino:2001ew} that  
the inclusive branching ratio for $B \to X_s \gamma$ is uncertain, 
depending on whether this ratio is
interpreted as the one involving the  pole masses, $\sqrt z = m_{c,{\rm
pole}}/m_{b,{\rm pole}}$, or as  $\sqrt z =
\bar{m}_c(\mu)/m_{b,{\rm pole}}$ involving the charm quark mass in the 
$\overline{\rm MS}$ scheme with $m_c < \mu < m_b$. Typical range for
the pole mass interpretation is $\sqrt z = 0.29 \pm 0.02$, 
while in the latter case the corresponding range is 
$\sqrt z = 0.22 \pm 0.04$~\cite{Gambino:2001ew}. 
This translates into an uncertainty of about~11\% in the inclusive decay 
rate. Not surprisingly, a corresponding sensitivity on~$\sqrt z$ is
also present in the decay rate for the exclusive radiative decays.
This has been shown through the $z$-dependence of the
matrix element squared for the exclusive decay $B \to K^* \gamma$ 
in Fig.~\ref{fig:mass-ratio-dependence} (left plot). 
To take into account the uncertainty in the decay rate from this
source, we use $\sqrt z = 0.27 \pm 0.06$.     
It is seen from Table~\ref{tab:xiKs-B0} that the decay rate 
for $B \to K^* \gamma$ is not sensitive to the variations in
the $K^{*0}$-meson wave-function
parameters ($f_\perp^{(K^*)}$, $a_{\perp 1}^{(K^*)}$,
and~$a_{\perp 2}^{(K^*)}$) in the indicated ranges, and hence the
derived errors on ~$\xi_\perp^{(K^{*0})} (0)$ from these sources are
small.  To get the overall theoretical error on
$\xi_\perp^{(K^{*0})} (0)$, we have added all the individual
theoretical errors (given in
rows 3 through 11) in quadrature. The errors from the experimental
input quantities (first two rows) are given separately.  
%
%
\begin{table}[tb]
\caption{Central values of the parameters and their $\pm 1 \sigma$
         errors used in
         estimating the quantity~$\xi_\perp^{(K^*)} (0)$ and its error 
         $\delta \xi_\perp^{(K^*)} (0)$ from the branching ratio  
         ${\cal B}_{\rm exp} (B^0 \to K^{*0} \gamma)$. The parameters
         in extracting $\xi_\perp^{(K^*)} (0)$ in the 
         $B^+ \to K^{*+} \gamma$ decay differ essentially in the first two
         entries, and are discussed in the text.}
\label{tab:xiKs-B0}
\begin{center}
\begin{tabular}{|l|l|l|} 
\hline \hline 
Parameter & Value & $\delta \xi_\perp^{(K^{*0})} (0)$ 
\\ \hline
${\cal B}_{\rm exp} (B^0 \to K^{*0} \gamma)$ 
       & $(4.44 \pm 0.35) \times 10^{-5}$ & $\pm 0.013$ \\ 
$\tau_{B^0}$               & ($1.546 \pm 0.018$) ps & $\pm 0.002$ 
\\ \hline
$\left | V_{tb} V_{ts}^* \right |$ & $0.0396 \pm 0.0020$ 
       & $\pm 0.016$ \\  
$m_{b, {\rm pole}}$        & ($4.65 \pm 0.10$) GeV & $\pm 0.006$ \\ 
$\sqrt z = m_c/m_b$        & $0.27 \pm 0.06$ & $^{+0.010}_{-0.003}$ \\ 
$f_B$                      & ($200 \pm 20$) MeV & $\pm 0.005$ \\ 
$\lambda_{B,+}^{-1}$       & ($3 \pm 1$) GeV$^{-1}$ & $\pm 0.017$ \\ 
$f_\perp^{(K^*)}$(1 GeV)   & ($185 \pm 10$) MeV & $\pm 0.003$ \\ 
$a_{\perp 1}^{(K^*)}$(1 GeV)  & $0.20 \pm 0.05$ & $\pm 0.002$ \\
$a_{\perp 2}^{(K^*)}$(1 GeV)  & $0.04 \pm 0.04$ & $\pm 0.001$ \\ 
$\mu / m_{b, {\rm pole}}$  & $0.5 - 2.0$        & $\pm 0.013$ \\
\hline 
$\xi_\perp^{(K^{*0})} (0)$        & 
\multicolumn{2}{|l|}{$0.269 \pm 0.028 ({\rm th}) \pm 0.013 ({\rm exp})$}     
\\ \hline \hline
\end{tabular}
\end{center}
\end{table}
%
%
The form factor~$\xi_\perp^{(K^{*+})} (0)$ extracted 
from the $B^+ \to K^{*+} \gamma$ branching ratio differs somewhat 
from the one presented in Table~\ref{tab:xiKs-B0} due to the difference 
in the~$B^+$- and~$B^0$-meson lifetimes and the corresponding
experimental branching ratios.  Thus, present data and the NLO expressions
in the  LEET approach yield the following values for~$\xi_\perp^{(K^*)}
(0)$: 
\begin{eqnarray} 
\xi_\perp^{(K^{*0})} (0) & = & 0.27 \pm 0.04~,  
\label{eq:xi-Ks-Branch} \\ 
\xi_\perp^{(K^{*+})} (0) & = & 0.24 \pm 0.04~.  
\nonumber 
\end{eqnarray}
The central value of~$\xi_\perp^{(K^{*0})} (0)$ is marginally larger 
than~$\xi_\perp^{(K^{*+})} (0)$ but they are consistent with each other 
within~$1 \sigma$. It has been argued recently in
Ref.~\cite{Kagan:2001zk} that the small difference may be accounted for
by taking into account the isospin-violating contributions from an
annihilation contribution in the penguin operator $O_6$ in the effective
weak Hamiltonian. With improved precision, it may become necessary to
include this contribution. As already stated, we have
ignored such isospin-violating contributions for the estimates
presented for $B \to K^* \gamma$ decays in this paper. We also note that
the above determination of
$\xi_\perp^{(K^{*0})} (0)$ and $\xi_\perp^{(K^{*+})} (0)$ are in
fair agreement with the one  presented in Ref.~\cite{Beneke:2001at}, 
($\xi_\perp^{(K^*)} (0) = 0.24 \pm 0.06$).

The non-perturbative parameter~$\xi_\perp^{(K^*)} (0)$ 
can also be extracted from the ratio of the decay rates 
for the exclusive decay $B \to K^* \gamma$ and the 
inclusive decay $B \to X_s \gamma$. In fact, one hopes that
some of the parametric uncertainties may be eliminated, or at
least reduced, 
in this ratio. Two particular parameters in point are the quark mass 
ratio~$\sqrt z$ and the product of the CKM elements $|V_{tb}V_{ts}^*|$. 
Also, in the experimental measurements some common systematic errors 
may be eliminated from the ratio. The current world average of the 
branching ratio for the inclusive $B \to X_s\gamma$ decay 
is~\cite{cleo,aleph,belle}:
\begin{equation}
{\cal B}_{\rm exp} (B \to X_s \gamma) = 
(3.22 \pm 0.40) \times 10^{-4} , 
\label{eq:Br(b->s+gam)-exp}
\end{equation}
which yields the following  exclusive to inclusive decay width ratio:
\begin{equation}
R_{\rm exp} (K^* \gamma/X_s \gamma) \equiv 
\frac{\bar {\cal B}_{\rm exp} (B \to K^* \gamma)}
     {{\cal B}_{\rm exp} (B \to X_s \gamma)} = 0.13 \pm 0.02 , 
\label{eq:Ks/Xs-exp}  
\end{equation}
where, as for the numerator, we have used an  experimental branching
ratio averaged over the
 $B^\pm \to K^{*\pm} \gamma$ and $B^0(\bar{B}^0) \to
K^{*0}(\bar K^{*0}) \gamma$ decays: $\bar {\cal B}_{\rm exp} 
(B \to K^* \gamma) = (4.22 \pm 0.28) \times 10^{-5}$. 

\begin{figure}[bt]
\centerline{
            \epsffile{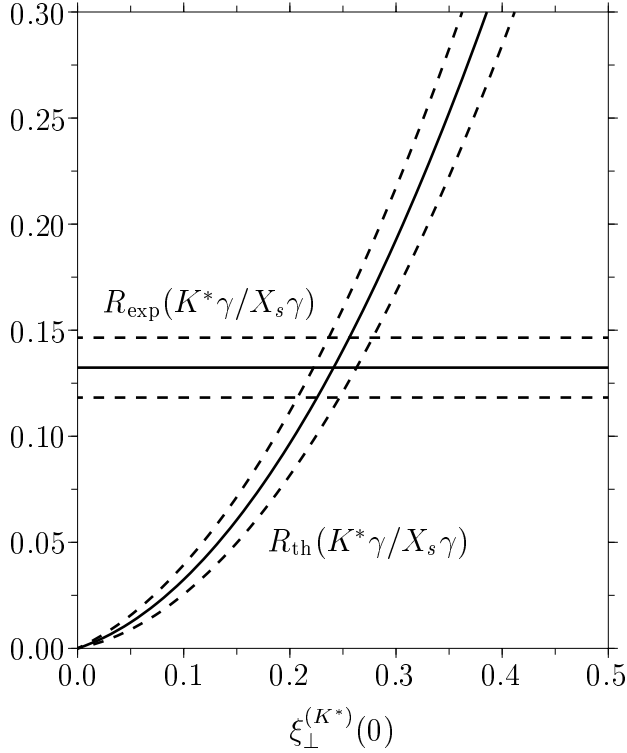}} 
\caption{The ratio of the exclusive $B \to K^* \gamma$ to the inclusive  
         $B \to X_s \gamma$ branching ratios, defined in
         Eq.~(\ref{eq:Br(b->s+gam)-th}), plotted as a function of the 
         LEET form factor~$\xi_\perp^{(K^*)} (0)$ and the current
         experimental measurement of this ratio. The solid lines are the
         central experimental 
         and theoretically predicted values and the dotted lines delimit 
         the $\pm 1 \sigma$ bands.} 
\label{fig:RatioKs}
\end{figure}

The theoretical expression for the inclusive branching ratio, 
${\cal B}_{\rm th} (B \to X_s \gamma)$, can be written 
as~\cite{Chetyrkin:1997vx,Kagan:1999ym,Gambino:2001ew}: 
\begin{equation} 
{\cal B}_{\rm th} (B \to X_s \gamma) = \tau_B \, 
\frac{G_F^2 \alpha m_{b, {\rm pole}}^5}{32 \pi^4}  
\left | V_{ts}^* V_{tb}\right |^2 
\left [ 
\left | 
C^{(0){\rm eff}}_7 (\mu) +  A^{(1)}_{\rm incl} (\mu)  
\right |^2 
+ B (\mu, \delta) 
\right ] ,
\label{eq:Br(B-Xs-gam)} 
\end{equation}
where $\delta$ is a lower cut on the photon energy in the 
bremsstrahlung corrections,  
$E_{\gamma, {\rm min}} = (1 - \delta) \, m_{b,{\rm pole}}/2$, 
in the massless limit of the final $s$-quark. 
The function~$ A^{(1)}_{\rm incl}$ 
is the $O (\alpha_s)$ corrections to the effective $b s \gamma$ vertex 
while $B (\mu, \delta)$ describe the bremsstrahlung corrections 
originated by the emission of a real gluon. To get   
the total branching ratio, we should 
integrate over all possible photon energies which corresponds to 
the limit $\delta \to 1$ in Eq.~(\ref{eq:Br(B-Xs-gam)}). In this 
limit, the vertex and bremsstrahlung corrections 
are~\cite{Chetyrkin:1997vx,Gambino:2001ew,Kagan:1999ym}: 
\begin{eqnarray}
 A^{(1)}_{\rm incl} (\mu) & = & \frac{\alpha_s (\mu)}{4 \pi} 
\left \{ C^{(1) {\rm eff}}_7 (\mu) + r_2 (z) \, C^{(0)}_2 (\mu)  
\right.
\label{eq:incl-vertex} \\ 
& - & \frac{2}{9} (39 + 4 \pi^2) \, C^{(0) {\rm eff}}_7 (\mu)
+ \frac{4}{27} (33 - 2 \pi^2 + 6 \pi i) \, C^{(0) {\rm eff}}_8 (\mu)
\nonumber \\
& + & 
\left . 
\frac{32}{81} \left [ 
13 C^{(0)}_2 (\mu) + 27 C^{(0) {\rm eff}}_7 (\mu) 
- 9 C^{(0) {\rm eff}}_8 (\mu) \right ] \ln \frac{m_b}{\mu} 
\right \} ,  
\nonumber \\ 
\left . B (\mu, \delta) \right |_{\delta \to 1} & = & 
\frac{\alpha_s (\mu)}{\pi} \sum_{i \le j = 2,7,8} 
\left . f_{ij} (\delta) \right |_{\delta \to 1} 
C_i^{(0) {\rm eff}} (\mu) C_j^{(0) {\rm eff}} (\mu) , 
\label{eq:incl-Bremss}
\end{eqnarray}
where the explicit forms of the 
functions $r_2 (z)$ and $f_{i j} (\delta)$ can be found in 
Refs.~\cite{Chetyrkin:1997vx,Kagan:1999ym,Gambino:2001ew}.
In the evaluation of the function~$f_{88} (\delta)$, which is divergent 
both in the $s$-quark massless limit and $\delta \to 1$, we take, 
as advocated in \cite{Kagan:1999ym}, $m_b/m_s \simeq 50$ and $\delta =
0.9$.  At NLO, the ratio $R(K^*\gamma/X_s \gamma)$ can be written as
follows:
\begin{eqnarray}
R_{\rm th} (K^* \gamma/X_s \gamma) & = & 
\frac{\bar {\cal B}_{\rm th} (B \to K^* \gamma)}
     {{\cal B}_{\rm th} (B \to X_s \gamma)} 
= \left [ \xi_\perp^{(K^*)}\right ]^2 
\left [ \frac{M}{m_{b,{\rm pole}}} \right ]^3  
\left \{ 1 + \frac{\alpha_s (\mu)}{\pi} \,
\left [ 1 + \frac{4 \pi^2}{9} \right ]
\right. \
\qquad 
\label{eq:Br(b->s+gam)-th} \\ 
& + & 
\left. 
\frac{2 \, {\rm Re} 
      \left [  A^{(1)K^*}_{\rm sp} (\mu_{\rm sp}) \right ]}
     {C^{(0) {\rm eff}}_7 (\mu)} 
- \frac{\alpha_s (\mu)}{\pi} \!\!\! \sum_{i \le j = 2,7,8} \!\! 
\left . f_{i j} (\delta ) \right |_{\delta \to 1}    
\frac{C^{(0)}_i (\mu) C^{(0)}_j (\mu)}
     {|C^{(0) {\rm eff}}_7 (\mu)|^2}   
\right \} .  
\nonumber  
\end{eqnarray}
The dependencies of this quantity on the charm-to-bottom quark mass 
ratio~$\sqrt z$ and on the form factor~$\xi_\perp^{(K^*)} (0)$ 
are presented in Figs.~\ref{fig:mass-ratio-dependence} 
(the right plot) and~\ref{fig:RatioKs}, respectively. It is seen that the
ratio $R_{\rm th} (K^* \gamma/X_s \gamma)$  
has a weaker dependence on the ratio~$\sqrt z$ 
than the branching ratio for the $B \to K^* \gamma$ decay, 
as this dependence is partially compensated in the last two terms in
Eq.~(\ref{eq:Br(b->s+gam)-th}).
%
%
%
%
The numerical analysis allows to estimate the nonperturbative 
quantity~$\xi_\perp^{(K^*)} (0)$ as: 
\begin{equation}
\xi_\perp^{(K^*)} (0) = 0.24 \pm 0.03~, 
\label{eq:xi-Ks-ratio}
\end{equation}
in which half the error is contributed by experiment. This  
coincides with the estimate of this quantity from the $B^+ \to
K^{*+} \gamma$ branching ratio presented in Eq.~(\ref{eq:xi-Ks-Branch}),
where the error is dominated by theory.  
The average of the three extracted values [Eqs.~(\ref{eq:xi-Ks-Branch}) 
and~(\ref{eq:xi-Ks-ratio})] is: 
\begin{equation}
\bar \xi_\perp^{(K^*)} (0) = 0.25 \pm 0.04 , \qquad 
\left [ \bar T_1^{(K^*)} (0, \bar m_b) = 0.27 \pm 0.04 \right ] .
\label{eq:xi-Ks-average}
\end{equation}
This estimate is significantly smaller than 
the corresponding predictions from the QCD sum rules analysis
$T_1^{(K^*)} (0) = 0.38 \pm 0.06$~\cite{Ali:2000mm,Ball:1998kk} 
and from the lattice simulations 
$T_1^{(K^*)} (0) = 0.32^{+0.04}_{-0.02}$~\cite{DelDebbio:1998kr}.
The reason for this mismatch is not obvious and this point deserves
further theoretical study. We shall make some comments in the 
concluding section.

\subsection{$B \to \rho \gamma$ Decays} 
\label{ssec:B-rho-gamma} 

After comparing the LEET-based approach with experiment in $B \to K^*
\gamma$ decays,  we now present the effect of including the hard-spectator
corrections on the branching ratios in $B \to \rho \gamma$ decays and
in the isospin-violating ratios and CP-asymmetries in the decay rates.

We recall that ignoring the perturbative QCD corrections to the penguin
amplitudes the ratio of the branching ratios for the charged and neutral
$B$-meson decays can be written as~\cite{Grinstein:2000pc,Ali:2000zu}
\begin{equation}
\frac{{\cal B} (B^- \to \rho^- \gamma)}
     {2 {\cal B} (B^0 \to \rho^0 \gamma)}
\simeq \left |1 + \epsilon_A {\rm e}^{i \phi_A} \,
\frac{V_{ub} V_{ud}^*}{V_{tb} V_{td}^*} \right |^2 ,
\label{eq:BR-ratio}
\end{equation}
where $\epsilon_A {\rm e}^{i \phi_A}$ includes the dominant
$W$-annihilation and possible sub-dominant long-distance contributions. 
Estimates in the framework of the light-cone QCD sum rules yield 
typically~\cite{Ali:1995uy}: $\epsilon_A \simeq - 0.30 \pm 0.07$ 
and $\epsilon_A \simeq 0.03 \pm 0.01$ 
for the decays $B^- \to \rho^- \gamma$ and $B^0 \to \rho^0 \gamma$, 
respectively, with which a recent calculation also
 agrees~\cite{Grinstein:2000pc}. However, analyses done within the HQET/LEET 
framework~\cite{Bosch:2001gv,Kagan:2001zk} show that the weak annihilation 
contribution in the $B^- \to \rho^- \gamma$ has an opposite 
sign. We have also checked it using
the HQET/LEET framework and agree
with the positive sign of $\epsilon_A$.  Taking this into account, we 
shall use the value $\epsilon_A \simeq + 0.30 \pm 0.07$ in further 
numerical estimations.  
The strong interaction phase~$\phi_A$ disappears in ${\cal O} (\alpha_s)$
in the chiral limit ~\cite{Grinstein:2000pc}. Henceforth we set $\phi_A =
0$;
the isospin-violating correction depends on the unitarity triangle
phase~$\alpha$ due to the relation:
\begin{equation}
\frac{V_{ub} V_{ud}^*}{V_{tb} V_{td}^*} =
- \left | \frac{V_{ub} V_{ud}^*}{V_{tb} V_{td}^*} \right |
{\rm e}^{i \alpha}
= F_1 + i F_2 ~.
\label{eq:UT-phase}
\end{equation}
The next-to-leading order vertex corrections for the branching
ratios of the exclusive decays $B^\pm \to \rho^\pm \gamma$ and
$B^0 \to \rho^0 \gamma$ can be derived from the corresponding
calculations of the inclusive decays
$B \to X_s \gamma$, discussed in the previous subsection,  and
$B \to X_d \gamma$~\cite{Ali:1998rr}. Ignoring the hard-spectator
corrections, but including the annihilation contribution, the result was
given in Ref.~\cite{Ali:2000zu}:
\begin{eqnarray}
{\cal B}_{\rm th} (B^\pm \to \rho^\pm \gamma) & = & 
\tau_{B^+} \, \Gamma_{\rm th} (B^\pm \to \rho^\pm \gamma) 
\label{eq:DecayWidth} \\
& = & 
\tau_{B^+} \, \frac{G_F^2 \alpha |V_{tb} V_{td}^*|^2}{32 \pi^4} \, 
m_{b,{\rm pole}}^2 \, M^3 \, 
\left ( 1 - \frac{m_\rho^2}{M^2} \right )^3 
\left [ \xi_\perp^{(\rho)} (0) \right ]^2 \, 
\nonumber \\ 
& \times &
\left \{ (C^{(0){\rm eff}}_7 + A^{(1)t}_R)^2 +
(F_1^2 + F_2^2) \, (A^u_R + L^u_R)^2
\right.
\nonumber \\
& + &
\left.
2 F_1 \, [ C^{(0){\rm eff}}_7 (A^u_R + L^u_R)
        + A^{(1)t}_R L^u_R ]
\pm 2 F_2 \, [ C^{(0){\rm eff}}_7 A^u_I - A^{(1)t}_I L^u_R ]
\right \}~,
\nonumber
\end{eqnarray}
where~$\xi_\perp^{(\rho)} (0)$ is the analogue of the quantity 
$\xi_\perp^{(K^*)} (0)$, discussed at length in the context 
of the decays $B \to K^* \gamma$, and 
$L^u_R = \epsilon_A \, C^{(0) {\rm eff}}_7$. Including the 
${\cal O} (\alpha_s)$ hard-spectator corrections to the matrix 
elements evaluated at the scale~$\mu$, the function~$A^{(1) t}$ 
is modified, and in addition the $u$-quark contribution~$A^u$ 
from the penguin can no longer be ignored. We decompose the 
amplitude~$A^{(1)t} (\mu)$ in its three contributing parts:
\begin{eqnarray}
A^{(1)t} (\mu) & = &  A^{(1)}_{C_7} (\mu) + 
 A^{(1)}_{\rm ver} (\mu) + A^{(1)\rho}_{\rm sp} (\mu_{\rm sp}), 
\label{eq:A1t}
\end{eqnarray}
where the functions~$ A^{(1)}_{C_7} (\mu)$
and~$ A^{(1)}_{\rm ver} (\mu)$ have been defined in
Eqs.~(\ref{eq:A1tb-C7}) and~(\ref{eq:A1tb-ver}), respectively,
in the context of the $B \to K^* \gamma$ decays.
The functions $A^{(1)\rho}_{\rm sp}$ and $A^u (\mu)$ are specific to the
decays $B \to \rho \gamma$, and both involve hard spectator contributions:  
\begin{eqnarray} 
A^{(1)\rho}_{\rm sp} (\mu_{\rm sp}) & = &
\frac{\alpha_s (\mu_{\rm sp})}{18 \pi} \, 
\frac{\Delta F_\perp^{(\rho)} (\mu_{\rm sp})}{\xi_\perp^{(\rho)} (0)} 
\left [
3 C^{(0) {\rm eff}}_7 (\mu_{\rm sp})
+ C^{(0){\rm eff}}_8 (\mu_{\rm sp})
+ C^{(0)}_2 (\mu_{\rm sp})
\left [ 1 - 
\frac{h^{(\rho)} (z, \mu_{\rm sp})}
     {\left < \bar u^{-1} \right >^{(\rho)}_\perp (\mu_{\rm sp})} 
\right ]
\right ] ,
\nonumber \\
A^u (\mu) & = & \frac{\alpha_s (\mu)}{4 \pi} \, C^{(0)}_2 (\mu) \,
\left [ r_2 (z) - r_2 (0) \right ]
- \frac{\alpha_s (\mu_{\rm sp})}{18 \pi} \, C^{(0)}_2 (\mu_{\rm sp}) \,
\frac{\Delta F_\perp^{(\rho)} (\mu_{\rm sp})}
     {\xi_\perp^{(\rho)} (0)} \, 
\frac{h^{(\rho)} (z, \mu_{\rm sp})}
     {\left < \bar u^{-1} \right >^{(\rho)}_\perp (\mu_{\rm sp})} ,
\qquad 
\label{eq:Au}
\end{eqnarray}
The terms proportional to $\Delta F_\perp^{(\rho)} (\mu_{\rm sp})$ above 
are the $O (\alpha_s)$ hard-spectator corrections which should be 
evaluated at the typical scale $\mu_{\rm sp} = \sqrt{\mu \Lambda_{\rm H}}$ 
of the gluon virtuality. The complex function~$r_2 (z)$ of the 
parameter $z = m_c^2 / m_b^2$, and the Wilson coefficients in the above 
equations can be found in Refs.~\cite{Greub:1996tg,Chetyrkin:1997vx}; 
the function~$h^{(\rho)} (z, \mu)$ and the dimensionless quantity 
$\Delta F_\perp^{(\rho)} (\mu)$ are defined through 
Eqs.~(\ref{eq:Di5-decomp}) and~(\ref{eq:DFperp}), respectively.  
The subscripts~$R$ and~$I$ in Eq.~(\ref{eq:DecayWidth}) denote the real 
and imaginary parts of~$A^{(1)t}$ and~$A^u$. The hard-spectator 
corrections contribute to both the real and imaginary parts of~$A^{(1)t}$
and~$A^u$. They do not depend on the charge of the spectator quark
in~$B^0$- or $B^\pm$-mesons, and hence are 
isospin-conserving. Isospin violations enter mainly via the
$W$-annihilation contribution, and they are suppressed in
$B^0$-meson decays as $L^u_R (B^0) \ll L^u_R (B^\pm)$.

To simplify the theoretical expressions somewhat, and in view of the
smallness of the branching ratios, we would like to present our numerical
results in this section in terms of the charge-conjugate averaged
branching ratios: 
\begin{eqnarray} 
\overline{\cal B}_{\rm th} (B^\pm \to \rho^\pm \gamma) & = &  
\frac{1}{2} \left [ 
{\cal B}_{\rm th} (B^+ \to \rho^+ \gamma) + 
{\cal B}_{\rm th} (B^- \to \rho^- \gamma)
\right ], 
\label{eq:ccBr} \\ 
\overline {\cal B}_{\rm th} (B^0 \to \rho^0 \gamma) & = & 
\frac{1}{2} \left [ 
{\cal B}_{\rm th} (B^0 \to \rho^0 \gamma) +
{\cal B}_{\rm th} (\bar B^0 \to \rho^0 \gamma)
\right ] .  
\nonumber  
\end{eqnarray}
Theoretical expressions for these quantities can be easily obtained from
the branching  ratios~(\ref{eq:DecayWidth}) by neglecting the
sign-dependent part. Unless otherwise stated, the results shown below
imply an average over the charge-conjugate states. 

\begin{figure}[bt]
\centerline{\epsfxsize=.45\textwidth
            \epsffile{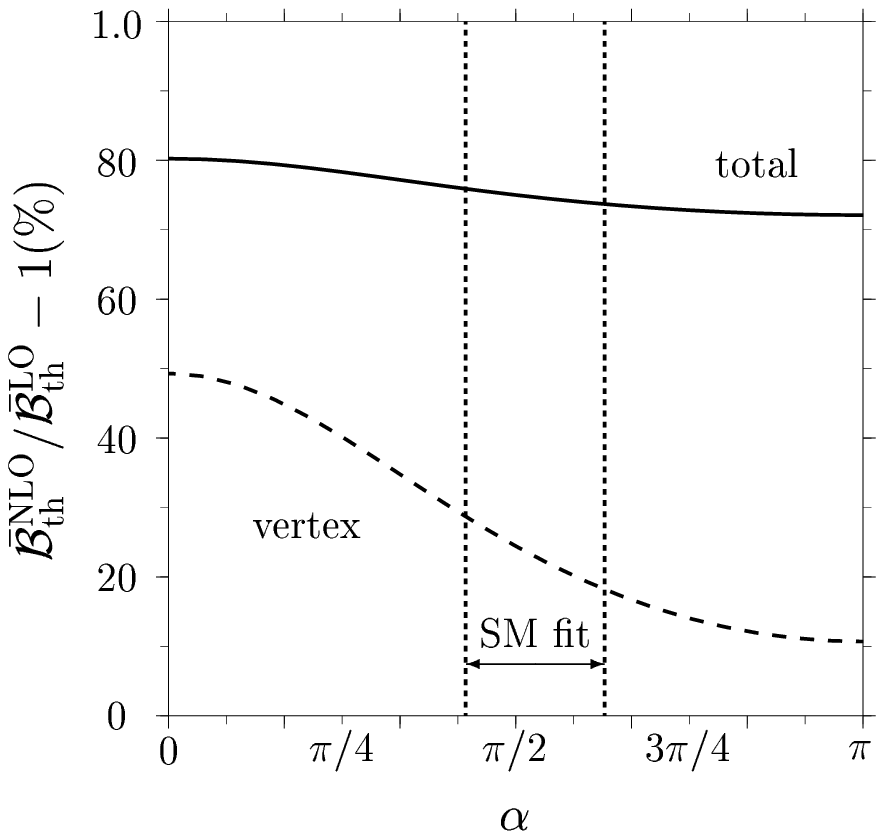} \qquad
            \epsfxsize=.45\textwidth
            \epsffile{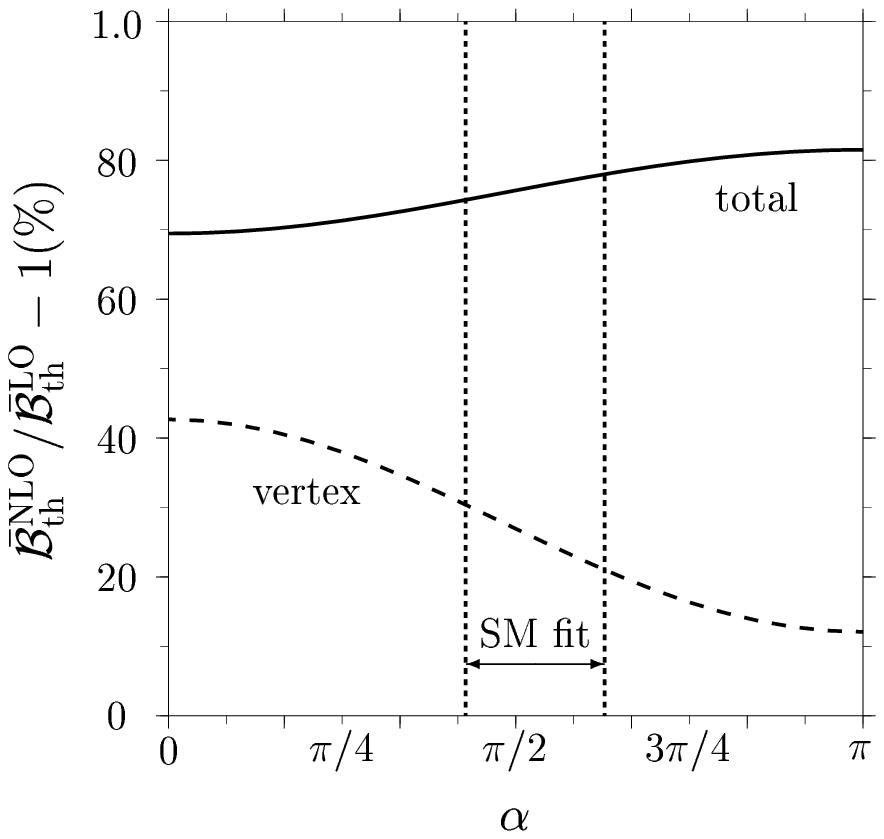}}
\caption{The relative NLO corrections (in percentage) to the
         charge-conjugate averaged branching ratios 
         $\bar {\cal B}_{\rm th} (B^\pm \to \rho^\pm \gamma)$
         (left figure) and 
         $\bar {\cal B}_{\rm th} (B^0 \to \rho^0 \gamma)$
         (right figure) as functions of the unitarity triangle
         angle~$\alpha$ without (dashed curves) and with (solid
         curves) the hard-spectator corrections. The $\pm 1 \sigma$
         allowed band of $\alpha$ from the SM unitarity fits is also
indicated.
}
\label{fig:Branch}
\end{figure}

To illustrate the effect of the NLO corrections on the
branching ratios for $B \to \rho \gamma$ decays, 
we show the relative NLO corrections 
$\bar {\cal B}_{\rm th}^{\rm NLO}/\bar {\cal B}_{\rm th}^{\rm LO} - 1$
to the $B^\pm \to \rho^\pm \gamma$ and $B^0 \to \rho^0 \gamma$
decay rates as functions of the~CP phase~$\alpha$ in
Fig.~\ref{fig:Branch}. All other input parameters are fixed to their
central values given earlier. What concerns the CKM parameters, we take
the ranges for the Wolfenstein parameters from the unitarity fits,
yielding \cite{Hocker:2001xe,Ali:1999we,Ali:2001ckm}:
\begin{equation}
\bar \rho = 0.20 \pm 0.07, \qquad 
\bar \eta = 0.39 \pm 0.07.
\label{eq.CKMfitter}
\end{equation}
They in turn lead to the ranges $\vert V_{ub}/V_{td}\vert = 0.49 \pm 0.09$
and $\alpha$ varying in the range $77^\circ \leq \alpha \leq 113^\circ$,
with $\alpha = 93^\circ$ as the central value. We will show this range
of $\alpha$ by a vertical band in most of the figures presented below.
We note from Fig.~\ref{fig:Branch} that the NLO vertex 
and the hard-spectator corrections are comparable, of order $(30 - 40)\%$
each, increasing the branching ratio altogether by about $(60 - 70)\%$ in 
$B \to \rho \gamma$ decays in the range of~$\alpha$
favored by the SM constraints. Note that the total NLO correction 
for the $B^0 \to \rho^0 \gamma$ decays has a very weak dependence on 
the angle~$\alpha$.  
%
%
%
%

We now proceed to calculate numerically the
branching ratios for the decays $B^\pm \to \rho^\pm \gamma$ and 
$B^0 \to \rho^0 \gamma$ with the help of Eqs.~(\ref{eq:BrBpm}) 
and~(\ref{eq:BrB0}). The theoretical ratio involving the theoretical decay 
widths on the r.h.s. of these equations can be written in the form  
\begin{equation} 
\frac{\overline {\cal B}_{\rm th} (B \to \rho \gamma)}
     {\overline {\cal B}_{\rm th} (B \to K^* \gamma)} = 
S_\rho \left | \frac{V_{td}}{V_{ts}} \right |^2 
\frac{(1 - m_\rho^2/M^2)^3}{(1 - m_{K^*}^2/M^2)^3} \, 
\zeta^2 \, \left[ 1 + \Delta R (\rho/K^*) \right ],
\label{eq:DWs-ratio}
\end{equation}
where~$\zeta$
is the ratio of the HQET form factors and $S_\rho = 1 (1/2)$ for 
$\rho^\pm$- ($\rho^0$-) meson. In the $SU (3)$-symmetry limit,  
$\xi_\perp^{(\rho)} (0) = \xi_\perp^{(K^*)} (0)$, yielding $\zeta = 1$.
$SU (3)$-breaking effects in the QCD form factors $T_1^{(K^*)} (0)$
and $T_1^{(\rho)} (0)$ have been evaluated within the QCD 
sum-rules~\cite{Ali:1994vd}. These can be taken to hold also for the
ratio of the HQET form factors. Thus, we take 
\begin{equation} 
\zeta = \frac{T_1^{(\rho)} (0)}{T_1^{(K^*)} (0)} 
\simeq 0.76 \pm 0.06. 
\label{eq:xi-QCD-SR}
\end{equation}
Taking this and Eq.~(\ref{eq:xi-Ks-average}) into account, we obtain: 
\begin{equation} 
\xi_\perp^{(\rho)} (0) = 0.190 \pm 0.034 ,     
\label{eq:xi-rho}
\end{equation}
which can be used in numerical analysis. 

\begin{figure}[bt]
\centerline{\epsfxsize=.45\textwidth
            \epsffile{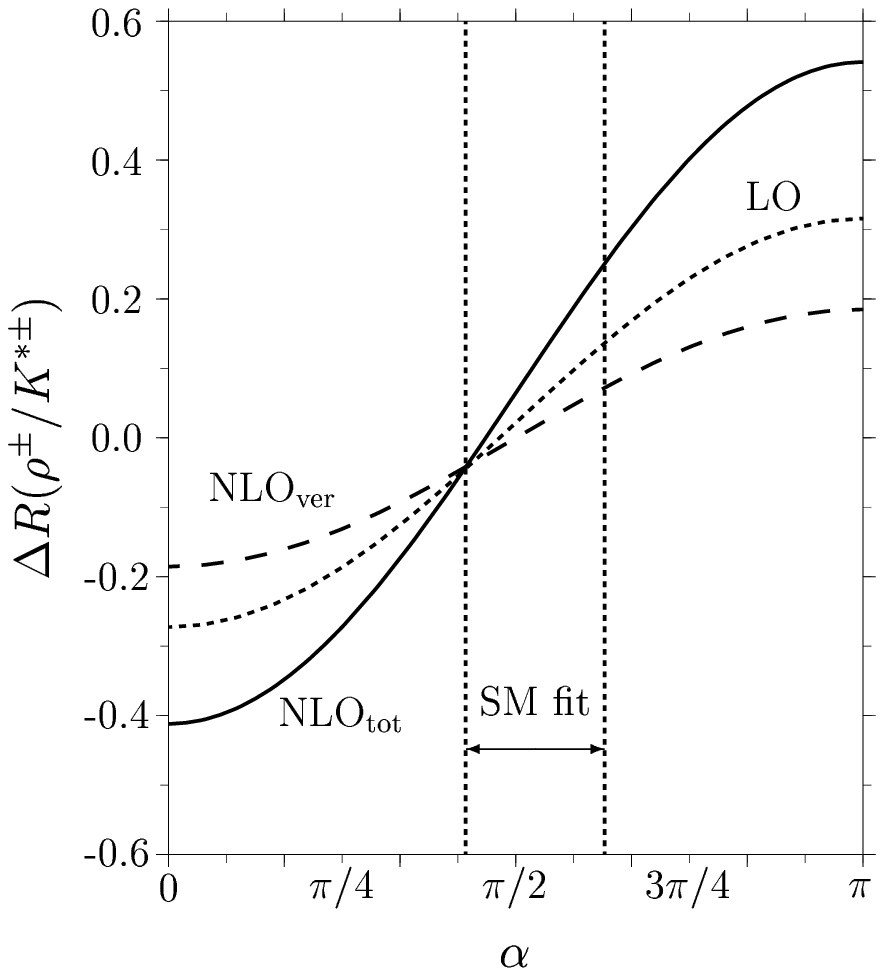} \qquad
            \epsfxsize=.45\textwidth
            \epsffile{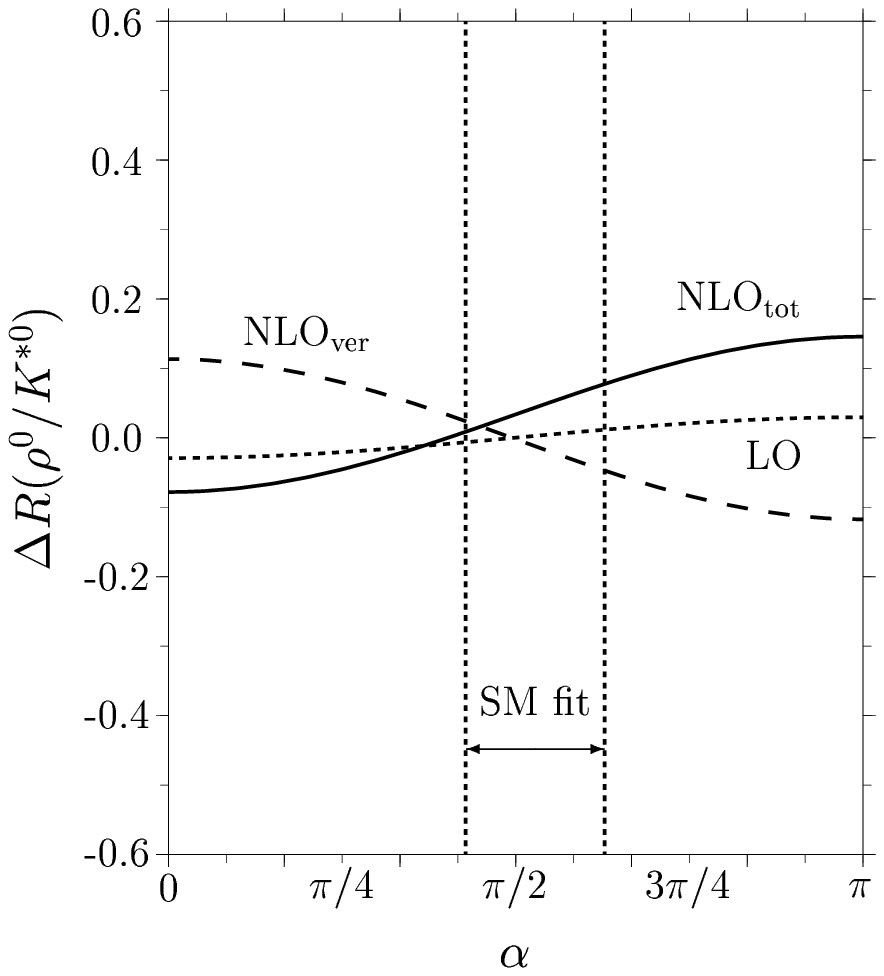}}
\caption{The quantities $\Delta R (\rho^\pm/K^{* \pm})$ and 
         $\Delta R (\rho^0/K^{*0})$ defined in Eq.~(\ref{eq:DWs-ratio}) as
         a function of the 
         unitarity triangle angle~$\alpha$ in the leading order~(LO) 
         [the dotted lines]
         and next-to-leading order without (NLO$_{\rm ver}$) [dashed
         lines] and with (NLO$_{\rm tot}$) [solid lines]
         the hard-spectator corrections. The $\pm 1 \sigma$ allowed
         band of $\alpha$ from the SM unitarity fits is also indicated.}
\label{fig:R}
\end{figure}

\begin{figure}[bt]
\centerline{\epsfxsize=.45\textwidth
            \epsffile{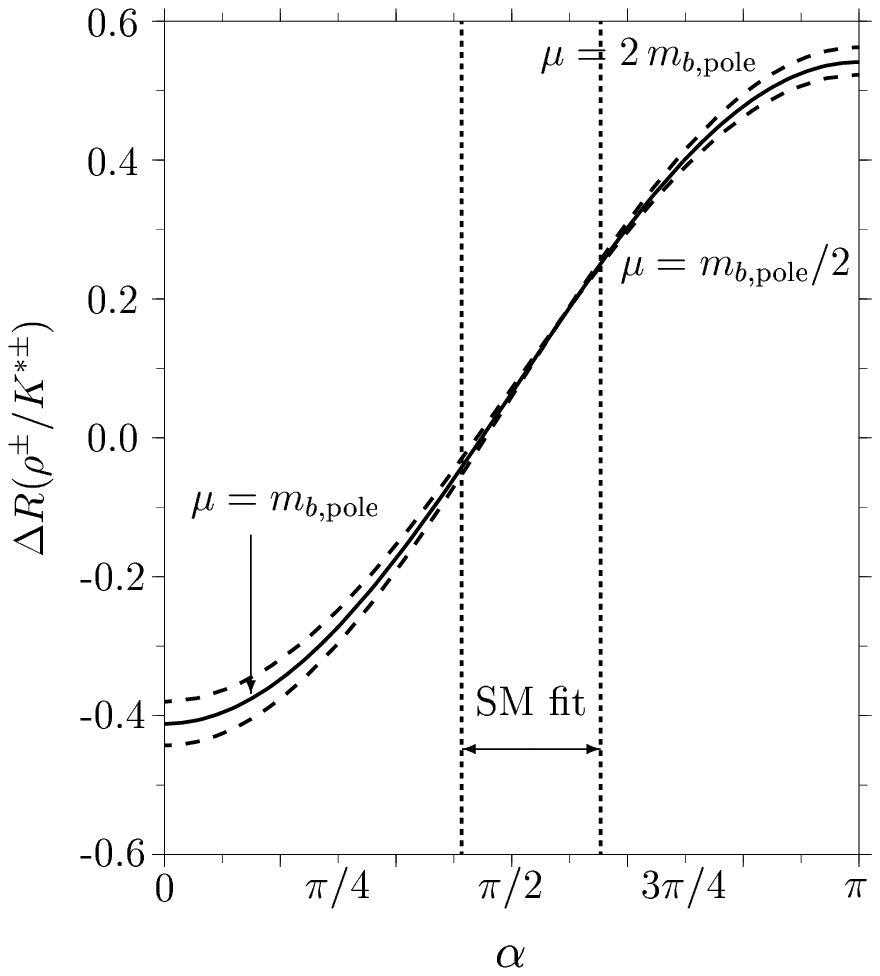} \qquad
            \epsfxsize=.45\textwidth
            \epsffile{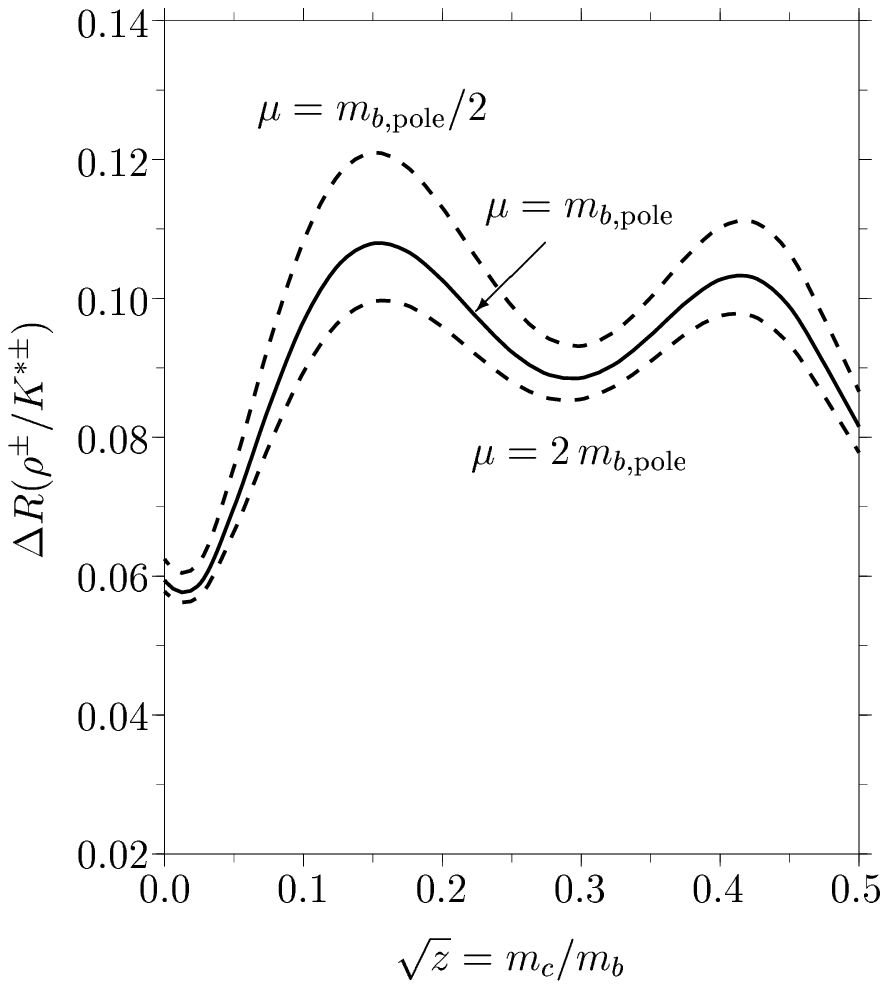}}
\caption{Scale dependence of the quantity $\Delta R(\rho^\pm/K^{*\pm})$
         defined in Eq.~(\ref{eq:DWs-ratio}) as a function of the
         unitarity triangle angle~$\alpha$ (left figure) and the quark
         mass ratio $\sqrt z = m_c/m_b$ (right figure). 
         The solid curves correspond to the  
         scale~$\mu = m_{b,{\rm pole}}$ and the dotted curves delimit 
         the  variation in the range $m_{b,{\rm pole}}/2 \le \mu \le 2
         m_{b,{\rm pole}}$. The $\pm 1 \sigma$ allowed 
         band of $\alpha$ from the SM unitarity fits is also indicated.}
\label{fig:Br-mu}
\end{figure}

The theoretical expression for dynamical function~$\Delta R(\rho/K^*)$ 
can be written as follows:
\begin{eqnarray} 
\Delta R (\rho/K^*) & = & 
2 \epsilon_A \, F_1 + \epsilon_A^2 (F_1^2 + F_2^2) 
\label{eq:RRho/Ks} \\ 
& + & 
\frac{2}{C^{(0) {\rm eff}}_7} \, {\rm Re} 
\left [ 
 A^{(1)\rho}_{\rm sp} -  A^{(1)K^*}_{\rm sp} 
+ F_1 (A^u + \epsilon_A A^{(1)t}) + \epsilon_A (F_1^2 + F_2^2) A^u 
\right ].  
\nonumber 
\end{eqnarray}
%
%
\begin{table}[tb] 
\caption{Central values of the partial amplitudes in  
         $B \to \rho \gamma$ and $B \to K^* \gamma$ decays. The  
         column WC $+$ VC is sum of the NLO corrections in the Wilson
         coefficients (WC) and vertex corrections (VC) which are estimated
         at the scale   $\mu = m_{b,{\rm pole}} = 4.65$~GeV; the column
         HSC contains the 
         hard-spectator contributions evaluated at the 
         scale~$\mu_{\rm sp} = 1.52$~GeV; and in the last column the 
         total amplitudes (WC+VC+HSC) are presented.} 
\label{tab:B-Rho-gam:amplitude}
\begin{center}
\begin{tabular}{|l|c|c|c|}  
\hline\hline
& WC + VC & HSC & total \\ \hline 
$A^{(1)t}(\rho \gamma)$ & $- 0.0428 - i \, 0.0177$ & 
$- 0.0763 - i \, 0.0274$ & $- 0.1191 - i \, 0.0451$ \\ 
$A^u(\rho\gamma)$      & $+ 0.0479 + i \, 0.0485$ & 
$- 0.0708 - i \, 0.0274$ & $- 0.0229 + i \, 0.0211$ \\
$ A^{(1)}(K^* \gamma)$ & $- 0.0428 - i \, 0.0177$ & 
$- 0.0711 - i \, 0.0256$ & $- 0.1139 - i \, 0.0433$ \\ 
\hline \hline 
\end{tabular}
\end{center}
\end{table}
%
%
For the numerical calculations of~$\Delta R (\rho/K^*)$, we recall 
that the vertex and the hard-spectator 
corrections are evaluated at different scales: for the former we use the
scale of the $b$-quark mass (pole mass $m_{b, {\rm pole}} = 4.65$~GeV in 
our analysis) and for the last the typical scale
is $\mu_{\rm sp} = 1.52$~GeV.  
The combined Wilson coefficient and vertex contributions (WC+VC),
the hard-spectator contributions (HSC), and the total contributions 
(WC+VC+HSC) to the  
functions~$A^{(1)t} (\rho \gamma)$, $A^{(1)} (K^*\gamma)$ 
and~$A^u (\rho \gamma)$ are presented in 
Table~\ref{tab:B-Rho-gam:amplitude}. They correspond to the central
values of the input parameters specified earlier. For the assumed input
parameters,  
it is seen that the vertex and hard-spectator contributions (WC$+$VC and
HSC) in~$A^{(1)t}(\rho\gamma)$ and~$A^{(1)}(K^*\gamma)$ are of the same
sign and comparable to each other. The small difference in the HSC 
parts is due to $SU(3)$-breaking effects in the ratio of 
$\Delta F_\perp^{(V)}$ [defined in Eq.~(\ref{eq:DFperp})] 
and~$\xi_\perp^{(V)} (0)$. For the $\rho$- and $K^*$-mesons, their 
central values are estimated to be:
\begin{displaymath}
\frac{\Delta F_\perp^{(\rho)} (\mu_{\rm sp})}
     {\xi_\perp^{(\rho)} (0)} \simeq 8.61,
\qquad
\frac{\Delta F_\perp^{(K^*)} (\mu_{\rm sp})}
     {\xi_\perp^{(K^*)} (0)} \simeq 7.85,
\end{displaymath}
where the values of~$\xi_\perp^{(K^*)} (0)$, $\xi_\perp^{(\rho)} (0)$
and $\Delta F_\perp^{(V)} (\mu_{\rm sp})$ are taken from
Eqs.~(\ref{eq:xi-Ks-average}), (\ref{eq:xi-rho}) and
Table~\ref{tab:VM-parameters}, respectively. The main uncertainty
($\sim 30\%$) in these ratios originates from the first negative
moment of the $B$-meson~$\lambda_{B,+}^{-1}$.
The contributions to~$A^u(\rho\gamma)$ from the 
vertex  and the hard-spectator corrections have opposite signs, 
and the real part of the sum is rather small. 
For the numerical estimates, one also needs to know the
CKM-functions~$F_1$ and~$F_2$ defined by Eq.~(\ref{eq:UT-phase}).
In the analysis the central value resulting from the CKM fits: 
$\sqrt{F_1^2 + F_2^2} \simeq |V_{ub}/V_{td}| = 0.49$ is used. 
The dynamical functions~$\Delta R (\rho^\pm/K^{*\pm})$ 
and~$\Delta R (\rho^0/K^{*0})$ for the $B^\pm \to \rho^\pm \gamma$ 
and $B^0 \to \rho^0 \gamma$ are presented in Fig.~\ref{fig:R}. 
%
%
%
%
It is seen that taking into account the hard-spectator corrections both
in the charged and neutral $B$-meson decays makes the branching rates
in the leading (LO) [the dotted lines in Fig.~\ref{fig:R}] and 
next-to-leading (NLO$_{\rm tot}$) [the solid lines in Fig.~\ref{fig:R}] 
close to each other. 
The dependences of the dynamical function~$\Delta R(\rho^\pm/K^{*\pm})$ 
on the CKM angle~$\alpha$ and on the quark mass ratio~$\sqrt z = m_c/m_b$  
are presented in Fig.~\ref{fig:Br-mu}. The solid lines correspond to 
the scale~$\mu = m_{b,{\rm pole}}$. It is seen that the dashed lines 
representing the  $\mu = m_{b,{\rm pole}}/2$ and 
$\mu = 2 m_{b,{\rm pole}}$ results are very close to each other. Notice
also the mild dependence of $\Delta R(\rho^\pm/K^{*\pm})$
on the quark mass ratio~$\sqrt z$.  
%
%
%
%
The main uncertainties in the dynamical functions come from the 
uncertainties in the CKM angle~$\alpha$ and the nonperturbative 
parameters~$\xi^{(\rho)}_\perp (0)$ and~$\xi^{(K^*)}_\perp (0)$ which 
can be seen, in particular, for the $B^\pm \to \rho^\pm \gamma$ decay 
from Table~\ref{tab:Br-B-rho-gamma}.  
%
%
\begin{table}[tb]
\caption{Input parameters and the assumed errors used in the evaluation of
         the function  $\Delta R (\rho^\pm/K^{*\pm})$, defined in
         Eq.~(\ref{eq:RRho/Ks}). The resulting errors $\delta R
         (\rho^\pm/K^{*\pm})$ are given in the last column. The central
         value of $\Delta R (\rho^\pm/K^{*\pm})$ and the $\pm 1
         \sigma$ errors obtained by quadrature are given in the last row.}
\label{tab:Br-B-rho-gamma}
\begin{center}
\begin{tabular}{|l|l|l|} 
\hline \hline 
Parameter & Value & $\delta R (\rho^\pm/K^{*\pm})$ 
\\ \hline
$\alpha$                   & $93^\circ {}^{+20^\circ}_{-16^\circ}$ & 
$ +0.1612 / -0.1321$ \\ 
$\xi^{(\rho)}_\perp (0)$     & $0.190 \pm 0.035$ & $+0.1155 / -0.0805$ \\
$\bar \xi^{(K^*)}_\perp (0)$ & $0.250 \pm 0.040$ & $+0.0623 / -0.0860$ \\ 
$f_\perp^{(\rho)}$(1 GeV)   & ($160 \pm 10$) MeV & $\pm 0.0332$ \\ 
$a_{\perp 2}^{(\rho)}$(1 GeV)  & $0.20 \pm 0.10$ & $\pm 0.0332$ \\
$\lambda_{B,+}^{-1}$       & ($3 \pm 1$) GeV$^{-1}$ & $\pm 0.0267$ \\ 
$f_\perp^{(K^*)}$(1 GeV)   & ($185 \pm 10$) MeV & $\pm 0.0244$ \\ 
$\epsilon_A$               & $0.30 \pm 0.07$ & $+0.0187 / -0.0163$ \\ 
$a_{\perp 1}^{(K^*)}$(1 GeV)  & $0.20 \pm 0.05$ & $\pm 0.0172$ \\ 
$\left | V_{ub} / V_{td} \right | = \sqrt{F_1^2 + F_2^2}$ & 
$0.49 \pm 0.09$ & $+0.0175 / -0.0153$ \\  
$a_{\perp 2}^{(K^*)}$(1 GeV)  & $0.04 \pm 0.04$ & $\pm 0.0106$ \\ 
$\sqrt z = m_c / m_b$ & $0.27 \pm 0.06$ & $+ 0.0109 / - 0.0011$ \\
$f_B$                      & ($200 \pm 20$) MeV & $\pm 0.0080$ \\
$\mu / m_{b,{\rm pole}}$      & $0.5 - 2.0$ & $+0.0055 / -0.0036$ \\ 
$m_{b, {\rm pole}}$        & ($4.65 \pm 0.10$) GeV & $\pm 0.0001$ \\ 
\hline 
$\Delta R (\rho^\pm/K^{*\pm})$ & 
\multicolumn{2}{|c|}{$0.090 \pm 0.202$}  
\\ \hline \hline
\end{tabular}
\end{center}
\end{table}
%
%
The function~$\Delta R(\rho^0/K^{*0})$ involving the neutral $B$-meson
decays can be similarly  calculated. The allowed ranges of the two
functions are estimated to be:
\begin{equation}
\Delta R(\rho^\pm/K^{*\pm}) = 0.09 \pm 0.20~, 
\qquad  
\Delta R(\rho^0/K^{*0}) = 0.04 \pm 0.14~. 
\label{eq:dyn-functions} 
\end{equation}
The central values of both these functions are close to zero, and  
they impart an uncertainty~$\sim 30\%$ and~$\sim 15\%$ to the 
ratios~(\ref{eq:DWs-ratio}) of the $B^\pm \to \rho^\pm \gamma$ 
to $B^\pm \to K^{*\pm} \gamma$ and $B^0 \to \rho^0 \gamma$ to 
$B^0 \to K^{*0} \gamma$ branching ratios, respectively. Note also 
that the function analized is increased with increasing the 
parameter~$\epsilon_A$ characretized the weak annihilation contribution. 
While the recent central value obtained at $\epsilon_A = + 0.3$ is
substantially larger than the one from the previous version of the 
paper (where $\epsilon_A = - 0.3$ was used), it remains too small 
to influence on the branching ratio of $B^\pm \to \rho^\pm \gamma$ 
decay. Nevertheless, the uncertainty in this function becomes 
significantly larger. 
  
\begin{figure}[bt]
\centerline{\epsfxsize=.45\textwidth
            \epsffile{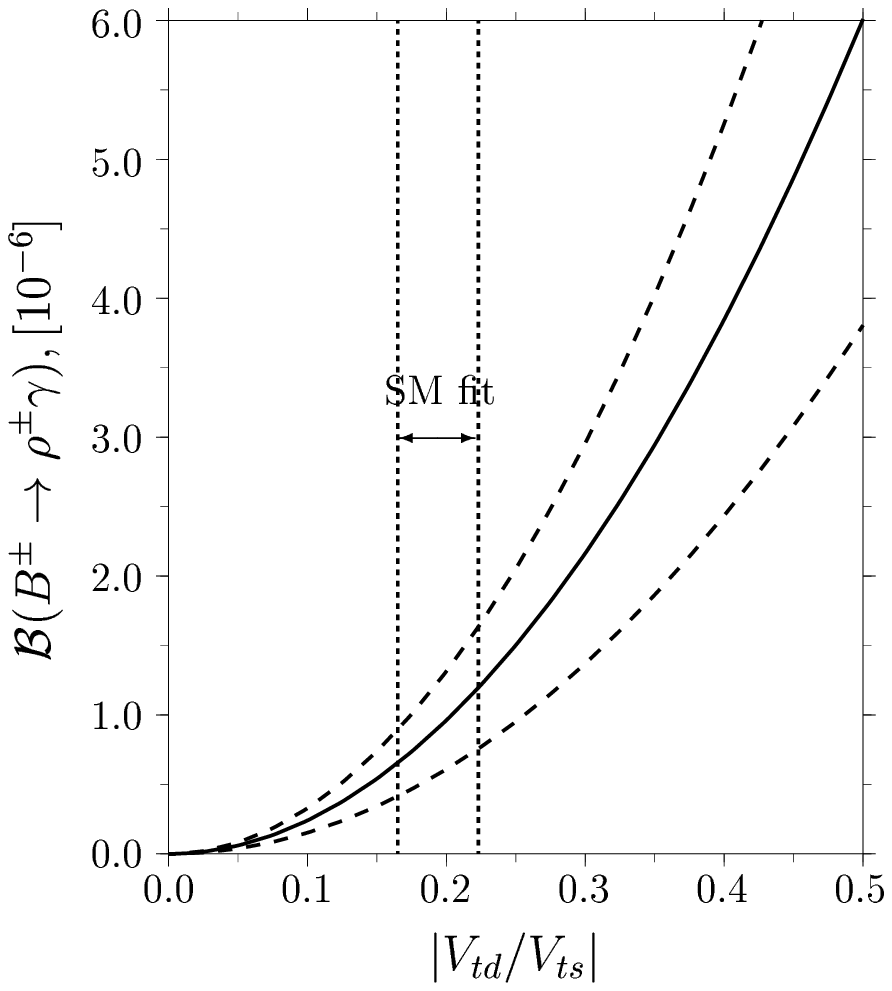} \qquad
            \epsfxsize=.45\textwidth
            \epsffile{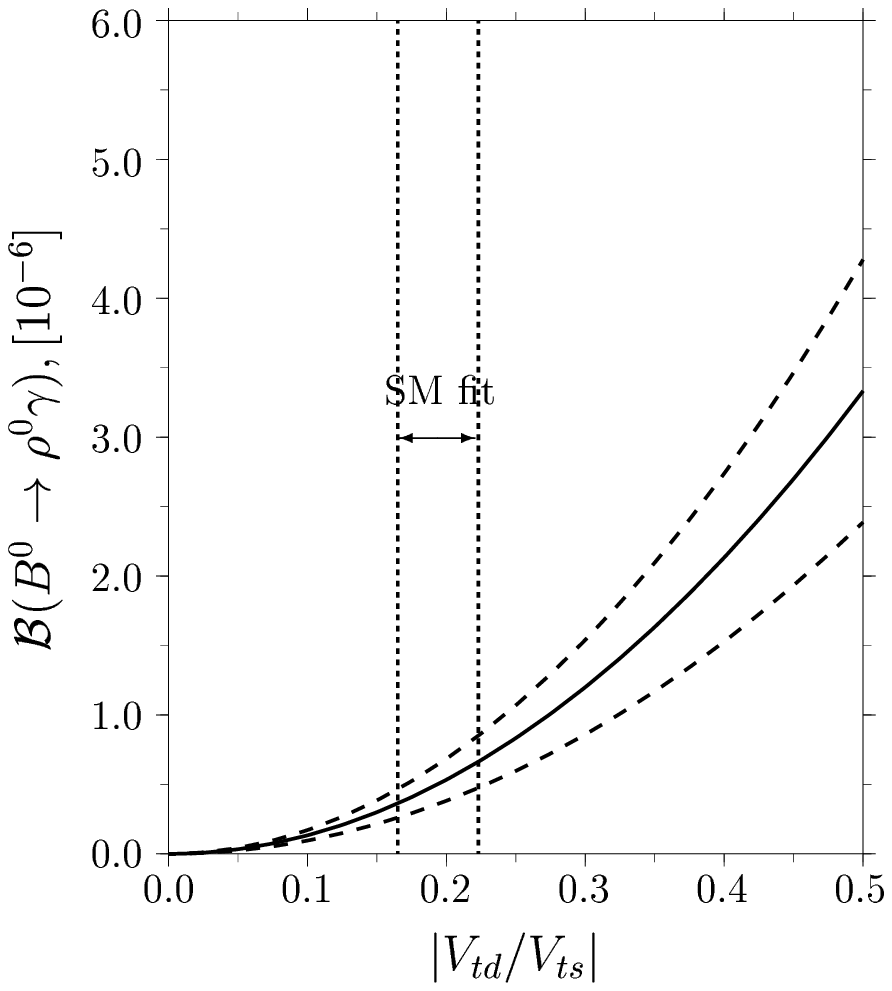}}
\caption{Branching ratios ${\cal B}(B^\pm \to \rho^\pm \gamma)$ (left
         figure) and ${\cal}(B^0 \to \rho^0 \gamma)$ (right figure) 
         in the NLO-LEET approach as functions of the  CKM matrix element
         ratio~$|V_{td}/V_{ts}|$. The solid curves represent the central
         values and the dashed curves delimit the theoretical $\pm 1
         \sigma$ variations;
         the vertical bands show the SM favoured range: 
         $|V_{td}/V_{ts}| = 0.194 \pm 0.029$.}
\label{fig:ranges}
\end{figure}

The product of the CKM matrix elements~$|V_{tb} V_{td}^*|$ from  
Eq.~(\ref{eq:DecayWidth}) can be estimated by using the CKM fits, 
which gives 
\begin{equation}
|V_{tb} V_{td}^*| = 0.0077 \pm 0.0011~. 
\label{eq:Vtd}
\end{equation}
Taking into account the value $|V_{tb} V_{ts}^*| = 0.0396 \pm 0.0020$ 
[Eq.~({\ref{eq:lam-ts}})], we get: 
\begin{equation}
\left | \frac{V_{td}}{V_{ts}} \right | = 0.194 \pm 0.029~, 
\label{eq:Vtd/Vts}
\end{equation}
which allows to predict the values for the branching ratios for  
$B \to \rho \gamma$ decays with the help of Eqs.~(\ref{eq:BrBpm}), 
(\ref{eq:BrB0}), and~(\ref{eq:DWs-ratio}). The branching ratios for  
$B^\pm \to \rho^\pm \gamma$ and $B^0 \to \rho^0 \gamma$ decays 
are presented in Fig.~\ref{fig:ranges}. The vertical bands in these 
plots delimit the $\pm 1 \sigma$ range for $\vert V_{td}/V_{ts}\vert$
given in Eq.~(\ref{eq:Vtd/Vts}). 
%
%
%
%
Due to the present analysis, the $B \to \rho \gamma$
branching ratios can be estimated as
\begin{eqnarray}
&& \bar {\cal B}_{\rm th} (B^\pm \to \rho^\pm \gamma) = 
(0.91 \pm 0.33 [{\rm th}] \pm 0.11 [{\rm exp}]) \times 10^{-6},
\label{eq:Br-predictions} \\
&& \bar {\cal B}_{\rm th} (\bar B^0 \to \rho^0 \gamma) = 
(0.50 \pm 0.18 [{\rm th}] \pm 0.04 [{\rm exp}]) \times 10^{-6},
\nonumber 
\end{eqnarray}
where the SM favoured range $77^\circ \le \alpha 
\le 113^\circ$~\cite{Hocker:2001xe,Ali:1999we,Ali:2001ckm} was used.
In the above estimates, the first error is defined by the uncertainties 
of the theory and the second is from the direct experimental data 
on the $B \to K^* \gamma$ branching ratios~(\ref{eq:Br-Ks-exp}).

These estimates can be compared with the ones obtained in the QCD sum rule
approach of Ref.~\cite{Ali:1995uy}: ${\cal B}(B^\pm \to \rho^\pm
\gamma)=(1.9 \pm 1.6) \times 10^{-6}$ and ${\cal B}(B^0 \to \rho^0
\gamma)=(0.85 \pm 0.65) \times 10^{-6}$, in which only the leading order
QCD corrections in $\alpha_s$ and annihilation contributions were taken
into account.  The central values of the estimates presented here are
typically a half of the corresponding values in Ref.~\cite{Ali:1995uy},
and the errors in our case are significantly smaller. As shown in
Eq.~(\ref{eq:dyn-functions}), the theoretical improvement discussed in
the present paper has only a marginal impact on the ratio
$R(\rho\gamma/K^*\gamma)$, used here and in Ref.~\cite{Ali:1995uy}.
The source of the larger values in Ref.~\cite{Ali:1995uy} is to be traced
back to the differences in the input values of the CKM parameters and the
experimental branching ratios for $B \to K^* \gamma$ in the two
calculations. Present measurements
have decreased the allowed range of the ratio $\vert V_{td}/V_{ts}\vert$,
compared to what was assumed in Ref.~\cite{Ali:1995uy}. Moreover,
the branching ratios ${\cal B}(B \to K^* \gamma)$ are now more precisely
measured and are found to be smaller  than the experimental values {\it en
vogue} in 1995. These two circumstances, in turn, reduce the central
values of the branching ratios ${\cal B}(B^\pm \to \rho^\pm \gamma)$ and
${\cal B}(B^0 \to \rho^0  \gamma)$, and the residual uncertainty is also
reduced. We also wish to point out 
that the estimate of the $B^- \to \rho^- \gamma$ decay rate 
presented in Ref.~\cite{Bosch:2001gv} allows a substantially larger
uncertainty:   ${\cal B} (B^- \to \rho^- \gamma) = (1.2 - 3.6) \times
10^{-6}$, reflecting the larger parametric uncertainties in the branching
ratio, as well as the fact that the NLO corrections increase the
individual branching ratios by typically 60\%. As shown in
Eq.~(\ref{eq:dyn-functions}), these corrections largely cancel in the
ratio. Hence, our predictions for ${\cal B}(B \to \rho \gamma)$ are 
typically half as large as the ones given in \cite{Bosch:2001gv}.
Finally, compared to the present experimental bounds (at 90\%C.L.)
\cite{Casey:2001mo}: ${\cal B}(B^\pm \to \rho^\pm \gamma) < 0.99 \times
10^{-5}$ and ${\cal B}(B^0 \to \rho^0 \gamma) < 1.06 \times 10^{-5}$, one
needs to cover an order of magnitude to reach the SM sensitivity in these
decays. This should be possible at the present B factories.

\section{Isospin-violating Ratios and CP-violating \newline  
         Asymmetries in $B \to \rho \gamma$ Decays}
\label{sec:asymmetries}

We now compute the isospin-violating ratios:
\begin{equation}
\Delta^{\pm 0} =
\frac{\Gamma (B^\pm \to \rho^\pm \gamma)}
     {2 \Gamma (B^0 (\bar B^0)\to \rho^0 \gamma)} - 1~.
\label{eq:IVR-def}
\end{equation}
These ratios deviate from zero (the isospin symmetry limit) due to
the interference of the short distance penguin amplitudes and long
distance tree amplitudes, where the latter are given by the lowest order
annihilation contributions, as the ${\cal O}(\alpha_s)$-contribution to
the annihilation amplitude vanishes in the chiral limit in the leading-twist
approximation~\cite{Grinstein:2000pc}. We present our numerical analysis
for the charge-conjugate averaged ratio:
\begin{equation}
\Delta = \frac{1}{2} \,
\left [ \Delta^{+ 0} + \Delta^{- 0} \right ] ,
\label{eq:CCAR-def}
\end{equation}
which is expressed in the NLO perturbative QCD as~\cite{Ali:2000zu}:
\begin{eqnarray}
\Delta_{\rm LO} & \simeq & 2 \epsilon_A
\left [ F_1 + \frac{1}{2} \, \epsilon_A \, (F_1^2 + F_2^2) \right ] ,
\label{eq:CCAR-LO} \\
\Delta_{\rm NLO} & \simeq &  \Delta_{\rm LO} -
\frac{2 \epsilon_A}{C^{(0) {\rm eff}}_7}
\left [
F_1 A^{(1)t}_R + (F_1^2 - F_2^2) A^u_R
+ \epsilon_A (F_1^2 + F_2^2) (A^{(1)t}_R + F_1 A^u_R)
\right ] ,
\qquad
\label{eq:CCAR-NLO}
\end{eqnarray}
where $\Delta_{\rm LO}$ is the leading-order charge-conjugate
averaged ratio. The NLO quantity ~$\Delta_{\rm NLO}$ is sensitive to the
hard-spectator corrections which are contained in the ${\cal O} (\alpha_s)$
functions~$A^{(1)t}_R$~(\ref{eq:A1t}) and~$A^u_R$~(\ref{eq:Au}).

\begin{figure}[bt]
\centerline{
            \epsffile{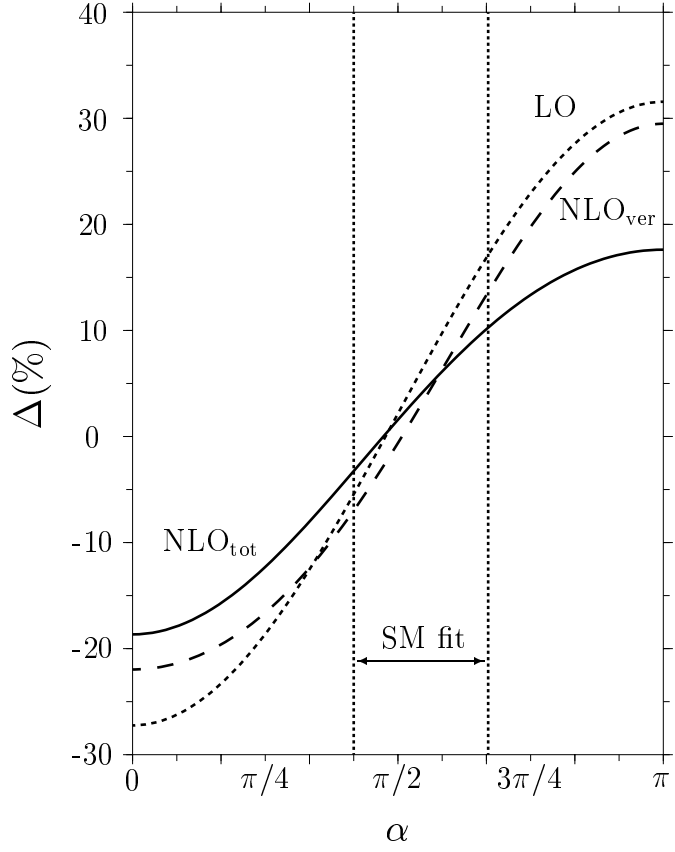}}
\caption{The charge-conjugate averaged ratio~$\Delta$ for $B \to \rho
         \gamma$ decays defined in Eq.~(\ref{eq:IVR-def}) as a
          function of the unitarity triangle
          angle~$\alpha$ in the leading order (dotted curve), 
          next-to-leading order without (dashed curve) and
         with (solid curve) hard-spectator corrections. The $\pm 1 \sigma$
         allowed band of $\alpha$ from the SM unitarity fits is also indicated.}
\label{fig:Delta}
\end{figure}

The ratio~$\Delta$ is shown as a function 
of the inner angle~$\alpha$ in Fig.~\ref{fig:Delta}. The solid curve
shows the complete ${\cal O} (\alpha_s)$-corrected ratio
including the vertex and hard spectator corrections, calculated using
Eq.~(\ref{eq:CCAR-NLO}), the dashed
curve shows this ratio when only the vertex corrections are
included, and the dotted curve shows the lowest order result, obtained
using Eq.~(\ref{eq:CCAR-LO}).
%
%
%
%
We note that taking into account the spectator corrections
slightly modifies the ${\cal O} (\alpha_s)$ vertex-corrected
result, obtained in Ref.~\cite{Ali:2000zu}. Thus, even if one
takes a generous error on this quantity, the theoretical precision
of~$\Delta_{\rm NLO}$ is not perceptibly influenced by the uncertainty 
in~$\Delta F_\perp^{(\rho)}$ entering in the hard spectator
correction. We also note that the region of~$\alpha$ where the
NLO corrections are large is not favored by the CKM unitarity
constraints in the SM, which yield typically $77^\circ \le \alpha \le
113^\circ$~\cite{Hocker:2001xe,Ali:1999we,Ali:2001ckm}. 

In Ref.~\cite{Bosch:2001gv} the charge-conjugate 
averaged ratio~$\Delta$ was found to have an opposite sign than the one
obtained in Ref.~\cite{Ali:2000zu}. Our result presented here, after 
we have reversed the sign of the annihilation contribution
characterized by $\epsilon_A$, agrees now with the one 
given in the former of these references. We would also like to point out that
the dependence of~$\Delta$ on the inner angle~$\alpha$, shown 
by us in presenting our numerical results,  emerges naturally from
Eq.~(\ref{eq:UT-phase}).

Finally, we discuss the direct CP-asymmetry in the $B^\pm \to \rho^\pm
\gamma$ decay rates:
\begin{equation}
{\cal A}_{\rm CP} (\rho^\pm \gamma) =
\frac{{\cal B} (B^- \to \rho^- \gamma) - {\cal B} (B^+ \to \rho^+ \gamma)}
     {{\cal B} (B^- \to \rho^- \gamma) + {\cal B} (B^+ \to \rho^+ \gamma)} .
\label{eq:CPasym-def}
\end{equation}
The CP-asymmetry arises from the interference
of the penguin operator~${\cal O}_7$ and the four-quark
operator~${\cal O}_2$~\cite{Soares:1991te,Greub:1995tb}.
The expression for the CP-asymmetry in $B^\pm \to \rho^\pm \gamma$ 
decays can be written as~\cite{Ali:2000zu}:
\begin{equation}
{\cal A}_{\rm CP} (\rho^\pm \gamma) = - 
\frac{2 F_2 \big ( A^u_I - \epsilon_A \, A^{(1) t}_I \big )}
     {C^{(0) {\rm eff}}_7 \left ( 1 + \Delta_{\rm LO} \right )} ,
\label{eq:CP-asym}
\end{equation}
where~$\Delta_{\rm LO}$ is the charge-conjugate averaged ratio in
the leading order~(\ref{eq:CCAR-LO}). Note that due to the charm quark
mass dependence of the hard-spectator corrections, which enters through
the function~$h^{(\rho)} (z)$~(\ref{eq:Di5-decomp}), the
functions~$A^{(1)t}$ and~$A^u$ are modified compared to their vertex
contributions. More importantly for the CP-asymmetry, the hard spectator 
contributions are complex. The resulting numerical changes in the
functions $A^{u}(\rho\gamma)$ and $A^{(1) t} (\rho\gamma)$, and in their
imaginary parts, are illustrated in Table~\ref{tab:B-Rho-gam:amplitude}. 
%
%
\begin{figure}[bt]
\centerline{\epsfxsize=.45\textwidth
            \epsffile{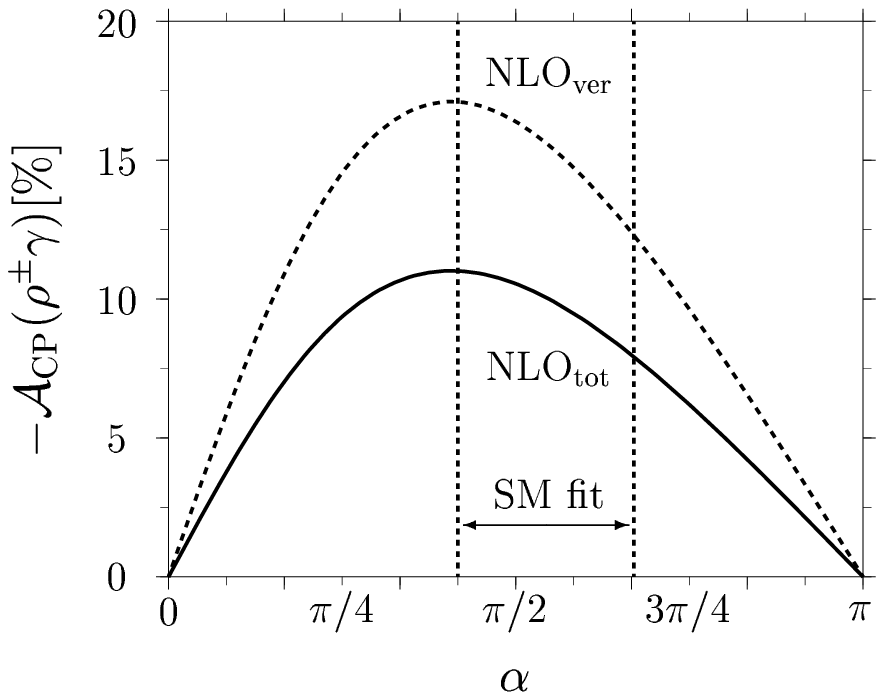} \qquad
            \epsfxsize=.45\textwidth
            \epsffile{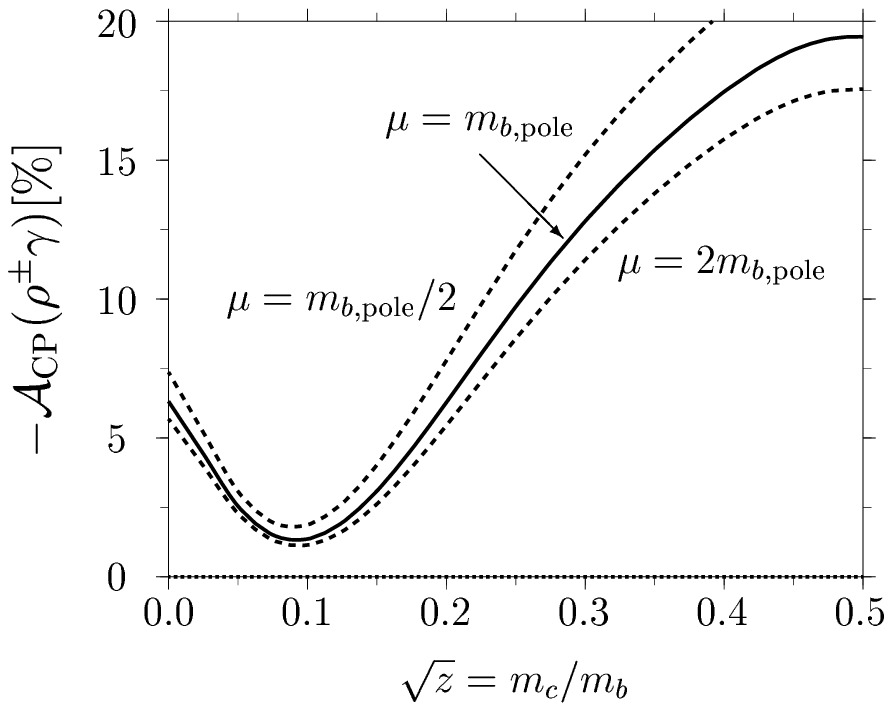}}
\caption{Left figure: Direct CP-asymmetry in the decays  
         $B^\pm \to \rho^\pm \gamma$ as a function of the unitarity 
         triangle angle~$\alpha$ without (dotted curves) and with 
         (solid curves) the hard-spectator corrections.
         The $\pm 1 \sigma$ allowed   
         band of $\alpha$ from the SM unitarity fits is also indicated. 
         Right figure: Direct CP-asymmetry in the decays 
         $B^\pm \to \rho^\pm \gamma$ as a function of the quark mass 
         ratio~$\sqrt z = m_c/m_b$; the scale dependence of the
         asymmetry is shown in the interval: 
         $m_{b,{\rm pole}}/2 \le \mu \le 2 m_{b,{\rm pole}}$.}  
\label{fig:ACPdir}
\end{figure}
%
%
The dependence on the angle~$\alpha$ of the CP-asymmetry
${\cal A}_{\rm CP} (\rho^\pm \gamma)$ is presented in the left plot in 
Fig.~\ref{fig:ACPdir}. It is seen that the CP-asymmerty is significantly  
suppressed by the hard-spectator corrections. In the SM favoured
interval for $\alpha$, $77^\circ \le \alpha \le 113^\circ$, the direct
CP-asymmetry ${\cal A}_{\rm CP} (\rho^\pm \gamma)$ takes its
maximum value, reaching about~$11\%$. The maximum value for the CP-asymmetry 
without the hard spectator corrections is about~$17\%$. Our numerical
result for ${\cal A}_{\rm CP} (\rho^\pm \gamma)$ 
agrees now with the corresponding CP-asymmetry calculated in
Ref.~\cite{Bosch:2001gv}. The CP-asymmetry shown in the left plot in
Fig.~\ref{fig:ACPdir} is drawn for $\sqrt z = m_c/m_b = 0.27$ and
$\epsilon_A = 0.3$. The $z$-dependence of 
${\cal A}_{\rm CP} (\rho^\pm \gamma)$ is shown in the right plot of
Fig.~\ref{fig:ACPdir}; the dependence on the scale~$\mu$ of
this quantity in the interval 
$m_{b,{\rm pole}}/2 \le \mu \le 2 m_{b,{\rm pole}}$ is also shown in 
this figure.  The sensitivity of the CP-asymmetry on the quark mass ratio
is quite marked. 
In Fig.~\ref{fig:ACPdirzero} we plot the corresponding CP-asymmetry
in the decays $B^0/\bar B^0 \to \rho^0 \gamma$, for which the
annihilation contribution is both colour and charge suppressed, and
setting $\epsilon_A = 0$ is not a bad approximation. 

%
\begin{figure}[bt]
\centerline{\epsfxsize=.45\textwidth
            \epsffile{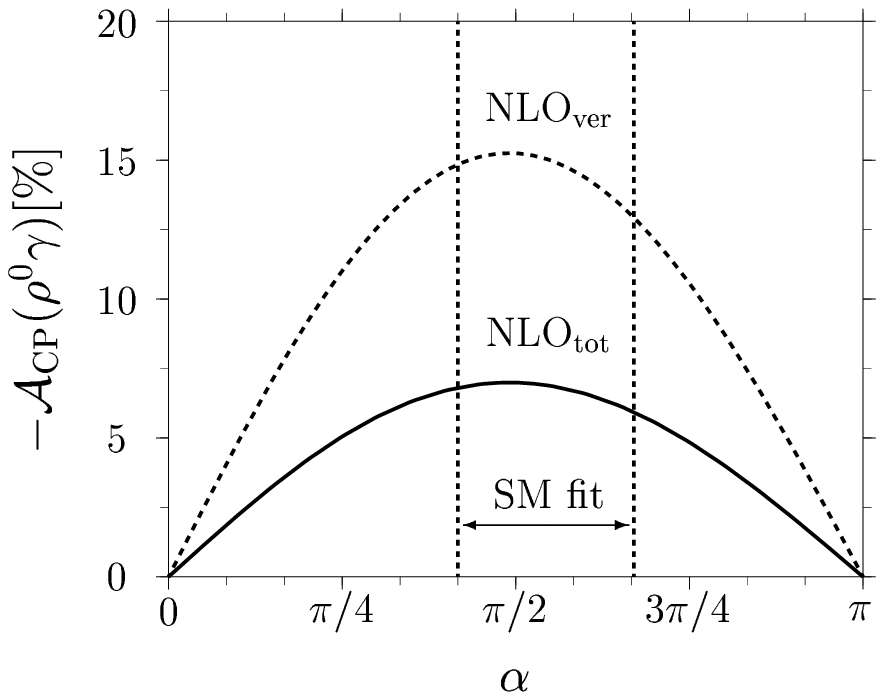} \qquad
            \epsfxsize=.45\textwidth
            \epsffile{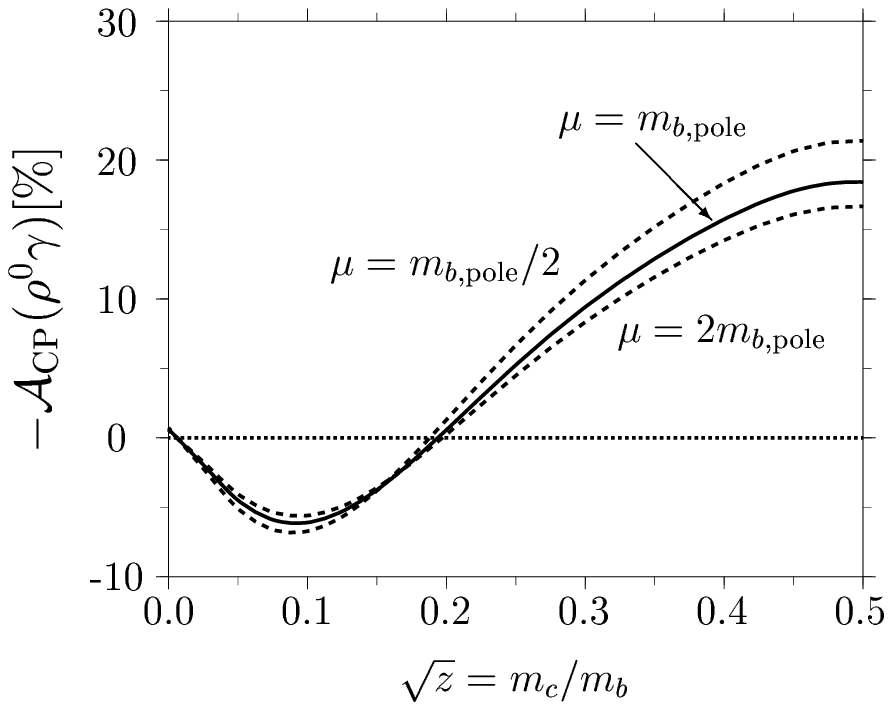}}
\caption{Left figure: Direct CP-asymmetry
         contribution $a_{\epsilon^\prime}$ in ${\cal A}_{\rm CP} (\rho^0
         \gamma) (t)$ in Eq.~(\ref{eq:CPzero-asym}) in the  
         decays $B^0 (\bar B^0) \to \rho^0 \gamma$ as a function of
         the unitarity
         triangle angle~$\alpha$ without (dotted curves) and with
         (solid curves) the hard-spectator corrections.
         The $\pm 1 \sigma$ allowed
         band of~$\alpha$ from the SM unitarity fits is also indicated.
         Right figure: Direct CP-asymmetry in the decays
         $B^0 (\bar B^0) \to \rho^0 \gamma$ as a function of the
         quark mass
         ratio~$\sqrt z = m_c/m_b$; the scale dependence of the
         asymmetry is shown in the interval:
         $m_{b,{\rm pole}}/2 \le \mu \le 2 m_{b,{\rm pole}}$.}
\label{fig:ACPdirzero}
\end{figure}
%
%

A word of caution in the interpretation of Fig.~\ref{fig:ACPdirzero}
is in order. We recall that the  CP-asymmetry
in the decays $B^0/\bar B^0 \to \rho^0 \gamma$  has an
additional complication due to the $B^0 - \bar B^0$ mixing,
which modulates the time-evolution of the CP-asymmetry 
${\cal A}_{\rm CP} (\rho^0 \gamma)$~\cite{Ali:1998gb}:
\begin{equation}
{\cal A}_{\rm CP} (\rho^0 \gamma) (t) = a_{\epsilon^\prime} 
\cos (\Delta M_d t) + a_{\epsilon + \epsilon^\prime} 
\sin (\Delta M_d t)~,
\label{eq:CPzero-asym}
\end{equation}    
where $\Delta M_d \simeq 0.49 \, {\rm ps}^{-1}$ is the mass difference
between the two mass eigenstates of the $B^0 - \bar B^0$ system, 
and the definitions of the direct~$a_{\epsilon^\prime}$ and 
mixing-induced~$a_{\epsilon + \epsilon^\prime}$ components can be seen 
in Ref.~\cite{Ali:1998gb}. The CP-asymmetry shown in 
Fig.~\ref{fig:ACPdirzero} is actually the quantity~$a_{\epsilon^\prime}$, 
which is given by Eq.~(\ref{eq:CP-asym}) with~$\epsilon_A$ set to zero 
for the decays $B^0 (\bar B^0) \to \rho^0 \gamma$.    

To summarize direct CP-asymmetries ${\cal A}_{\rm CP} (\rho^\pm \gamma)$
and ${\cal A}_{\rm CP} (\rho^0 \gamma)$, we remark that the hard spectator
corrections decrease the corresponding asymmetries obtained from the
vertex contributions alone. The degree of this suppression depends on the
annihilation contribution, parametrized here by the quantity
$\epsilon_A$. In addition to this, the CP-asymmetries depend on both the
quark mass ratio~$\sqrt z$ and the scale~$\mu$.   
Hence, without knowing precisely the values of these quantities, it is
difficult to quantify ${\cal A}_{\rm CP} (\rho^\pm \gamma)$ and
${\cal A}_{\rm CP} (\rho^0 \gamma)$. However, in all likelihood, 
$\vert {\cal A}_{\rm CP} (\rho^\pm \gamma)\vert$ and
$\vert {\cal A}_{\rm CP} (\rho^0 \gamma)\vert$  are less than~10\% 
in the SM.


\section{Summary and Concluding Remarks}
\label{sec:concl}

We summarize our main results and offer some remarks on the underlying
theoretical framework.
We have computed the hard-spectator corrections in~$O (\alpha_s)$ and
leading order in $\Lambda_{\rm QCD}/M$ to the decay widths for
$B \to K^* \gamma$ and $B \to \rho \gamma$,  
in the leading-twist approximation, using the 
Large Energy Effective Theory. This is then
combined with the existing contributions from the vertex
corrections and the annihilation amplitudes to
arrive at the NLO expressions for the corresponding decay rates. The
annihilation contributions are important only in the $B \to \rho \gamma$
decays due to the favourable CKM structure. The matrix elements for
these decays in $O(\alpha_s)$, and to leading power in $\Lambda_{\rm
QCD}/M$, are finite,
also including the intermediate charm quark contributions from the penguin
diagrams, correcting an earlier version of this paper, and providing an
explicit proof of the factorization Ansatz of Eq.~(\ref{eq:fact-formula})
advocated by Beneke et al.~\cite{Beneke:1999br,Beneke:2001wa}. For
radiative decays $B \to V \gamma$, these proofs  have also been provided
in the meanwhile in Refs.~\cite{Bosch:2001gv,Beneke:2001at}. We have made
extensive comparisons of our derivations and numerical estimates with the
ones presented in these papers, pointing out the agreements and
some numerical differences related to the hard-spectator corrections. The
NLO corrections in the decay rates are substantial, with the branching 
ratios increasing typically by $60 - 70 \%$ in the NLO approximation. 
Approximately, a half of this enhancement is due to the vertex corrections, 
which are common between the corresponding inclusive and exclusive 
radiative decay rates.

The branching ratios for the decays $B \to K^* \gamma$ in the NLO
accuracy are then compared with
current data to determine the form factor $\xi_\perp^{(K^*)}(0)$ in
the LEET approach. For this purpose we have used the measured values 
of the branching ratios for $B \to K^* \gamma$ and $B \to X_s
\gamma$, getting  $\xi_\perp^{(K^*)}(0) = 0.25 \pm 0.04$,  
taking into account various parametric uncertainties and experimental
errors. Converting it to the form factor in the full QCD, using a
relation correct to leading order 
in~$\alpha_s$ and $\Lambda_{\rm QCD}/M$~\cite{Beneke:2001wa}, yields
$T_1^{(K^*)} (0,\bar m_b) = 0.27 \pm 0.04$, to be compared with   
the estimates $T_1^{(K^*)} (0,\bar m_b)= 0.38 \pm 0.06$
\cite{Ball:1998kk,Ali:2000mm}, and
$T_1^{(K^*)} (0,\bar m_b)= 0.32^{+0.04}_{-0.02}$
\cite{DelDebbio:1998kr}, obtained using  the LC-QCD sum
rule and Lattice-QCD approaches, respectively. Thus, the form factor
in the LEET-based factorization approach, combined with current data, is
found to be typically $(15 - 30)\% $ smaller than the
ones in the LC-QCD/Lattice-QCD
approaches. Another way to judge the same issue is to use the central 
value of the form factor $\xi_\perp^{(K^*)}(0) = 0.35$ extracted from the
LC-QCD sum rules as input and
calculate the branching ratio for $B \to K^* \gamma$ in the LEET
approach. This yields 
${\cal B}(B \to K^* \gamma) = (7.3 \pm 2.7) \times 10^{-5}$, 
compared to the current experimental value of
${\cal B}_{\rm exp}(B \to K^* \gamma) = (4.22 \pm 0.28) \times 10^{-5}$,
where we have averaged the theory and experiment over the $B^\pm$- and  
$B^0/\bar{B}^0$-decay modes. Thus, the LEET-based branching ratio is found
to be significantly larger than data, though the underlying parametric
uncertainties can be used to reduce this difference to some extent [see,
Eq.~(\ref{eq:Br-Ksgam})]. This discrepancy, while not overwhelming, is
discomfortingly  large for a precision test of the SM using the exclusive
decay rates. Improved data may bridge this gap, bringing
in line the LEET-based branching ratios with data, or equivalently the 
form factors in this approach with the ones in the other two QCD
methods. This remains to be seen. Of course, also in the sum rule and
lattice approaches to QCD, the desired theoretical accuracy on the form
factors is not yet reached. However, despite its tentative nature, the
current mismatch between the LEET-based and
the other two QCD results in the $B \to V \gamma$ sector may after all
turn out to be symptomatic of a generic problem afflicting the LEET
approach, having to do with the inadequacy of power corrections in 
exclusive two-body decays in its present formulation. This, despite the
fact that the underlying framework,  as formulated in
Refs.~\cite{Beneke:1999br,Beneke:2001at,Bosch:2001gv} and
also applied in this paper, rests firmly on perturbative QCD
factorization, with the argument made even more persuasive by an all order
proof of factorization in the strong coupling
\cite{Beneke:2000ry,Bauer:2001cu}.

What concerns the decay rates in the 
$B \to \rho \gamma$ decays, we have argued that a more reliable route
theoretically is to  calculate them via the ratio of the branching ratios 
$ {\cal B} (B \to \rho \gamma)/{\cal B} (B \to K^*\gamma)$, defined
in Eq.~(\ref{eq:DWs-ratio}), using experimental measurement of the
$B \to K^* \gamma$ decay rates. This ratio depends essentially on the
CKM matrix element squared $|V_{td} / V_{ts}|^2$, given the
non-perturbative  quantity $\zeta = \xi_\perp^{(\rho)} (0) /
\xi_\perp^{(K^*)} (0)$  and a dynamical function called 
$\Delta R(\rho/K^*)$, involving the vertex, hard-spectator and 
annihilation contributions, which is derived and analyzed in this 
paper. Taking into account various parametric uncertainties, we 
find that $\Delta R(\rho/K^*)$ is constrained in the range 
$| \Delta R(\rho/K^*) | \leq 0.30$, with the central value around 
$\Delta R(\rho/K^*) \simeq 0.09$. This quantifies the statement that 
the ratio ${\cal B} (B \to \rho \gamma)/{\cal B} (B \to K^*\gamma)$ 
is stable  against~$O(\alpha_s)$ and $1/M$-corrections. As shown in this
paper, the same does not hold for the individual branching ratios 
${\cal B}(B \to \rho \gamma)$ and ${\cal B}(B \to K^* \gamma)$.
 Thus, apart from the dependence on the CKM factor $|V_{td} / V_{ts}|$, 
whose determination is the principal interest in measuring the ratio
${\cal B} (B \to \rho \gamma)/{\cal B} (B \to K^*\gamma)$, the 
dynamical uncertainties are estimated not to exceed~$\pm 20\%$. Using the 
current branching ratio ${\cal B}_{\rm exp} (B^\pm \to K^{*\pm} \gamma) = 
(3.82 \pm 0.47) \times 10^{-5}$, the $SU(3)$-breaking estimate 
from the LC-QCD sum rule~\cite{Ali:1994vd}, yielding $\zeta = 0.76\pm
0.06$, and $\Delta R(\rho/K^*)$, calculated in this paper, we find 
${\cal B} (B^\pm \to \rho^\pm \gamma) = (0.91 \pm 0.44) \times 10^{-6}$, 
where the present range for the CKM ratio as determined from the
unitarity fits, $|V_{td} / V_{ts}| = 0.19 \pm 0.03$, is folded in the error. 
The corresponding branching ratio for the neutral $B$-meson decay mode 
in the same approach is estimated to be: ${\cal B} (B^0 \to \rho^0
\gamma) = (0.50 \pm 0.22) \times 10^{-6}$. The isospin-violating
ratios~$\Delta^{\pm 0}$ and its charge-conjugate average~$\Delta$ 
for the decays $B \to \rho \gamma$ are found to be likewise stable 
against the NLO and $1/M$-corrections, In the expected range of the 
CKM parameters, we find $\vert \Delta \vert \leq 10\%$. 

The CP-asymmetries ${\cal A}_{\rm CP} (\rho^\pm \gamma)$
and ${\cal A}_{\rm CP} (\rho^0 \gamma)$ receive 
contributions from the hard-spectator corrections which tend to decrease 
their corresponding values estimated from the vertex corrections
alone. Unfortunately, the 
predicted values of the CP-asymmetries are sensitive to both the choice of
the scale, as already pointed out in Ref.~\cite{Bosch:2001gv}, and the quark 
mass ratio~$\sqrt z = m_c/m_b$. In addition, annihilation contributions
play an important role numerically. In line with the QCD sum rule
estimates for the magnitude of the
annihilation contribution~\cite{Ali:1995uy,Khodjamirian:1995uc}, but
with the sign in the charged $B$-meson decays as determined in  
Refs.~\cite{Bosch:2001gv,Kagan:2001zk},
we have taken $\epsilon_A = 0.3$ and $\epsilon_A \simeq 0$
for the decays $B^\pm \to \rho^\pm \gamma$ and 
$B^0/\bar B^0 \to \rho^0 \gamma$, respectively.  
Typical values for both $|{\cal A}_{\rm CP} (\rho^\pm \gamma)|$ 
and the corresponding direct CP-asymmetry component in 
$|{\cal A}_{\rm CP} (\rho^0 \gamma)|$ lie around ($5 - 10$)\%, 
but the parametric uncertainty in both cases is rather large. 

In conclusion, exclusive radiative decays 
$B \to \rho \gamma$ and $B \to K^* \gamma$ (and their semileptonic 
counterparts) provide an excellent laboratory to test the underlying 
theory (LEET) and ideas on perturbative non-factorizing corrections 
put forward by Beneke et al.~\cite{Beneke:1999br} in the context of 
$B \to \pi \pi$ and $B \to \pi K$ decays. These latter decays are more
involved due to the presence of two strongly interacting particles in
the final state and the underlying framework is more tractable in
radiative and semileptonic decays.
Precise measurements of the radiative and semileptonic decay branching
ratios and the related isospin and CP-asymmetries will test this
theoretical framework. We have given a fairly detailed
phenomenological profile of the radiative decays $B \to K^*
\gamma$ and $B \to \rho \gamma$ in this paper, and have pointed out some
phenomenological issues in this approach related to understanding
the current experimental branching ratios for $B \to K^* \gamma$ decays,
which will have to be resolved on our way to a completely quantitative
theory of exclusive radiative decays.

\section*{Acknowledgements}

We would like to thank L.T.~Handoko for collaboration in the early
stage of this work. We  thank
Christoph Greub, David London, Enrico Lunghi, and  N.V.~Mikheev for
helpful discussions.  A.A. would like to thank Cai-Dian L\"u for his 
kind hospitality at the Institute for High Energy Physics, Beijing. 
A.P. would like to thank the DESY theory group and Professor Reinhold 
R\"uckl for their kind hospitality in Hamburg and W\"urzburg, respectively. 
The work of A.P. was supported in part by the Russian Foundation for 
Basic Research under the Grant No.~01-02-17334, and in part by the 
German Academic Exchange Service~DAAD. We thank Martin Beneke,
Stefan Bosch and Gerhard Buchalla for pointing out several errors in the
earlier version of this paper, which we have  corrected and the
results modified accordingly. Finally, we thank Amjad Gilani for finding a
numerical error in Fig. 7 of this paper. The revised figures and numerical
estimates are presented in the text.

\end{document}